%% file: GDR3_microlensing.tex
\documentclass{aa}
\usepackage{graphicx}
\usepackage{txfonts}

\usepackage{hyperref}
\include{GEDR3_master_include}

\def\gaia{\textit{Gaia}\xspace}
\newcommand\gdr[1]{\gaia~DR#1}
\newcommand\egdr[1]{\gaia~EDR#1}

\def\gmag{$G$\xspace}
\def\gbp{$G_{\rm BP}$\xspace}
\def\grp{$G_{\rm RP}$\xspace}
\def\bpminrp{$G_{\rm BP}-G_{\rm RP}$\xspace}
\providecommand{\Msun}{\ensuremath{\,{M}_{\odot}}\xspace}
\def\deg{\ensuremath{^\circ}\xspace}
\providecommand{\pc}{\ensuremath{\,\rm pc}\xspace}
\providecommand{\kpc}{\ensuremath{\,\rm kpc}\xspace}

\def\ML{M_{\rm L}\xspace}
\def\piL{\pi_{\rm L}\xspace}
\def\piS{\pi_{\rm S}\xspace}
\def\piG{\pi_{\rm G}\xspace}
\def\murel{\mu_{\rm rel}\xspace}
\def\tE{t_{\rm E}\xspace}
\def\thetaE{\theta_{\rm E}\xspace}
\def\Fs{F_{\rm S}\xspace}
\def\Fb{F_{\rm B}\xspace}
\def\fs{f_{\rm S}\xspace}
\def\piEN{\pi_{\rm EN}\xspace}
\def\piEE{\pi_{\rm EE}\xspace}
\def\piE{\pi_{\rm E}\xspace}
\def\tmax{t_{\rm max}\xspace}

\hypersetup{
    colorlinks=true,
    linkcolor=blue,
    filecolor=blue,      
    urlcolor=blue,
    citecolor=blue,
    pdftitle={\gdr{3}: Microlensing Events from All Over the Sky}
    }

\titlerunning{\gdr{3}: Microlensing Events from All Over the Sky} 

\begin{document} 

\title{\gaia Data Release 3}
\subtitle{Microlensing events from all over the sky}

\author{{\L}.~Wyrzykowski\inst{\ref{instOAUW}}\fnmsep\thanks{\email{lw@astrouw.edu.pl}}
\and
K.~Kruszy{\'n}ska\inst{\ref{instOAUW}}
\and
K.~A.~Rybicki\inst{\ref{instOAUW}, \ref{Weiz}}
\and
          B.~Holl\inst{\ref{Sauverny},\ref{Ecogia}}
\and
          I.~Lec\oe ur-Ta\"ibi\inst{\ref{Ecogia}}
\and
          N.~Mowlavi\inst{\ref{Sauverny},\ref{Ecogia}}
\and
          K.~Nienartowicz\inst{\ref{Ecogia},\ref{Sednai}}
\and
          G.~Jevardat de Fombelle\inst{\ref{Ecogia}}
\and
          L.~Rimoldini\inst{\ref{Ecogia}}
\and
          M.~Audard\inst{\ref{Sauverny},\ref{Ecogia}}
\and
          P.~Garcia-Lario\inst{\ref{ESAC}}
\and
P.~Gavras\inst{\ref{rhea}}
\and
D.~W.~Evans\inst{\ref{Cambridge}}
\and
S.~T.~Hodgkin\inst{\ref{Cambridge}}
\and
          L.~Eyer\inst{\ref{Sauverny},\ref{Ecogia}}
}
\authorrunning{Wyrzykowski et al.}

 \institute{
 Astronomical Observatory, University of Warsaw, Al. Ujazdowskie 4, 00-478 Warszawa, Poland\label{instOAUW}
 \and
 Department of Particle Physics and Astrophysics, Weizmann Institute of Science, Rehovot 76100, Israel\label{Weiz}
\and
Department of Astronomy, University of Geneva, Chemin Pegasi 51, 1290 Versoix, Switzerland \label{Sauverny}
             \and
             Department of Astronomy, University of Geneva, Chemin d'Ecogia 16, 1290 Versoix, Switzerland\label{Ecogia}
             \and
             Sednai Sarl, 1204 Geneva, Switzerland \label{Sednai}
             \and
             European Space Astronomy Centre (ESA/ESAC), Villanueva de la Canada, 28692 Madrid, Spain \label{ESAC}
             \and
             RHEA for European Space Agency (ESA), Camino bajo del Castillo, s/n, Urbanizacion Villafranca del Castillo, Villanueva de la Ca{\~n}ada, 28692 Madrid, Spain\label{rhea}
            \and    
             Institute of Astronomy, University of Cambridge, Madingley Road, CB3 0HA, Cambridge, UK \label{Cambridge}
}

\date{Received 11 April 2022 / Accepted 13 June 2022}

 
  \abstract
   {One of the rarest types of variability is the phenomenon of gravitational microlensing, a transient brightening of a background star due to an intervening lensing object. Microlensing is a powerful tool for studying the invisible or otherwise undetectable populations in the Milky Way, including planets and black holes.}
   {We describe the first \gaia catalogue of candidate microlensing events, give an overview of its content, and discuss its validation. }
   {The catalogue of Gaia microlensing events was composed by analysing the light curves of around 2 billion sources of \gdr{3} from all over the sky covering 34 months, between 2014 and 2017.}
   {We present 363 Gaia microlensing events and discuss their properties. Of these, 90 have never been reported before and have not been discovered by other surveys. The contamination of the catalogue is assessed to 0.6\%-1.7\%.}
   {}

\keywords{Catalogues -- Methods: data analysis -- Stars: variables: Microlensing}

\maketitle


\section{Introduction \label{sec:introduction}}
The gravitational microlensing phenomenon occurs when the light of a background source star is bent by the foreground lens object. Its foundations lie in Einstein's theory of gravity \citep{1936Einstein}, although even Einstein himself was not convinced that the effect could be detectable at all. In part, Einstein was right because such almost-perfect cosmic alignments are extremely rare and the typical angular separations of the images in a single point lens in the Milky Way are of the order of milliarcseconds, and the sources are therefore hard to resolve. 

It was only 1986 when Bohdan Paczy\'nski first proposed to look for temporal brightening of stars from the Large Magellanic Cloud (LMC) in order to probe the dark matter halo of the Milky Way for massive astrophysical compact halo objects (MACHOs) \citep{1986Paczynski}. Following this proposal, multiple dedicated large-scale  microlensing surveys were launched, namely
MACHO \citep{1992ASPC...34..193A, 1993NYASA.688..612B}, OGLE \citep{1992AcA....42..253U}, EROS \citep{1993Msngr..72...20A}, MOA \citep{1997vsar.conf...75A, 2001Bond}, and KMTNet \citep{KMTNet}, which monitored the Large and Small Magellanic Clouds (LMC and SMC) and also the Galactic bulge and disc. The microlensing era began in earnest  with the first microlensing event discoveries \citep{AlcockNatureMACHO, Udalski1993BinaryOGLE}. 

The hopes for solving the dark matter problem through microlensing were first raised by the MACHO group \citep{1997Alcock, Alcock2001, 2000Alcock, 2005Bennett} who claimed there was an excess of microlensing events towards the LMC. If those events could be attributed to compact objects with masses below 1\Msun, they would compose up to 20\% of the dark matter halo. However, a later study based on the LMC and SMC EROS data \citep{2007Tisserand} did not see the excess of events and hence placed strong limits on the MACHO content in the halo.
Independent verification of this discrepancy came from the OGLE project \citep{2009Wyrzykowski, 2010Wyrzykowski, 2011aWyrzykowski, 2011bWyrzykowski}, which concluded that the MACHOs can be almost certainly ruled out for masses all the way up to about 10\Msun. This left only a tiny window for dark matter in the form of compact objects in the mass range between 10 and 30\Msun, the window recently populated by numerous detections using gravitational wave (GW) signals from merging black holes \citep{2021LigoVirgo}. This, in turn, raised new hopes that at least some of the GWs from black holes are due to dark matter in the form of primordial black holes \citep{1975Carr, 2016Carr, 2016Bird, 2017Garcia-Bellido}.

Microlensing as a tool was also useful in other directions on the sky. The lensing probability increases with the density of sources \citep{1994KiragaPaczynski, 2002Evans} and is the highest towards the Galactic bulge region of the Milky Way. Tens of thousands of microlensing events have been discovered to date by dedicated surveys continuing their operation through recent decades, such as OGLE \citep{2015Udalski} and MOA \citep{2008SakoMOA}, and initiated recently, namely KMTNet \citep{KMTNet}. Among those events, as predicted by \cite{1991MaoPaczynski}, signatures for planets accompanying the lensing star were discovered \citep[e.g.][]{2004Bond, 2005Udalski, 2006Beaulieu, 2012Cassan, Suzuki2016, Bennett2021, 2021Poleski} as well as candidates for free-floating planets \citep[e.g.][]{2011Sumi, 2017Mroz, 2018Mroz}. Moreover, the large samples of microlensing events allowed the selection of promising candidates for very rare cases of black hole and neutron star lenses \citep{2000Gould, 2002Mao, 2002Bennett, 2016Wyrzykowski, 2019Wiktorowicz, 2020WyrzykowskiBH, 2021Mroz, 2022SahuBH, 2022LamBH}. Other directions not towards the bulge were also systematically probed by dedicated microlensing surveys \citep[e.g.][]{2001Derue, 2009Rahal} with the most comprehensive and broad being the  recent study by \cite{Mroz2}.

In the two decades or so, microlensing events have also been   serendipitously discovered by large-scale photometric surveys not aiming at microlensing events, such as ASAS-SN \citep{ASASSN}, ZTF \citep{ZTF}, and amateur observers \citep[e.g.][]{2008GaudiTagoEvent, 2018Nucita, 2019Fukui}.
However, it was the \gaia space mission\citep{2016GaiaMission} that was the first to deliver candidate microlensing events  from all over the sky at a wide range of magnitudes. \gaia Science Alerts (GSA; \citet{Hodgkin2021}), was initiated at the very beginning of the operation of the \gaia mission in 2014 to identify ongoing astrophysical temporal phenomena in order to alert the astronomical community and to provide an opportunity to study them in detail. This included a very broad range of events, including supernovae, cataclysmic variables, tidal-disruption events, but also microlensing events. With its first microlensing event discovered in 2016, GSA has so far reported more than 400 candidate microlensing events. This includes spectacular cases such as Gaia16aye \citep{2020Wyrzykowski16aye}, a rotating binary lens event, Gaia18cbf \citep{2022Kruszynska}, one of the longest microlensing events ever found, and Gaia19bld \citep{2022Rybicki-bld,2022Bachelet}, a very bright event with the lens mass measured using space parallax, the finite source effect, and optical interferometry \citep{2021Cassan}.

\gaia has been collecting photometric, astrometric, and spectroscopic measurements of nearly 2 billion objects from all over the sky since 2014. The study and classification of objects with photometric time-series until 2017 is one of the main results of \gdr{3} \citep{DR3-DPACP-185}. 
In this work, we report the findings of a systematic search for microlensing events in the form of brightening episodes in the photometric time-series data from \gaia from years 2014-2017. 
We perform a statistical study of the collection, present its main properties and potential application for Galactic studies, and report findings from our investigations of individual lensing objects. 
The uniqueness of the \gaia microlensing catalogue is twofold. Firstly, for the first time, the entire sky has been homogeneously monitored for microlensing events. Secondly, \gaia has collected time-series of simultaneous three-band photometry for the events as well as astrometric and spectroscopic measurements, which will become fully available in the next \gaia data releases, leading to full solutions for the microlensing events reported here.

The paper is organised as follows. In Section \ref{sec:micro} we introduce the microlensing models, and in Section  \ref{sec:data} we describe the \gaia data and its pre-processing. Section \ref{sec:method} presents the methods used in order to build the catalogue. An overview of the results and the properties of the catalogue are presented in Section  \ref{sec:results}, while the validation of the sample and a discussion on individual cases are presented in Section  \ref{sec:discussion}.

The time-series data used in the paper as well as the parameters of the events are available in the online \gaia archive
at http://gea.esac.esa.int/archive/ which contains the table 
{\verb vari_microlensing } with its fields described in Appendix \ref{app:fields}. The detailed \gdr{3} documentation and catalogue data model are available at http://gea.esac.esa.int/archive/documentation/.

\section{Microlensing models}
\label{sec:micro}

In microlensing, a compact massive object distorts the space-time, forcing the light rays of a background source to change trajectory. In the case of a single point lens, this results in the appearance of two magnified images of the source \citep{Paczynski1996}, typically separated by less than 1 milliarcsecond \citep{2019Dong}. Each lensing configuration, that is, the distance to the lens (or lens parallax, $\piL$), the distance to the background source (or source parallax, $\piS$), and the lensing mass ($\ML$), defines an Einstein ring, with its angular radius:
\begin{equation}
    \thetaE = \sqrt{\frac{4G \ML}{c^2}(\piL-\piS)}~.
\end{equation}

Because of the relative proper motion of the observer, the lens, and the source ($\murel$), the source only becomes lensed temporarily when crossing the Einstein ring. As in most cases the two images are not resolvable, the total observed amplification ($A$) of the source light is the sum of the light of both images \citep{Paczynski1996} and is described only by one parameter, the projected separation between the source and the lens in units of the Einstein radius as a function of time ($u(t)$):
\begin{equation}
\label{eq:ampl}
A(t) = \frac{u(t)^2 + 2}{u(t) \sqrt{u(t)^2 + 4}}~.
\end{equation}

In the case of the simplest linear motion of the lens with respect to the source, the separation can be computed as
\begin{equation}
\label{eq:microu0}
    u(t) = \sqrt{\tau(t)^2 + \beta(t)^2}~, ~~~
    \tau(t) = \frac{t-t_\mathrm{max}}{t_E}, \,\, \beta(t) = u_0~,
\end{equation}
where $u_0$ is the minimal lens--source separation (impact parameter) at the moment of time $\tmax$.
The duration of the event, which is often referred to as the timescale of the event, or the Einstein time, $\tE$, depends on the size of the Einstein radius and the relative proper motion,  $\tE=\thetaE/\murel$. 
The bell-like shape of the photometric amplification curve resulting from Equation  \ref{eq:ampl}, is now often referred to as the `Paczynski curve'.

A standard photometric microlensing event is therefore defined as a temporal increase in brightness of a background point-like object or star. In practice, the total observed brightness of the source is a combination of the light of a background source of flux $\Fs$ amplified $A$-times and possible additional light $\Fb$ from unresolved sources within the resolution of the instrument, which might include or be the light of the lens:
\begin{equation}
\label{eq:mag}
\begin{aligned}
    mag(t) ={} & mag_0-2.5\log_{10}(A(t)\times \Fs + \Fb) = \\ 
    & mag_0 - 2.5 \log_{10}(\fs \times(A(t)-1)+1)~,
\end{aligned}
\end{equation}
where we introduce a blending parameter, $\fs=\frac{\Fs}{\Fs+\Fb}$, which defines a fraction of the light of the  source in the total observed flux, that is, $\fs=1$ for a non-blended source and a dark lens.
The $mag_0$ parameter is the observed brightness of the source and all the blends and is called the baseline magnitude.  As the microlensing effect is achromatic, the amplification does not depend on the wavelength. However, only source light is amplified and the spectral energy distribution (SED) of the blend(s) can,  in general, be different from that of the source; therefore, any modelling of multi-band data requires independent set of $mag_0$ and $\fs$ parameters.

In the case of events with timescales of longer than about 20 days, the annual motion of the  observer around the Sun can play a significant role and may distort the linearity of the projected lens--source motion \citep{2000Gould} in a measurable way.
In order to account for this effect, which is called microlensing parallax, equation \ref{eq:microu0} should be modified with the following: 
\begin{equation}
\label{eq:microparallax1}
    \tau(t) = \frac{t-\tmax}{\tE} + \delta t, \,\, \beta(t) = u_0 + \delta \beta,
\end{equation}
where
\begin{equation}
\label{eq:microparallax2}
    (\delta \tau, \delta \beta) = \piE \mathbf{\Delta s} = (\mathbf{\piE} \cdot \mathbf{\Delta s}, \mathbf{\piE} \times \mathbf{\Delta s})
\end{equation}
is the displacement vector due to parallax and $\mathbf{\Delta s}$ is the positional offset of the Sun in the geocentric frame \citep{2004Gould}. 
The microlensing model with annual parallax has to include a new parameter, the microlensing parallax vector, connected with the parallaxes of the lens, the source, and the Einstein radius, $\mathbf{\piE}=(\piL-\piS)/\thetaE$, decomposed into east and north directions, $\piEE$ and $\piEN$, respectively. 

The microlensing parallax is one of the crucial components needed for determination of the mass and the distance of the lens:
\begin{equation}
        M=\frac{\thetaE}{\kappa \piE}~,~~\kappa = \frac{4G}{c^2 \mathrm{au}} \approx 8.144 \frac{\mathrm{mas}}{\Msun}~,
        \label{eq:mass}
\end{equation}
\begin{equation}
    \piL = \thetaE \piE + \piS.
    \label{eq:dist}
\end{equation}


\section{Data}
\label{sec:data}

The results in \gdr{3} are based on the first 34~months of \gaia observations collected  between 25 July 2014 (JD=2456863.9375) and 28 May 2017 (JD=2457901.86389), and spanning a period of 1038~days. 
On average, in that period each star was scanned by \gaia about 40~times. However, this number strongly depends on the region of the sky following \gaia's scanning law \citep{2016GaiaMission}. In particular, the bulge region, where  most of microlensing events are expected, was observed typically only about 20~times.
Moreover, the data points were not evenly distributed, with pairs separated by 106~minutes because observations were conducted using two of \gaia's telescopes. 
There were also parts of the sky with an increased number of observations, located near the ecliptic latitudes of $\pm 45^{\deg}$, where the number of individual epochs per star often exceeded 100. 

In this work, we study the photometric \gmag, \gbp, and \grp time series in magnitudes and in Barycentric-corrected Julian days, the processing of which is described in detail in  \cite{Riello2021} and \cite{DR3-DPACP-162}. The Microlensing Specific Object Studies (MSOS) pipeline was run together with other variability studies on all \gdr{3} sources.
The photometric data were pre-filtered for outliers, as described in \cite{DR3-DPACP-162}, 
and only sources with at least ten observations (transits) in \gmag-band were investigated further. The \gbp and \grp band time series were investigated together with the \gmag-band only if there were at least 30 data points available in each band. 

As our selection criteria relied on the goodness of the microlensing model fit and the light curves of microlensing events often change over a wide magnitude range, we rescaled the photometric error bars provided with each \gaia data point in order to more closely reflect the expected scatter at each magnitude (e.g. \citealt{2009Wyrzykowski, 2016Skowron}). We used \egdr{3} mean \gmag versus its standard deviation to derive the expected lowest possible error for each \gmag measurement with its error bar of $\sigma^{old}_G$, assuming 30~observations on average per object, using the following formulae:

\begin{equation}
\label{equ:errorscalling}
\sigma^{exp}_G = 
    \begin{cases}
         \sqrt{30}\times10^{0.17\cdot13.5-5.1}, & ~~ \mathrm{for}~ \gmag<13.5~\mathrm{mag}\\
    \sqrt{30}\times10^{0.17\cdot G-5.1}, & ~~ \mathrm{for}~ \gmag>=13.5~\mathrm{mag}\\
    \end{cases}
    \end{equation}
\begin{equation}
    \label{equ:errorscalling2}
    \sigma^{new}_G =  \sqrt{\left(\sigma^{old}_G\right)^2 + \left(\sigma^{exp}_G\right)^2} \\
.\end{equation}
    
Despite our best efforts to identify and mask the data points outlying significantly from the overall trend of the \gbp and \grp light curve, some still remain in the data. In order to avoid the microlensing model being incorrectly driven by those outliers, we rescaled error bars in \gbp and \grp time series by a constant factor of 10. This allowed us to continue using the \gbp and \grp data points in the microlensing model and minimised the impact of the outliers. 


\section{Methods \label{sec:method}}

The microlensing events in \gdr{3} were selected among 1\,840\,947\,142 light curves processed by the \gaia DPAC Variability Processing Unit CU7 \citep{DR3-DPACP-162}. There were two samples created, Sample A based on an automated procedure and selection criteria tuned to \gaia light curves (163 events), and Sample~B, comprising an additional 233~events selected from the literature, with 33 overlapping with Sample A. The summary of all events, together with the coordinates in equatorial and Galactic frames, the baseline magnitude, and the sample name are presented in Table \ref{tab:all} of Appendix \ref{app:table}. Each \gaia source has a unique source identifier (sourceid), but we also introduce here alias names for the \gdr{3} microlensing events in the format GaiaDR3-ULENS-NNN, where NNN is the order number obtained after sorting our catalogue by event baseline magnitudes. Hereafter, we also refer to the events using their numbers (e.g. \#001).

The parameters of all 363~events are contained in the \gdr{3} {\verb vari_microlensing } table, and these are listed and described in Table \ref{tab:parameters} of Appendix \ref{app:fields}.

\subsection{Sample~A}

The selection of rare variability types such as microlensing events (typically less than 1 in a million stars, \citealt{Paczynski1996}) requires a very efficient approach, optimised to work on a large volume of data to be analysed. This was already an issue when the first microlensing searches were conducted on millions of stars in the 1990s (e.g. \citealt{1996Alcock, 2003Afonso, 1994UdalskiTau}), as well as in later times, when the number and length of light curves grew significantly in surveys like OGLE, MOA, and ZTF (e.g. \citealt{2000Wozniak, 2009Wyrzykowski, 2011aWyrzykowski, 2013Sumi, 2015Wyrzykowski, 2016Wyrzykowski, 2020Navarro1,2020Navarro2, Mroz1, Mroz2,2020MrozZTF, 2021RodriguezZTF}). 
The common approach among those studies was to first identify all potentially variable stars exhibiting a single episode of brightening, and then study the narrowed-down sample in more detail with more computationally expensive tests. Here, we also followed this philosophy. The main filtering process was divided into two parts, the first called Extractor, and the second being the Microlensing Specific Object Studies (MSOS). Details of the procedure are described in \gdr{3} Documentation, and below we provide a brief outline. 

The Extractor studied \gmag, \gbp, and \grp data points as well as statistical parameters derived from these time series. It conducted a series of simple and quick tests, in particular, testing whether or not there is a significant number of points brighter than the median magnitude; testing whether or not there is a brightening episode present in all three pass-bands; checking for a unique rise and fall; and checking whether or not the light curve outside of the episode can be considered constant. The Extractor is described in more detail in \gaia DR3 Documentation. 

Before the next stage, MSOS, there were 98\,750\,637 sources left for further investigation, which is less than 10\% of the original input list. 
The second part of the pipeline performed a microlensing model fit to the data with the re-scaled error bars, as described in Section \ref{sec:data}. The modelling was divided into substages, dubbed {`Levels'}. 

Level~0 was the simplest Paczynski model without blending and no annual parallax effect included (Equation  \ref{eq:microu0} and Equation  \ref{eq:mag} with $f_S\equiv1$). This model is a good approximation for many microlensing events, especially in less crowded regions of the sky and for faint or dark lenses. This model is also the most mathematically stable. It contains three parameters common for each band (\texttt{paczynski0\_u0}, \texttt{paczynski0\_te}, \texttt{paczynski0\_tmax}), and one additional parameter per band (\texttt{paczynski0\_g0}, \texttt{paczynski0\_bp0}, \texttt{paczynski0\_rp0}). 

Level~1 was the standard microlensing model (i.e. with linear relative motion, using Equation  \ref{eq:microu0}) which included blending parameters for \gmag-band, \gbp, and \grp if enough data points were available in those bands (Equation  \ref{eq:mag} and $f_S$ defined for each band). The blended model fitting is less stable, in particular for light curves with poorly sampled wings of the microlensing event.  
The fitting of microlensing parameters in  Level~1 (\texttt{paczynski1\_u0}, \texttt{paczynski1\_te}, \texttt{paczynski1\_tmax}), and baselines in each band (\texttt{paczynski1\_g0}, \texttt{paczynski1\_bp0}, \texttt{paczynski1\_rp0}) was initialised with the values obtained at Level~0. Each band had also additional blending parameters initialised for the fitting at 1 (\texttt{paczynski1\_fs\_g}, \texttt{paczynski1\_fs\_bp}, \texttt{paczynski1\_fs\_rp}).

Level~2 model included annual parallax (Eqs \ref{eq:microparallax1} and \ref{eq:microparallax2}), but excluded blending. Blending and microlensing parallax both cause modifications to the standard microlensing light curve, particularly in the region of the wings of the curve. In the case of sparse \gaia light curves, the degeneracy between these two parameters can be severe and can prevent the quick convergence of the minimisation procedure.

This Level~2 model contained the standard microlensing parameters as in Level~0 (\texttt{paczynski2\_u0}, \texttt{paczynski2\_te$, $paczynski2\_tmax}) and two additional ones describing the parallax vector in north and east directions  (\texttt{paczynski2\_parallax\_north}, \texttt{paczynski2\_parallax\_east}), as well as baseline magnitudes for each of the bands (\texttt{paczynski2\_g0}, \texttt{paczynski2\_bp0}, \texttt{paczynski2\_rp0}). 
The Level~2 model was fit only for sources with Level~0 timescale \texttt{paczynski0\_te}$>20$~days, as the microlensing parallax is not going to be detectable in shorter events. 

Finally, all the parameters derived for the models at all Levels, together with their goodness of fit ($\chi^2/dof$) and other statistical parameters (e.g. skewness of the magnitude distribution, Abbe value \citep{1941Neumann}, amplitudes, etc.) computed for the light curves during the Variability Processing \cite{DR3-DPACP-162}, were used to derive the membership score and define a set of filters, which were used to select candidates for microlensing events.
Details of the selection criteria are described in the \gdr{3} Documentation and briefly in the Appendix. 
This automated procedure returned 324~candidate events, which were then visually inspected by three independent experienced microlensers and the final sample was composed of 163~events ({\bf Sample~A}).
We note that the Level~2 parameters, along with parameters of the Extractor used in the sample selection, were not included in the {\verb vari_microlensing } table of the \gdr{3} release. 

\subsection{Sample~B}

In the time period of \gdr{3}~(2014-2017), there were other large-scale surveys in operation designed to detect microlensing events. We identified events primarily from the OGLE-IV survey, published separately for the bulge region \citep{Mroz1} and the southern Galactic disc \citep{Mroz2}. 
From the most numerous OGLE-IV sample (6251 events), we selected 1527 events that had reported timescales of longer than 20~days and their maximum was between JD=2,456,863.9375 and JD=2,457,901.86389, which is in the DR3 data span. 

In the \gdr{3} data, we also found two bright microlensing events reported by the ASAS-SN survey \citep{ASASSN}, namely ASASSN-16oe \citep{2016Asassn16oe1, 2016Asassn16oe2} and ASASSN-V~J044558.57+081444.6 \citep{AntiCenter2017}.
There were also seven~microlensing events identified by the \gaia Science Alerts system \citep{Hodgkin2021} that occurred during \gdr{3} and were added to this sample: Gaia16aua, Gaia16aye \citep{2020Wyrzykowski16aye}, Gaia16bnn, Gaia17ahl, Gaia17aqu, Gaia17bbi, and Gaia17bej. 

The cross-match of the selected events with DR3 within 0.5 arcsec yielded a total of 1074~sources; these have been identified as \gaia sources. Their time series in \gmag, \gbp, and \grp were then processed through the Extractor and MSOS Pipeline and 404 of them were successfully recovered. However, the cuts and filters used for defining Sample A were not applied for this sample. Instead, a visual inspection of the candidates was conducted and 233~events were selected. We required that the event be clearly visible in the \gaia data and contain at least two points within the brightening so that the parameters could be determined robustly. 
As {\bf Sample~B} was composed independently from {\bf Sample~A}, the samples overlapped in 33~cases. The \gdr{3} catalogue contains the same set of parameters for both samples. 

\subsection{Cross-match with known events}\label{sec:xm}

Sample~A of events was obtained in an independent procedure and contained 33~events which also ended up in Sample~B. In order to assure we identified all possible cross-matches between \gaia-discovered events and previously known events, we performed an additional match on the coordinates between Sample~A and all available catalogues of candidate microlensing events. 
The result of the cross-match is provided in 
Table  \ref{tab:xm} in the Appendix, which gathers all matches between \gaia-combined Samples~A and~B and other microlensing surveys and catalogues containing microlensing events. This includes OGLE-IV bulge \citep{Mroz1} and disc \citep{Mroz2} events, the OGLE-IV Early Warning System webpages \citep{2015Udalski}, KMT-Net \citep{KMTNet}, MOA \citep{MOA}, ASAS-SN \citep{ASASSN}, and GSA-reported microlensing events \citep{Hodgkin2021}.
 Of 363 events, a total of 273 were alerted by one or more of the above-mentioned surveys (227 cases matched to OGLE-IV alone), and therefore 90 of the events reported here are completely new.

\section{Results \label{sec:results}}
We compiled a catalogue of 363~candidates for microlensing events among nearly 2~billion stars monitored by \gaia in the years 2014-2017. Below we describe the properties of the catalogue.

\subsection{Events location}

\begin{figure}
    \centering
    \includegraphics[width=\hsize]{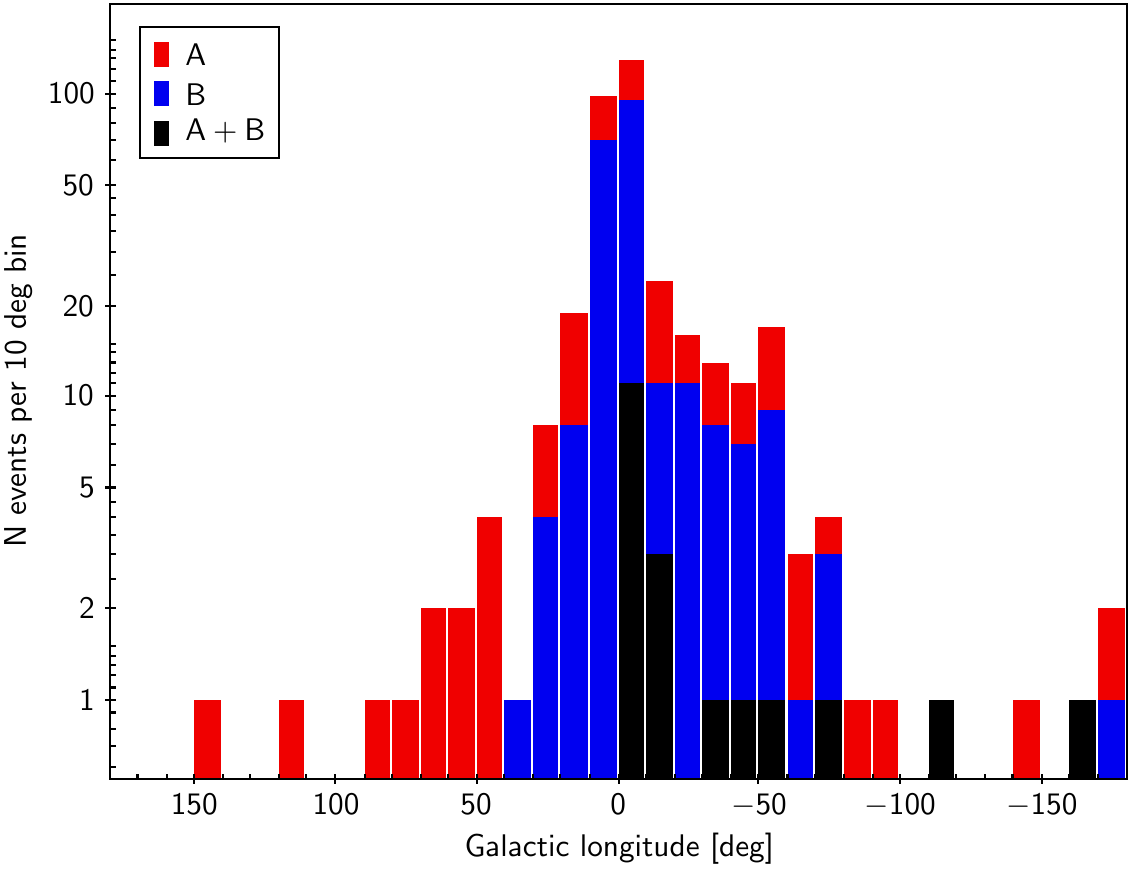}
    
    \includegraphics[width=\hsize]{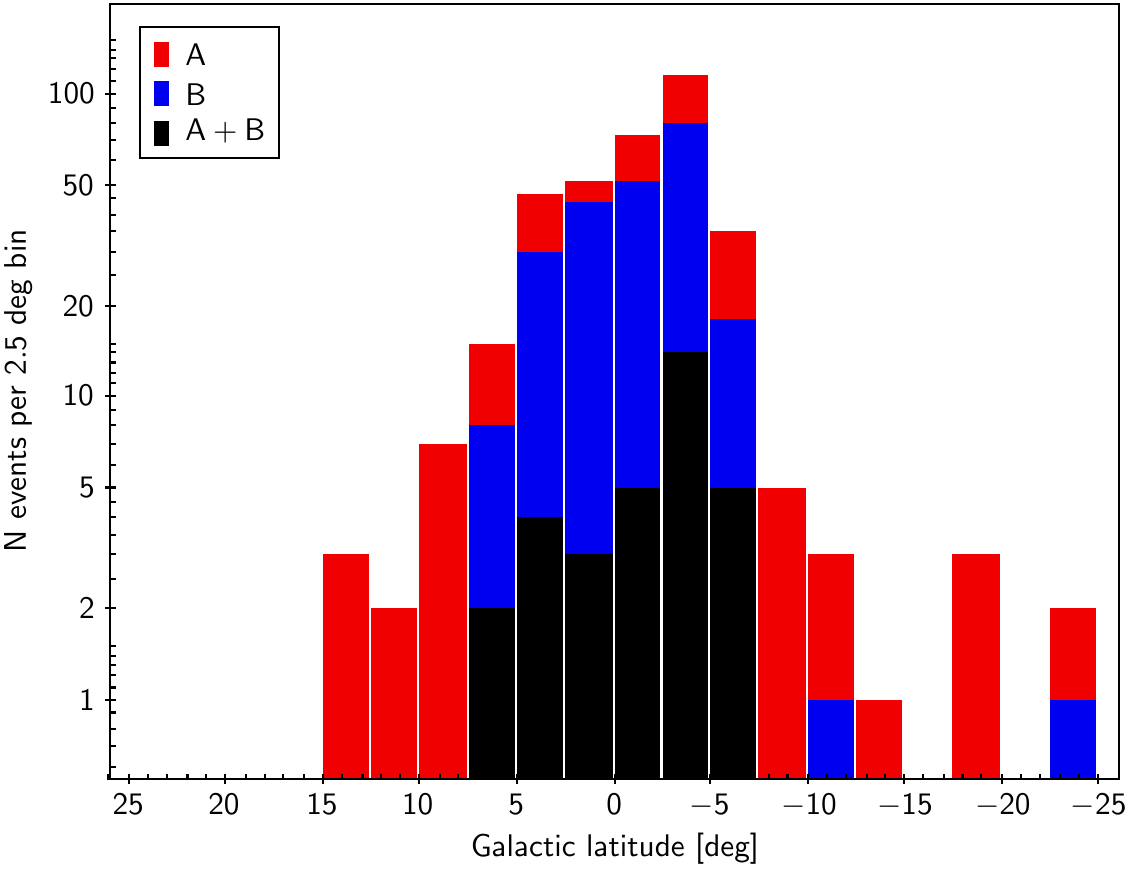}
    \caption{Distribution of Gaia microlensing events over the sky in Galactic longitude (top panel) and latitude (bottom panel). Events from samples A and B and their overlap are marked with different colours. }
    \label{fig:histgal}
\end{figure}

As expected, the majority of the events, that is 228~events (63\%), are located towards the Galactic bulge, defined hereafter in Galactic coordinates as $|l|<10$\deg (see Figure  \ref{fig:histgal}), with the highest concentration of events being found at $|b|<7$\deg. This is also where most of the events  identified before \gdr{3} (from Sample~B) are found. The density of events then roughly follows the density of stars, concentrating on the Galactic plane to within 5 to 15 degrees. 
The Disc sample, defined hereafter as events with Galactic longitudes $|l|>10$\deg, contains 135~events (37\%) and is mostly concentrated within $|l|<50$\deg. 

\begin{figure*}
\centering
 \includegraphics[width=\hsize]{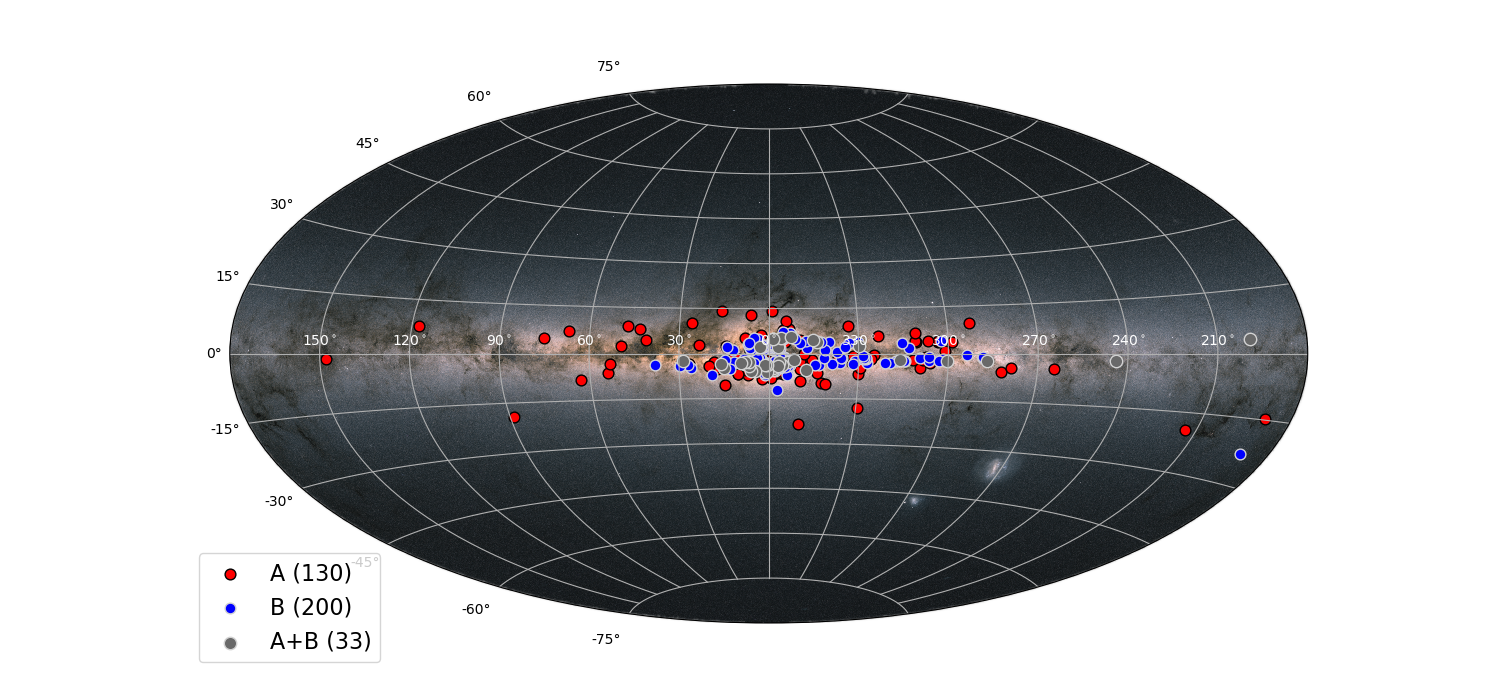}
  \caption{Distribution of the microlensing events on the sky in Galactic coordinates in degrees over-plotted on the \egdr{3} sky map.}
  \label{fig:mainmap}
\end{figure*}

\begin{figure*}
\centering
 \includegraphics[width=\hsize]{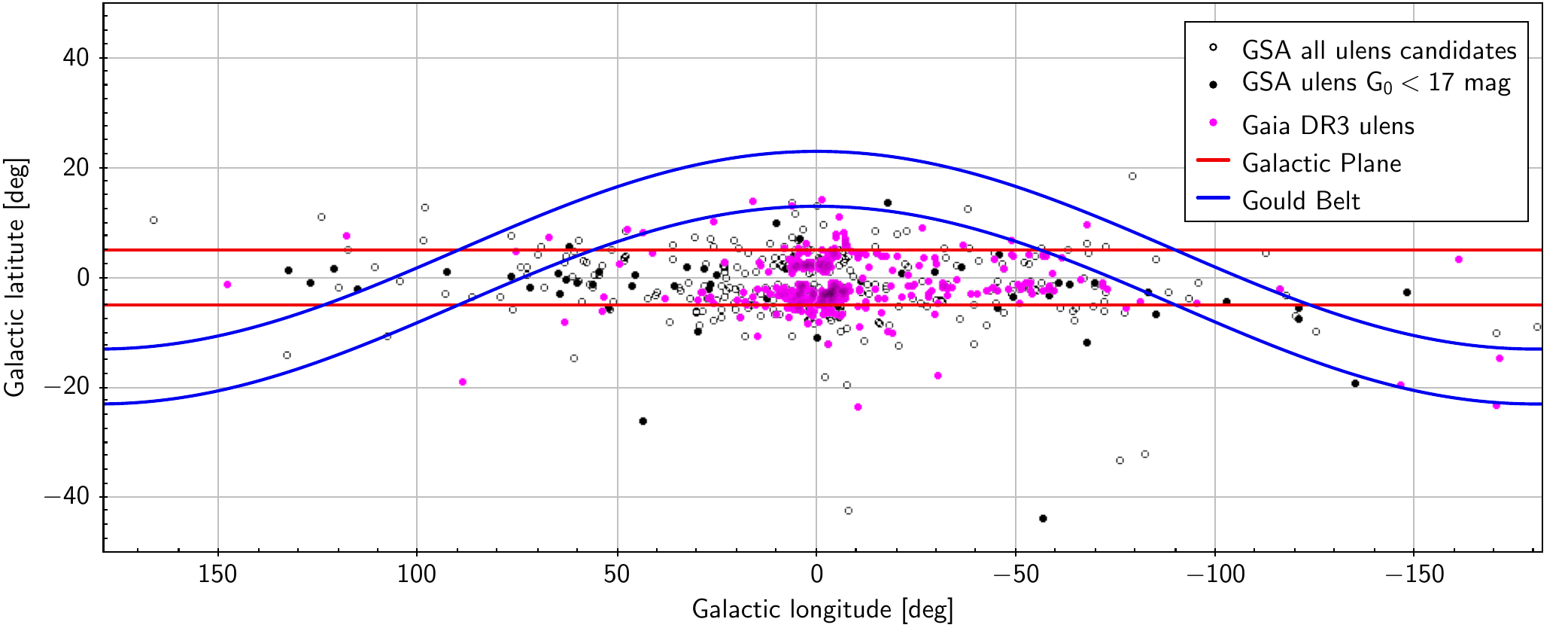}
  \caption{\gdr{3} microlensing candidates together with 343~candidates for microlensing events identified by GSA in the years 2016-2021. The red belt denotes the Galactic plane, while the blue lines roughly mark the location of the Gould Belt.}
  \label{fig:gouldbelt}
\end{figure*}

However, there are also individual events at very high Galactic longitudes, reaching as far as the Galactic anticentre (Figure \ref{fig:mainmap}).
The events with their Galactic coordinates not following the Galactic plane are located in the regions where the density of stars potentially acting as background sources is very low, and therefore the microlensing rate is also very low. 
Field microlensing events are not unexpected \citep{1998Nemiroff, 2008Han} and their rate in a survey like \gaia (assuming survey depth of $V\sim19$~mag) is about 0.001~events per year per degree$^2$.
The presence of events with Galactic latitudes of about 20\deg is therefore not surprising. 

In Figure \ref{fig:gouldbelt}, we plot the locations of all \gdr{3} events together with 343~candidate microlensing events identified by GSA \citep{Hodgkin2021} in the years 2016-2021. There were only seven events in common between \gdr{3} and GSA (listed in Sample~B description). In Figure \ref{fig:gouldbelt}, we emphasise those of GSA candidates, which are brighter at the baseline than \gmag$=17$ mag.
The red lines mark the $\pm$5\deg from the Galactic plane, while the blue lines roughly mark the extent of the Gould Belt, a great circle on the sky drawn by an over-density of young stars and active star formation regions \citep{1987Taylor, 2009Bekki}. We note that the distribution of microlensing events in the southern hemisphere shows a possible over-density that coincides with the Gould Belt. This could suggest that at least some of the events in those parts of the sky could be due to lenses originating from the enhanced stellar density in the Belt, with sources from the background Galactic disc. 

We find no events in or near the Magellanic Clouds. Despite initial claims of an excess of events and their possible origin in low-mass halo, dark-matter compact objects \citep{1997Alcock, 2000Alcock, 2005Bennett}, the event rate in that direction has now been established to be very low and consistent with events caused by the lenses from within the Clouds \citep{1994Sahu, 2007Tisserand, 2011bWyrzykowski}.

\subsection{Properties of the events}

\begin{figure}
    \centering
    \includegraphics[width=\hsize]{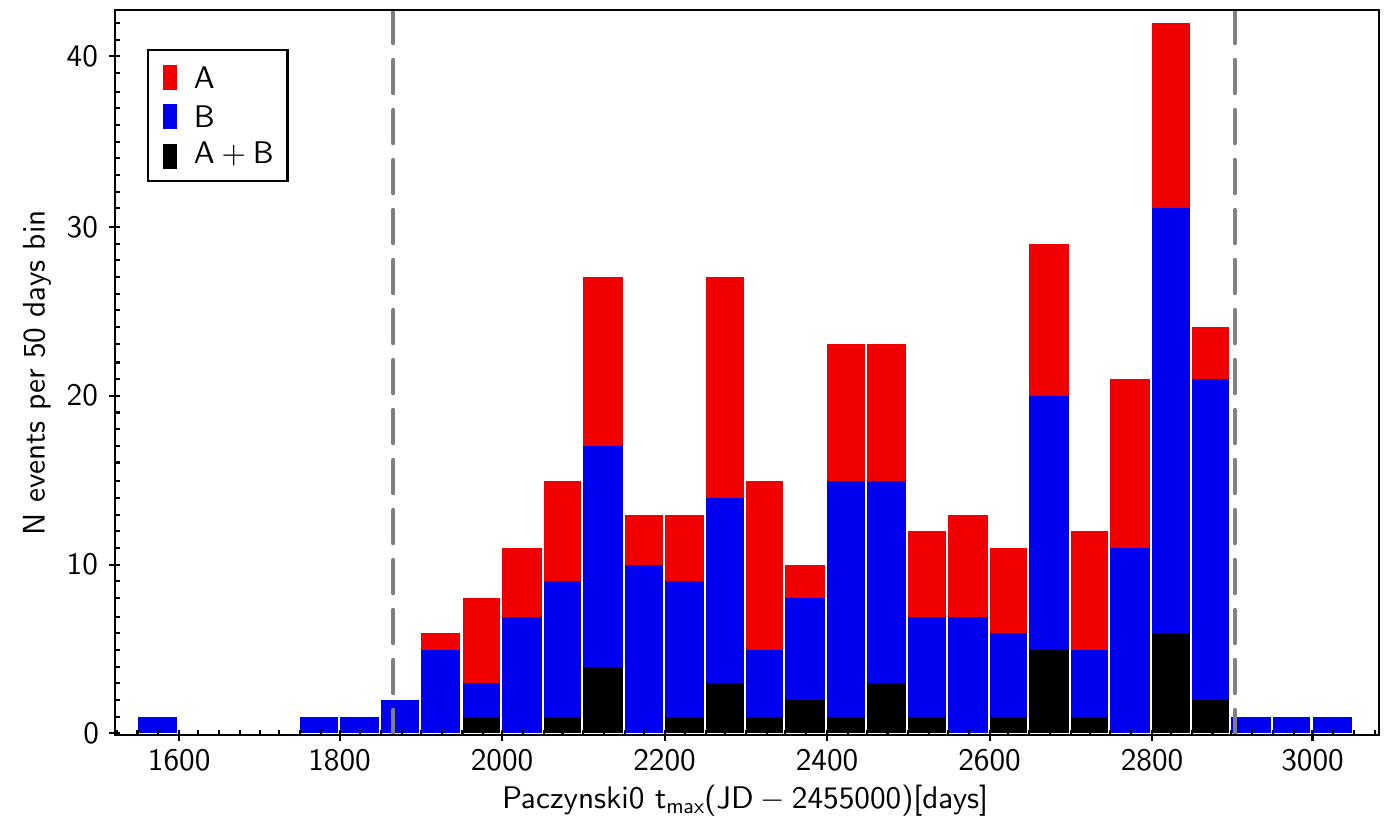}
    \caption{Distribution of the Level~0 parameter for the time of the maximum ($t_{max}$) of \gdr{3} microlensing events. 
    The dashed lines indicate the \gdr{3} data time-span.}
    \label{fig:histt0}
\end{figure}

\begin{figure}
    \centering
    \includegraphics[width=\hsize]{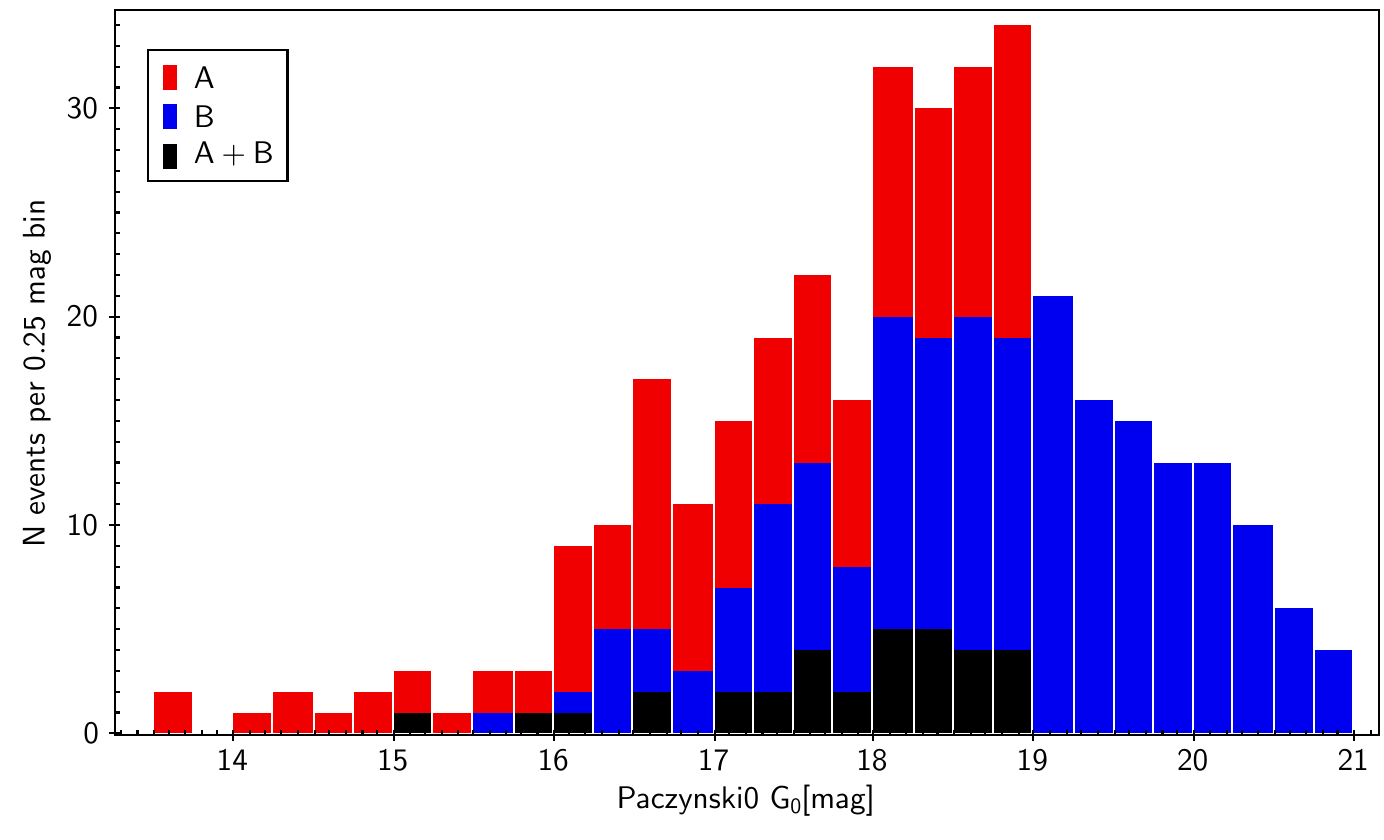}
    \caption{Distribution of Level~0 \gmag baseline magnitudes of \gdr{3} microlensing events.}
    \label{fig:histmag0}
\end{figure}

\begin{figure}
    \centering
    \includegraphics[width=\hsize]{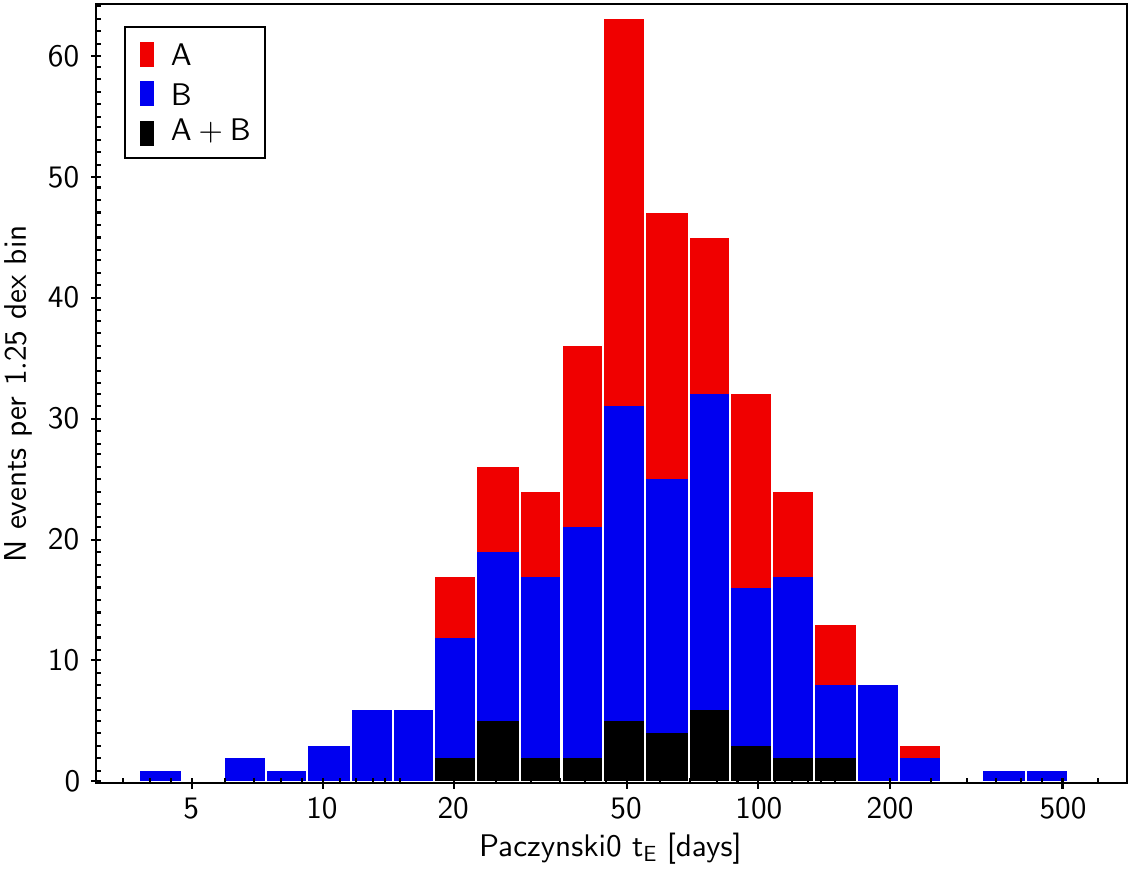}
    \caption{Distribution of timescales for \gdr{3} microlensing events obtained in the simplest Level~0 microlensing fit.}
    \label{fig:histte}
\end{figure}

Figure \ref{fig:histt0} shows the distribution of the times of the maximum ($\tmax$) of events from Sample~A,~B, and their overlap~(A+B) from the Level~0 microlensing model fit. The application of filtering on the moment of the maximum of an event resulted in the cuts seen for Sample~A, where events found automatically have been restricted to fall within the \gdr{3} data time-span. This filter was not applied for Sample~B (events from the literature with sufficient \gaia data to derive their parameters) and there are some events with their maxima outside of the \gdr{3} time-span.
All of these are the events with timescales longer than 100~days, and therefore their light curves are still well defined within the \gdr{3} data time window.

The distribution of \gmag baseline magnitudes found in the fitting procedure at Level~0 ({\verb paczynski0_g0 }) is shown in Figure \ref{fig:histmag0}. 
The full sample contains events with a baseline spanning a very wide range of magnitudes, from 13 to 21~mag. However, due to filtering of events fainter than 19~mag in the baseline, the Sample~A does not contain fainter events, while the Sample~B contains 98~events between 19 and 21~mag in \gmag, with the faintest baseline being 20.97~mag for GaiaDR3-ULENS-363/OGLE-2017-BLG-0116/ KMT-2017-BLG-1029 (sourceid=4067300223660786304). 
At the other end of the magnitude scale, there is a number of events with very bright baselines, with as many as 65~events brighter than \gmag$=17$~mag. The brightest in the sample (sourceid = 6059400613544951552) has a baseline of \gmag=13.58~mag and therefore it holds an alias name GaiaDR3-ULENS-001.
Such bright events are going to be the most suited for \gaia astrometric time-series studies, as shown in \cite{2018Rybicki}.
Of eight events brighter than \gmag$=15$ mag, six are new events that have not been seen by any other survey (see Section \ref{sec:xm}).

Figure \ref{fig:histte} shows the distribution of Level~0 timescales ({\verb paczynski0_te }) for Sample~A, Sample~B, and the overlap between the samples. The Level~0 timescale, when multiplied by a factor of about 6, is a good indicator of the event duration, that is, the time during which the magnification caused a significant deviation from the baseline. 
For Sample~A, the shortest timescale is about 20~days and the longest is 250~days; these ranges of timescales are set at the filtering process to generate that sample. 
Events in Sample~B were not restricted by their timescales and there are single and most likely unrealistic cases as short as 1.5~days and as long as 5000~days. 
However, most events have a timescale of about 50 to 70~days.

\subsection{Source and lens distance from \gaia}

\begin{figure*}
    \centering
    \includegraphics[width=\hsize]{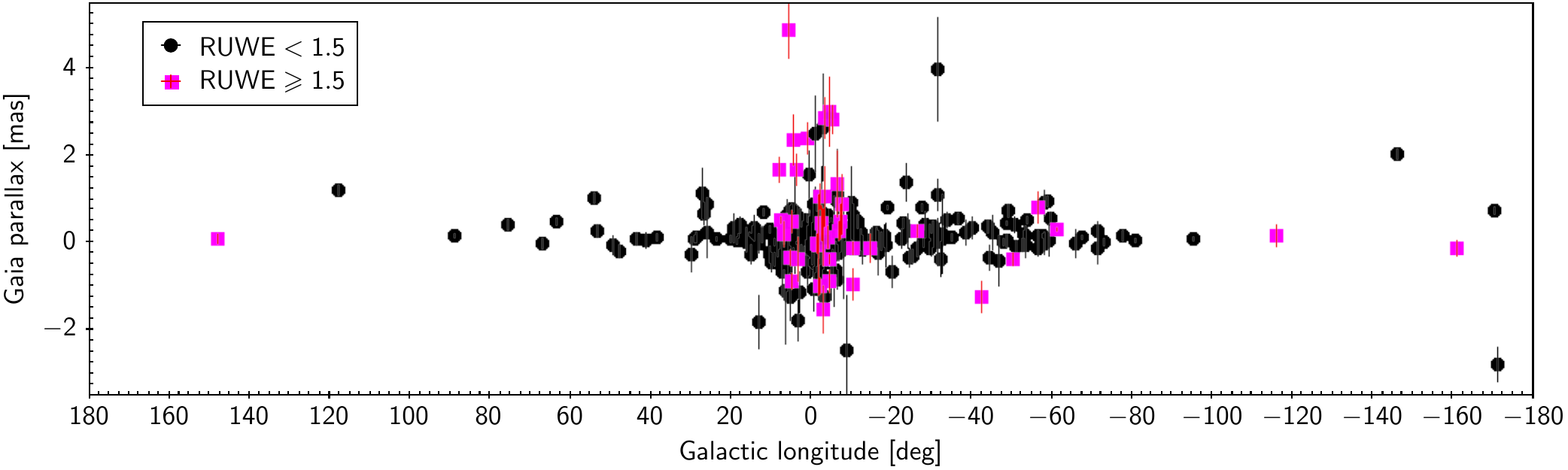}
    \caption{\egdr{3} geometric parallaxes for \gaia sources with microlensing events, split by the reliability of the parallax using the RUWE parameter. Circles denote RUWE$<1.5$, squares denote RUWE$\geqslant1.5$. }
    \label{fig:parallax-gall}
\end{figure*}

\begin{figure*}
    \centering
    \includegraphics[width=\hsize]{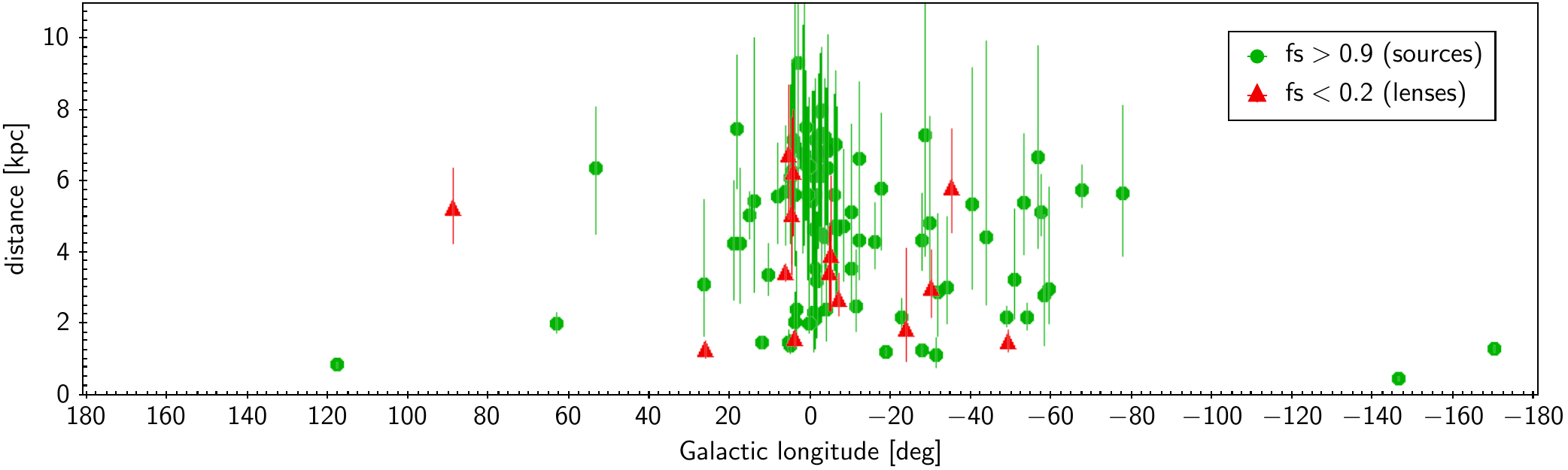}
    \caption{\egdr{3} geometric distances from \cite{2021Bailer-Jones} for sources and lenses in \gaia events, filtered on RUWE$<1.5$ and \texttt{parallax\_over\_error}$>1$ and divided into source-dominated events (\texttt{paczynski1\_fs\_g}$>0.9$, green circles) and lens-dominated events (\texttt{paczynski1\_fs\_g}$<0.2$, red triangles. }
    \label{fig:distance-gall}
\end{figure*}

Figure \ref{fig:parallax-gall} shows the parallaxes measured by \egdr{3} for sources where microlensing events were identified. We split the A and B samples into sources with reliable and unreliable astrometric measurements. We used the RUWE parameter \citep{2021Lindegren} as an indicator for sources where the parallax value should not be trusted (other indicators include Astrometric Excess Noise from \gaia or astrometric fidelity from \citealt{2022Rybizki}).
Most of the non-reliable parallax measurements are located in the densest parts of the Galactic plane, in the bulge. The negative parallaxes are also very common there because most sources are at a relatively large distance and the precision of their parallaxes is insufficient because of crowding and a very low number of epochs, which in \gdr{3} is typically less than 30. 

Among 363~events, there were 346 sources for which a distance was computed by \cite{2021Bailer-Jones} using \egdr{3} parallaxes.
Within that sample, we selected those sources that had RUWE$<1.5$ and \texttt{parallax\_over\_error}$>1$ to guarantee the geometric parallax measurement was robust and the astrometry was not affected by any effects, such as binarity \citep{2020Belokurov} or astrometric microlensing \citep{2000DominikSahu, 2002Belokurov, 2018Rybicki, 2018Kluter}.

However, in order to correctly utilise \gaia distances in applications to microlensing events, the blending parameter in \gmag-band, \texttt{paczynski1\_fs\_g}, is required. This parameter informs us about the amount of light that is magnified in a microlensing event, with values close to 1 indicating no or very little additional constant light. The extra light might originate from an unassociated star or, most naturally, from the lens itself. In the densest stellar regions of the Galaxy, towards the bulge, the blend is likely to be a random star unrelated to the event and located within the resolution of the instrument. In turn, in the less crowded Galactic plane disc region, it can be quite safely assumed that the blended light comes from the lens. 

Relying on the hypothesis that the blending light is the light of the lens, we can use \gaia parallaxes to derive some of the parameters of the microlensing events. In a general case, the parallax measured by \gaia, $\piG$, is a linear combination of source parallax, $\piS$, and lens parallax, $\piL$: $\piG = \piS \fs + \piL (1-\fs)$. This is also valid analogously for the proper motions. 
If the blending parameter is near 1, meaning most of the light seen in the baseline is coming from the lensed source, then the astrometric parallax and proper motion signal are primarily induced by the motion of the source. On the other hand, with the blending parameter value approaching 0, the dominant light is that of the blend (lens). Intermediate values of $\fs$ are mixed cases and \gaia parallaxes can be used to derive lens parallax only if the source distance is measured with other means. In the case of microlensing events, the fact that  only the source light is amplified during the event can be used in spectroscopic follow-up observations carried out on an ongoing event near its maximum brightness and hence the spectroscopic distance to the source can be estimated (e.g. \cite{2022Bachelet}). 

Figure  \ref{fig:distance-gall} shows distances as derived by \cite{2021Bailer-Jones} based on \egdr{3} data for events with extreme values of their blending parameter, \texttt{paczynski1\_fs\_g}, as a function of Galactic longitude. Only objects with RUWE$<1.5$ and \texttt{parallax\_over\_error}>1 were considered. The sample was divided into 89 source-dominated events (\texttt{paczynski1\_fs\_g}$>0.9$) and 14 lens-dominated events (\texttt{paczynski1\_fs\_g}$<0.2$). The bulk of sources in events located towards the Galactic bulge are clearly located in the bulge itself at distances of 6-9~\kpc. There is also a number of sources at much shorter distances of about 2-4~\kpc. In the disc fields, the distances are spread from 1 to 8~\kpc. Of particular importance are the three events at large Galactic longitudes (\#32, \#137 and \#153), with source distances estimated to $\sim$1~\kpc. If the blending parameter in those cases was determined correctly, this suggests that very nearby sources indicate even closer lenses. This would be somewhat surprising for events located in the very low-stellar-density regions of the sky, unless these events can be associated with the over-density of the Gould Belt (Figure \ref{fig:gouldbelt}). Two of these three 
(\#32, 3292310414861319936 at 1300~\pc and
\#137, 3012224323198211968 at 500~\pc)
are indeed located on the sky within the southern part of the Gould Belt.
The northern event (\#153, 2214532279378994816 at 900~\pc) is positioned far from both the Gould Belt and fairly far from the Galactic plane (b$\sim$7~\deg), and therefore seems to be a genuine field event \citep{2008Han}.
We note that the light curve of event \#137 shows a strong signature of microlensing parallax and a more detailed analysis of this event is necessary to derive both the microlensing parallax and the correct blending parameter. Moreover, event \#137 could actually be a young stellar object (YSO; see Section  \ref{sec:contam}); in such a case its distance would agree with its association to the Gould Belt. 

There are 14~events in which the blending parameter indicated strong domination of the light from the lens in the microlensing curve. In such cases, it is possible that the \gaia parallax is the actual measurement of the parallax of the lens and the distance to the lens can be directly obtained. Figure \ref{fig:distance-gall} shows their distances from \cite{2021Bailer-Jones} as a function of Galactic longitude. There is a range of lens distances in that sample, from objects at 1-2~\kpc to 6~\kpc, suggesting that lensing originates from both the Galactic disc and the bulge. We note that, if the microlensing parallax $\piE$ was detected and measured in such microlensing events, and the source distance $\piS$ was known from spectroscopy for example, the size of the Einstein radius, and therefore the mass of the luminous lens (Equation \ref{eq:mass}), could be determined, as $\thetaE = \frac{\piL-\piS}{\piE}$. The microlensing parallax determination for the events in this sample requires a detailed and dedicated study, which is beyond the scope of this paper. 

\subsection{Colour--magnitude diagrams}

Colours and colour--magnitude diagrams (CMDs) of microlensing events can be useful in studying the populations of sources, lenses, and individual events. In a genuine microlensing event, the colour, in principle, should not change throughout the event, unless there is blending with an object that has a different colour from that of the source. The colour of a \gaia microlensing event could be obtained from the \gbp and \grp time series, but there were only 155~events for which there are at least 30~data points available in both bands. 
Moreover, in a couple of cases, the fitting procedure for \gbp or \grp time series failed to find the correct values of the baselines (e.g. \#284, 4061109904516035840). 
On the other hand, there were 355 events for which \gdr{3} \bpminrp colours were derived from integrated \gbp and \grp fluxes from the \gaia on-board spectrophotometers \citep{2016GaiaMission}, which were computed as means based on the entire \gdr{3} time series. Any chromatic changes in microlensing events would naturally affect the colour, but in most cases it can be assumed as a good proxy of the colour.

\begin{figure}
    \centering
    \includegraphics[width=\hsize]{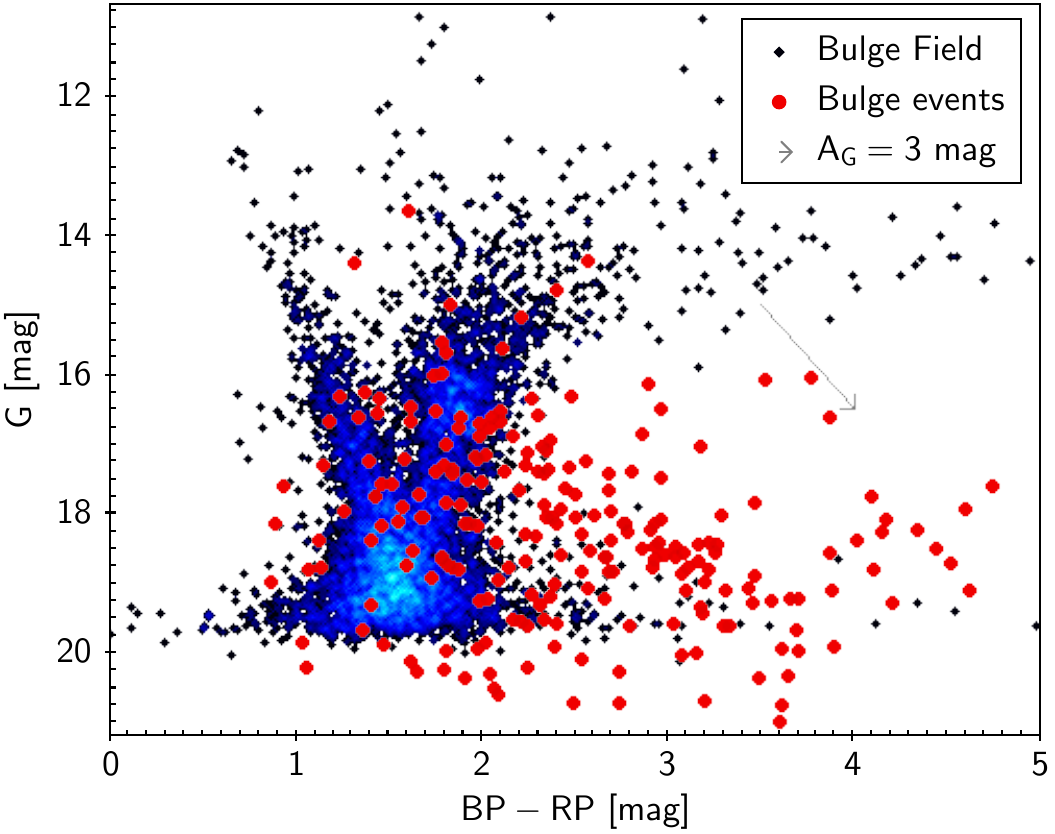}
    
    \includegraphics[width=\hsize]{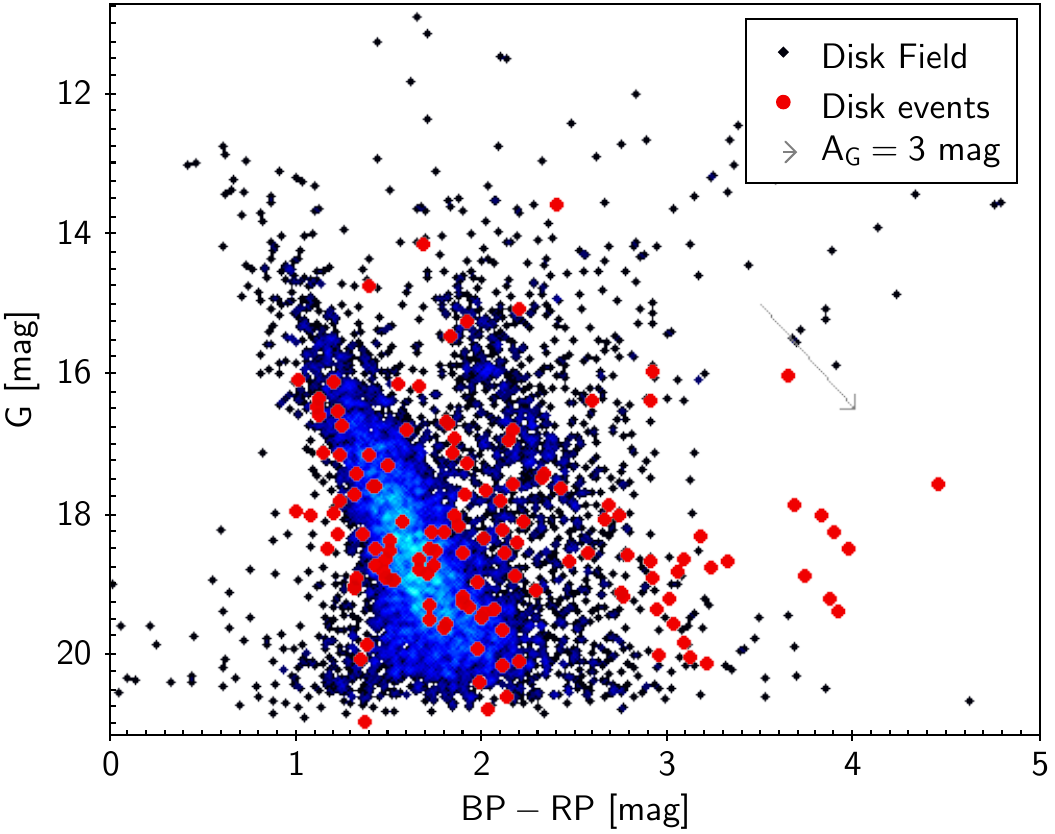}
    \caption{Colour--magnitude diagrams as observed by \gaia (black/blue points) for a typical bulge field (upper panel) and typical disc field (lower panel). The positions of microlensing events are displayed based on their average \bpminrp colour from \gdr{3} and their baseline \gmag magnitude from Level~0 fit. }
    \label{fig:cmd_observed}
\end{figure}

The plots in Figure \ref{fig:cmd_observed} show the observed colour--magnitude diagrams for the bulge ($|l|<10$\deg) and disc ($|l|\geqslant10$\deg) regions. The background shows distributions of mean magnitudes and colours of stars in typical fields for each of the regions. Over-plotted are the microlensing events for which the \bpminrp colours were available in \gdr{3} (355~sources). The selection cut for Sample~A required the median BP minus median RP to be between 1 and 4~mag and baseline magnitude brighter than \gmag$=$19~mag (see Appendix \ref{app:selection}); therefore, all faint and very red events belong to Sample~B. For the bulge region, there are many events located outside of the main locus of stars and their density roughly follows the direction of the interstellar extinction. An additional probable reason for the objects located at red colours is the blending of the source with a much redder lens or blend, or vice versa. For the disc, as expected, the effect of the extinction is not as pronounced as in the bulge and the events roughly follow the two main loci where the disc stars are concentrated.

The colour and the baseline magnitude are affected by  blending in a similar way to the \gaia parallax, as discussed earlier. This means that the CMD locations of events can only be interpreted correctly in
extreme cases of blending parameters. Therefore, for the events with all blending parameters (\texttt{paczynski1\_fs\_g}, \texttt{paczynski1\_fs\_bp}, \texttt{paczynski1\_fs\_rp}) larger than about 0.9  it can assumed that their observed quantities are related to the microlensed source star. In the opposite case, where blending is smaller than about 0.2, the magnitudes and colours would refer to the blend and possibly to the lens itself. 

\begin{figure}
    \centering
    \includegraphics[width=\hsize]{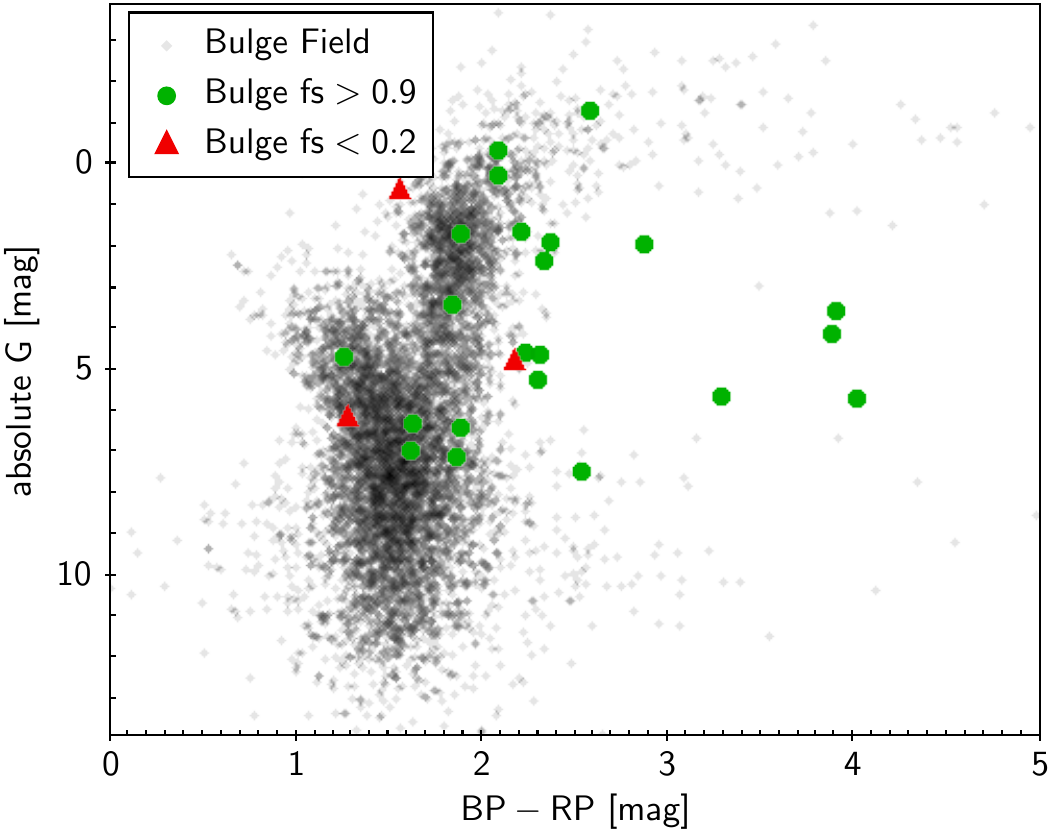}
    
    \includegraphics[width=\hsize]{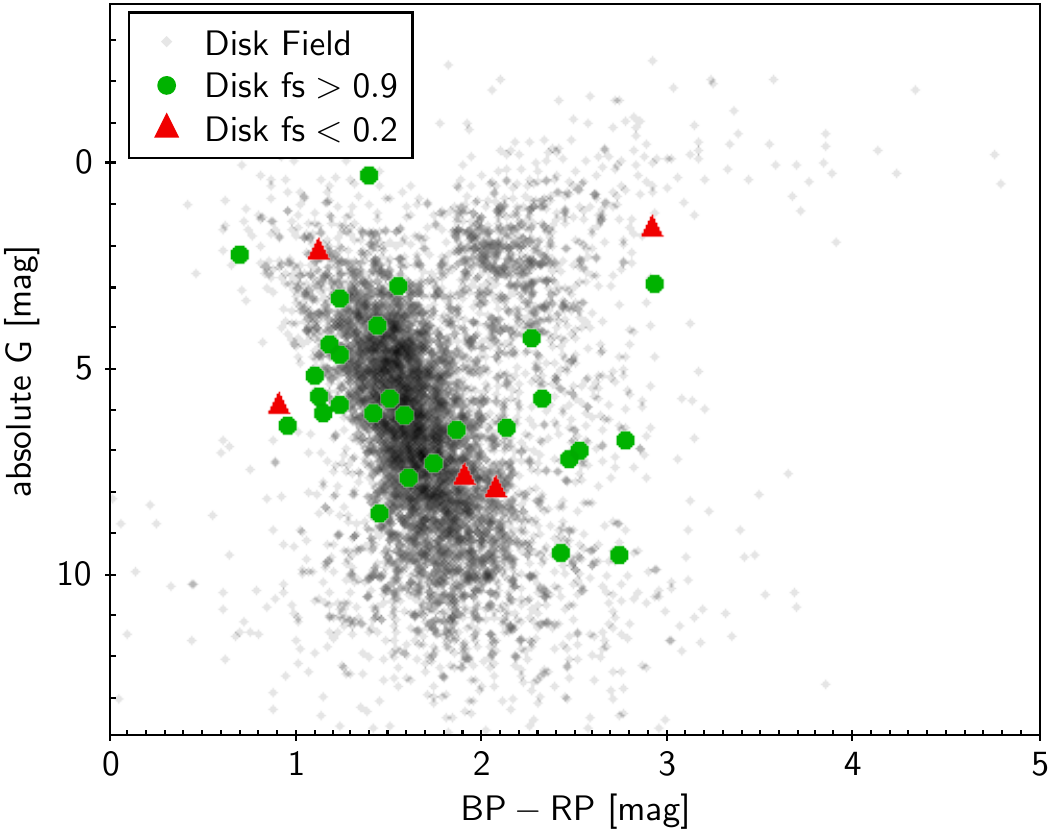}
    \caption{Colour--magnitude diagrams with absolute \gmag derived using \egdr{3} parallaxes. The same typical stellar field was used for bulge (upper plot) and disc (lower plot) as in Figure \ref{fig:cmd_observed} (black/blue points), with the cut on $parallax\_over\_error>1$.
    The baseline magnitudes from Level~1 were used to show microlensing events for \gmag-band as well as for the colour computation. Baseline magnitudes most likely indicate the source for non-blended events ($fs>0.9$) and the lens for highly blended events ($fs<0.2$). }
    \label{fig:cmd_abs}
\end{figure}

We computed the absolute
magnitudes and colours of the sources and lenses that are only defined by the \texttt{paczynski1\_fs\_g} parameter, which also have well-defined parallaxes from \gaia, requiring RUWE$<1.5$ and \texttt{parallax\_over\_error}$>1$. 
Figure \ref{fig:cmd_abs} shows the absolute magnitude for a selection of stars from typical bulge and disc fields, requiring \texttt{parallax\_over\_error}$>1$ with over-plotted microlensing events and their sources and lenses, depending on the blending parameter. In the bulge region, there are sources that can be clearly associated with the bulge stars; this includes both red clumps and main sequence stars. There are also redder sources, of which the vast majority are probably the red clump giants affected by the high level of interstellar extinction. In the disc, the majority of sources are associated with the main sequence stars, with only one or two sources belonging to the red clump giant population. 
As for the luminous lens candidates, in the bulge region there are cases of main sequence stars, red clump stars, and possibly a reddened red clump lens. For the disc, there are both main sequence star lenses and one possibly due to a red clump star.

\subsection{Timescale distributions}

\begin{figure*}
    \centering
    \includegraphics[width=\hsize]{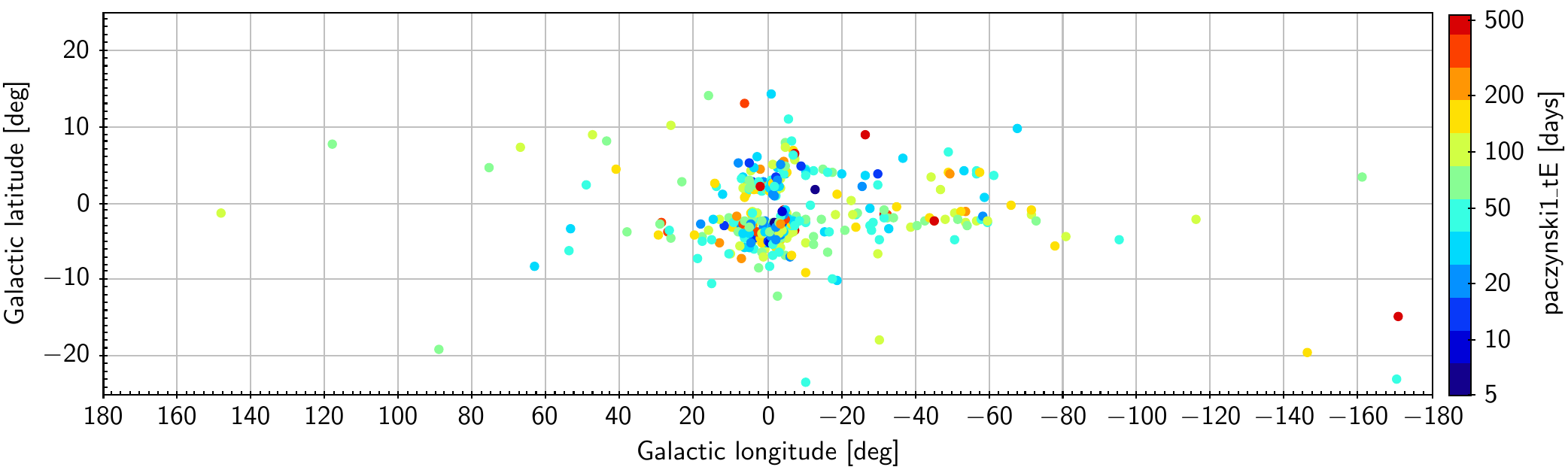}
    \caption{Map of \gaia microlensing events in Galactic coordinates, with colour encoding their Level~1 timescale.}
    \label{fig:gal-te}
\end{figure*}

The Einstein time ($tE$), or event timescale, is the only parameter of the standard Paczy{\'n}ski microlensing model that carries information about the physical parameters of the event (see Section  \ref{sec:introduction}). 
For an individual event, the measurement of its timescale is only an indication of the size of the Einstein radius ($\thetaE$) and the relative proper motion ($\murel$). When combined with the measurement of the microlensing parallax, it can already be used to guess the mass of the lens by employing expected and typical values for the proper motion; see for example \cite{2005Bennett, 2011Skowron, 2011Batista, 2016Wyrzykowski}. However, in the case of an individual event, the complete solution of the event, that is, computation of the lens mass and its distance, can only be achieved if the angular Einstein radius $\theta_E$ is measured indirectly; for example through a finite source effect in single \citep{2011Zub, 2004Yoo} or binary lenses \citep{Alcock2001Nature, 2002An}, or directly using high-angular-resolution imaging and interferometry \citep[e.g.][]{2007Kozlowski, 2019Dong, 2021Cassan} or astrometric microlensing \citep[e.g.][]{2014Sahu,2017Sahu, 2017Kains, 2018Rybicki, 2022SahuBH, 2022LamBH}.

If the timescales are derived for a larger sample of events, they can be used to study the properties of the population of the lens. 
As shown already in \cite{1994KiragaPaczynski}, the total optical depth for microlensing, which is measured from a distribution of event  timescales,  depends on the total mass of the lenses.
The subsequent catalogues of microlensing events towards the Galactic bulge \citep[e.g.][]{1994Udalski, 2003Afonso, 2001Popowski, 2013Sumi, 2015Wyrzykowski, Mroz1} as well as in the direction of the Galactic disc \citep[e.g.][]{2009Rahal, 2020MrozZTF, Mroz2, 2021RodriguezZTF} can help us to compare the Galaxy models with the observations \citep[e.g.][]{2002Evans, 2016Wegg}. 

The \gaia microlensing events reported here form a fairly small sample compared to the thousands of events reported by other longer and more sensitive surveys. In Sample~A, which was produced by an automated procedure and human vetting, the measured timescales span from 20 days up to about 300 days if no blending is modelled, and up to 2000 days if the blending is included. This is clearly a tail of the bulk of the distribution, as in the direction of the central part of the bulge the mean timescale is about 28 days \citep{2005Wood, 2015Wyrzykowski, Mroz1}. As noted in \cite{Mroz1}, the timescales of events tend to increase with increasing Galactic longitude, with 22~days near $l\sim0$\deg to 32~days at $l\sim8$\deg. The increase in timescale is due to the fact that in the directions far from the bulge, both lenses and sources are located in the disc and often have similar transverse velocities. Moreover, the increase in timescale is not symmetric around $l=0$, as it reflects the presence of the inclined Milky Way bar, causing longer timescales at negative longitudes \citep{1994Stanek, 2009Kerins, 2015Wyrzykowski, 2016Wegg, 2016Awiphan, Mroz1}.

We tried to see those trends in the mean timescales of \gaia events. 
Figure  \ref{fig:gal-te} shows the timescales of all events on a map in Galactic coordinates. 
We used a smaller sample of events that overlapped with those of OGLE in order to derive a rough corrected timescale distribution. 
\cite{Mroz1} and \cite{Mroz2} computed the detection efficiency curves as a function of $t_E$ for their OGLE-IV bulge and disk samples, respectively. This function shows the change in the sensitivity of the survey to events at different timescales. Only efficiency-corrected timescale distributions are suitable for comparison with the Galaxy model predictions, as survey-specific effects, such as cadence, survey duration, and seasonal and weather gaps, can limit the sensitivity to different timescales. For the \gaia events sample it would not be trivial to derive such a function, as this would require a full and detailed simulation of the spacecraft observations and the events detection pipeline. Moreover, part of our pipeline still involved humans, and therefore such a process would be hard to reproduce.  

\begin{figure}
    \centering
    \includegraphics[width=\hsize]{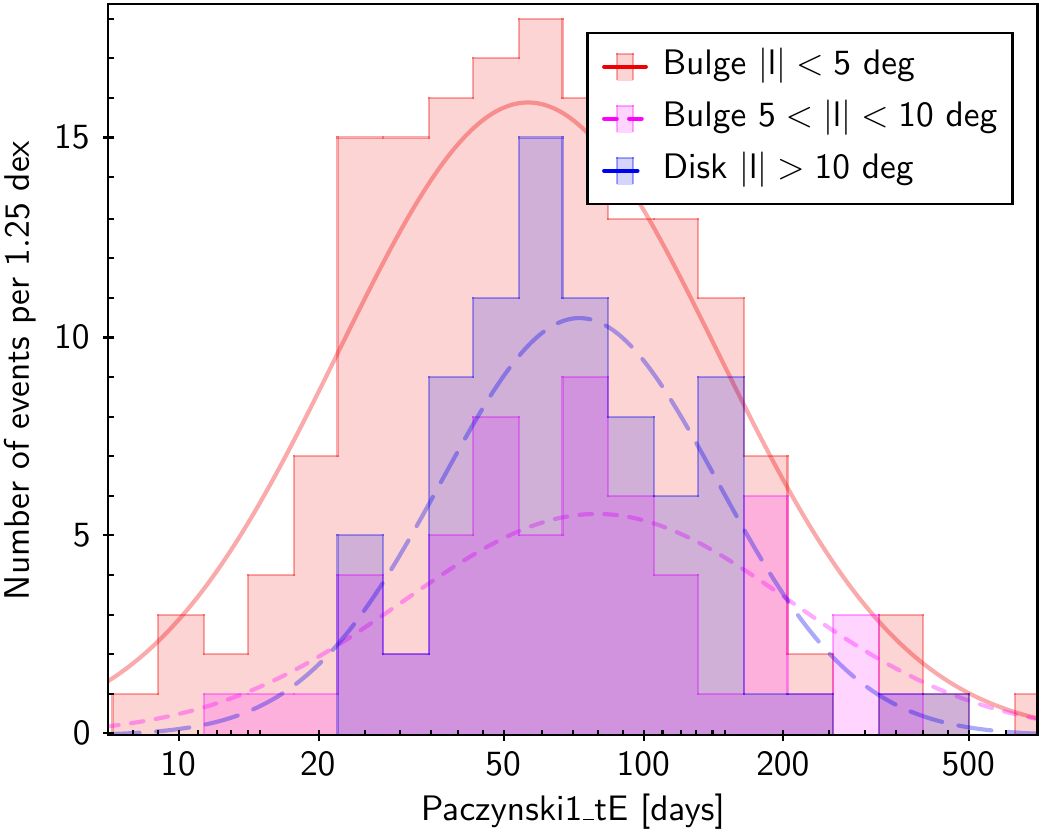}
    \caption{Distributions of Sample~A microlensing events in different Galactic longitude bins together with the Gaussian model fits.}
    \label{fig:te-gauss}
\end{figure}

Nevertheless, in order to estimate the mean timescale distribution for the all-sky sample of events found in \gaia,  we compared the efficiency-corrected timescale distributions of OGLE events with our events in the bulge ($|l|<10$\deg) and in disc separately. We applied a magnitude cut at $G_0<$19~mag to guarantee that we compare similar populations of sources and we find a rough ratio between mean timescales of 2.5 and 2, for bulge and disc, respectively. 
We computed the mean timescale for the blended model (Level~1) for events from Sample~A by modelling their distributions in log-space with a Gaussian. 
We divided the sample into three major sky sections: inner bulge ($|l|<5$\deg), outer bulge ($5\deg<|l|<10$\deg), and the remaining events as disc ($|l|>10$\deg). The distributions and their Gaussian fits are shown in Figure \ref{fig:te-gauss}. We find means of 56, 79, and 72~days, for the inner bulge, outer bulge, and disc, respectively. Applying the correction derived from the comparison of \gaia and OGLE-IV samples (2.5 for bulge fields and 2 for disc), we obtain mean timescales varying from 22 to 32 days between inner and outer bulge and about 36 days for the disc events. This is in a close agreement with the findings for much larger samples of events found by OGLE \citep{2015Wyrzykowski, Mroz1, Mroz2} and MOA \citep{2013Sumi} surveys.
We note that our sample is far too small to attempt whole-sky mean timescale computations. 

\section{Discussion}
\label{sec:discussion}
The \gdr{3} catalogue of microlensing events contains 363  unique sources spanning a wide range of baseline magnitudes, amplitudes, and timescales. 
The validation of the catalogue can be assessed most efficiently by comparing the \gaia detections with those from the OGLE-IV survey \citep{2015Udalski}, which operated and observed both the bulge and southern Galactic plane within the time-span of the DR3 data collection (2014-2017). In the direction of the bulge, OGLE-IV conducted a monitoring of stars to about $I<21$~mag with varying cadence, spanning from tens of observations per night for the densest parts of the bulge ($\pm$ 5~degrees in Galactic longitude and latitude around the Galactic centre) to observations separated by a couple of nights in the regions of the bulge with the lowest stellar density. More details on the OGLE-IV observing strategy can be found on the OGLE webpages\footnote{https://ogle.astrouw.edu.pl/} and in \cite{2015Udalski}. 
The Galactic plane has been observed by the OGLE-IV survey since 2013 with much sparser cadence spanning from a couple of days to weeks and with the depth of about $I<20$~mag, but covering the entire southern Galactic plane to about $\pm 7$\deg in Galactic latitude. More details on the Galactic plane survey can be found in \cite{Mroz2}.

\subsection{Completeness}

\begin{figure}
    \centering
    \includegraphics[width=\hsize]{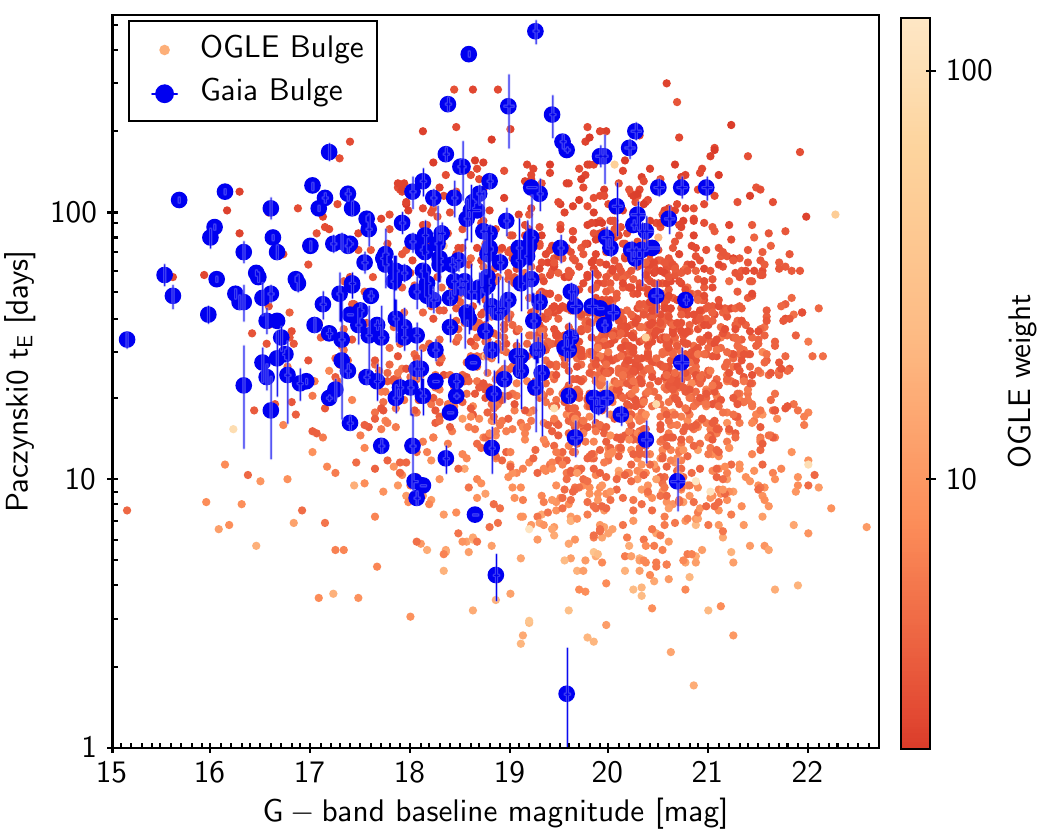}
    
    \includegraphics[width=\hsize]{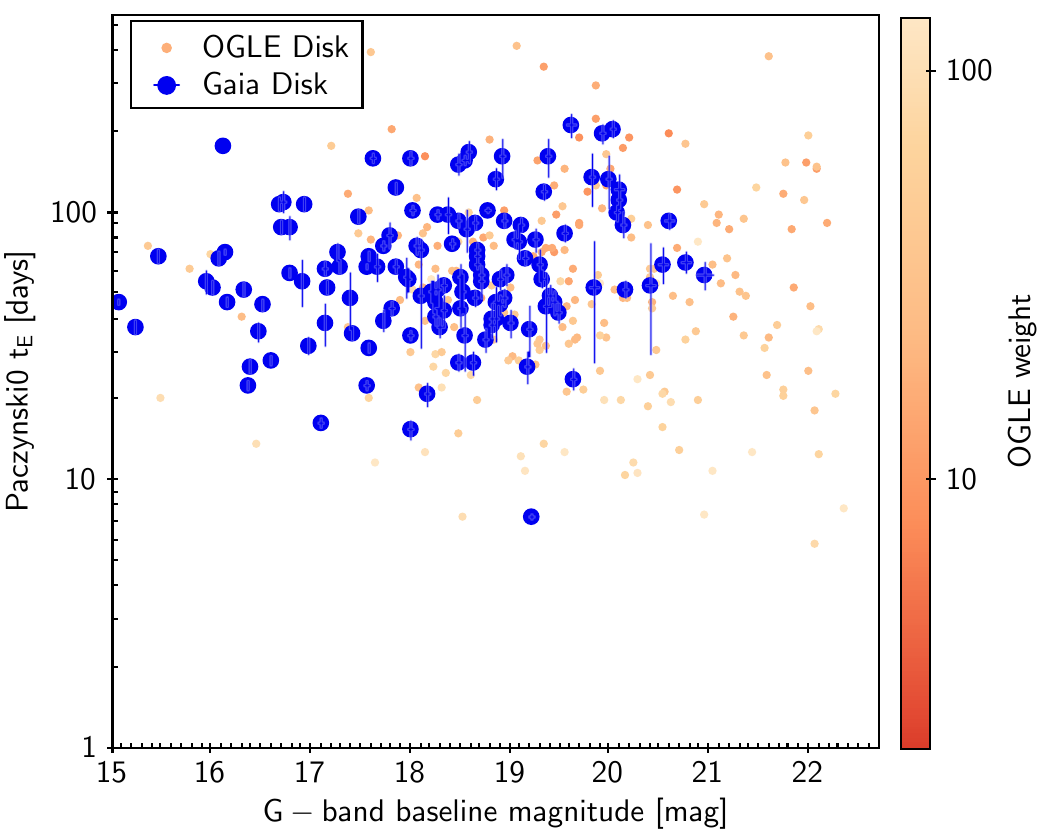}
    \caption{Comparison of \gaia event timescales and baseline magnitudes with OGLE-IV events for the bulge (top) and disc (bottom) samples. Colour shading for OGLE points indicates the weight, which reflects the detection efficiency of the OGLE survey \citep{Mroz1, Mroz2}. Only OGLE events that occurred during the DR3 time range are shown.}
    \label{fig:complet-te-mag}
\end{figure}

We use OGLE-IV bulge and disk samples to assess the completeness of the \gaia microlensing events catalogue. 
In total, there were 227 events from Samples~A and B that overlapped with the OGLE-IV catalogues, with 33 events found independently in \gaia data (Sample~A and A+B) and 194 events solely in Sample~B. 
Figure \ref{fig:complet-te-mag} shows the timescales and baseline magnitudes of bulge ($|l|<10$\deg) and disc ($|l|\geqslant10$\deg) events. \gaia events clearly coincide with the brighter and longer section of the diagram. The OGLE bulge events are dominated by baselines fainter than \gmag=20 mag and their typical timescales are about 30 days. In \gaia, the bulk of the events is bright, which is a result of one of the cuts on Sample~A (\gmag<19 mag). All events fainter than \gmag=19 mag are from Sample~B. 
In the disc, the OGLE and \gaia samples overlap more closely than in the case of the bulge, but still the difference in the baseline magnitude coverage is different, with OGLE reaching fainter baselines. 

\begin{figure}
    \centering
    \includegraphics[width=0.9\hsize]{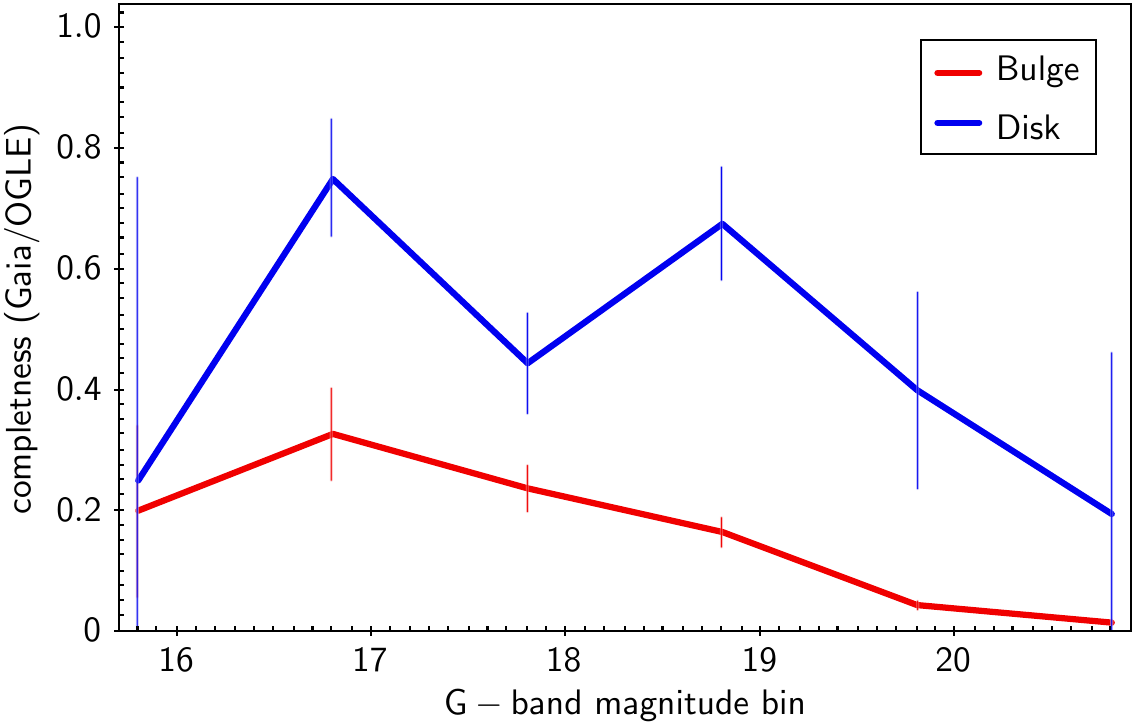}
    \caption{Completeness of the \gdr{3} microlensing events sample as a function of \gmag-band baseline magnitude. Completeness was derived by identifying matches of \gaia and OGLE-IV event catalogues and was derived separately for the bulge and disc sections.}
    \label{fig:complet-mag}
\end{figure}

We calculated the completeness in magnitude space by comparing bulge and disc samples of Gaia events with the OGLE-IV events, limiting the latter sample to events found during the time-span of \gdr{3} data collection. Figure  \ref{fig:complet-mag} shows completeness as a function of baseline magnitude in \gmag. For the bulge events, the function peaks at about 17 mag at about 30\%, dropping to a couple of percent at the faintest end. For the disc, the completeness is significantly better as it reaches nearly 80\% at 17 mag and has another peak at about 19 mag at $\sim$70\%. The poor performance in the bulge is primarily induced by the low completeness of \gaia itself in the bulge region due to high stellar density \citep{2021Fabricius}.
Additionally, the bulge is one of the least frequently scanned parts of the sky (typically about 30 epochs within \gdr{3}), and therefore the detection efficiency of microlensing events is low  there. In contrast, Galactic disc regions of the sky receive up to the maximum cadence possible with \gaia, with the numbers of epochs varying from about 70 to 140 in DR3.
There is also a possibility that the OGLE-IV survey is not fully complete in the disc regions, as their cadence is sparser and the depth is shallower than in the bulge. In such a case, \gaia completeness would actually be lower than that measured above.

\begin{figure}
    \centering
    \includegraphics[width=0.9\hsize]{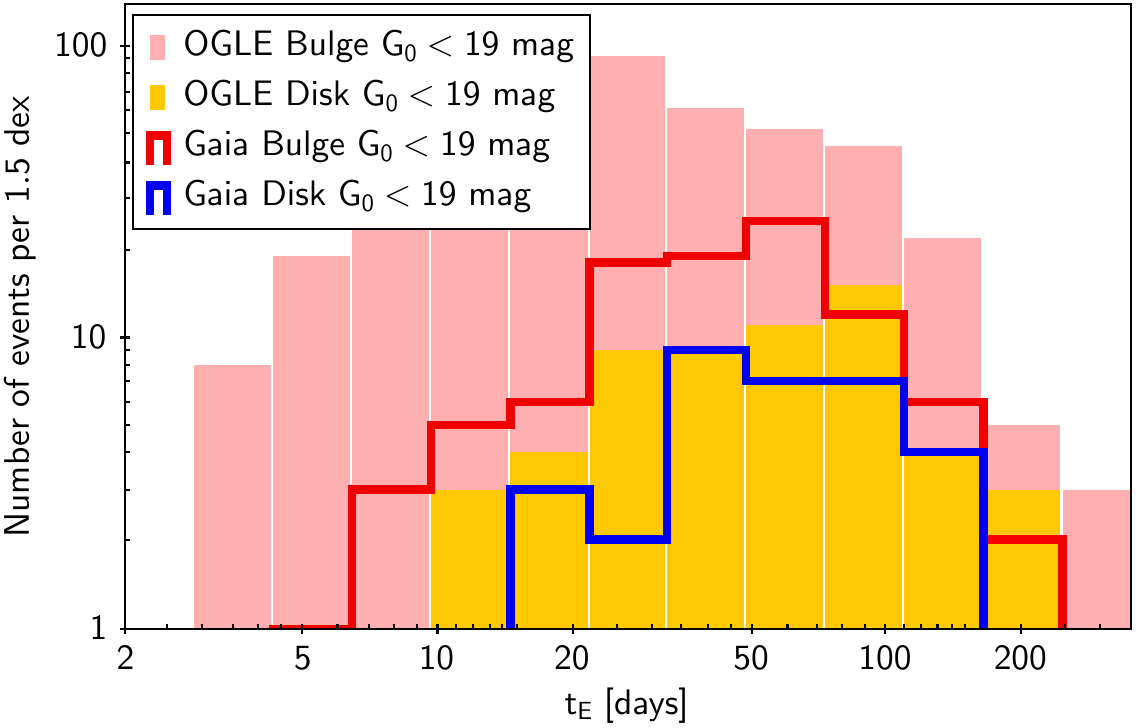}
    \caption{Comparison of the distributions of event timescales for \gdr{3} microlensing events found also by the OGLE-IV in the bulge and disc regions, cut at the baseline magnitude of \gmag of 19~mag. Parameter \texttt{paczynski0\_te} was used for \gaia events and $t_\mathrm{E(best)}$ for OGLE-IV events. }
    \label{fig:complet-te}
\end{figure}

The completeness of the event timescales can also be derived by comparing with the OGLE bulge and disk samples. In order to ensure a fair comparison, we limited OGLE events from \cite{Mroz1} and \cite{Mroz2} to those with their maxima within the \gdr{3} time-span and we cut OGLE and \gaia samples on a baseline magnitude of \gmag=19~mag. 
Figure \ref{fig:complet-te} shows the histograms of such trimmed OGLE and \gaia samples, again for the bulge and disc events. 
In the bulge, the completeness varies from 0\% for events with timescales shorter than 7 days to about 50\% for events with timescales of about 60~days and approximately 30\% for events with timescales of 100~days. 
For the disc, the numbers of events from both \gaia and OGLE-IV are fairly small, but the completeness is significantly higher than in the bulge. At about 30~days and 140~days, nearly 100\% of events were recovered by \gaia, while for the shortest timescales, recovery was only about 20\%. The main reason for this very low recovery rate of short events is again \gaia's scanning law, which generates a scanning pattern in which \gaia observes the same spot on the sky typically after about 30~days.
The detection efficiency drops for the longest timescales in both bulge and disc cases because of the limited time-span of the \gdr{3} data (about 1000~days). 

\subsection{Parameter accuracy}

\begin{figure}
    \centering
    \includegraphics[width=0.9\hsize]{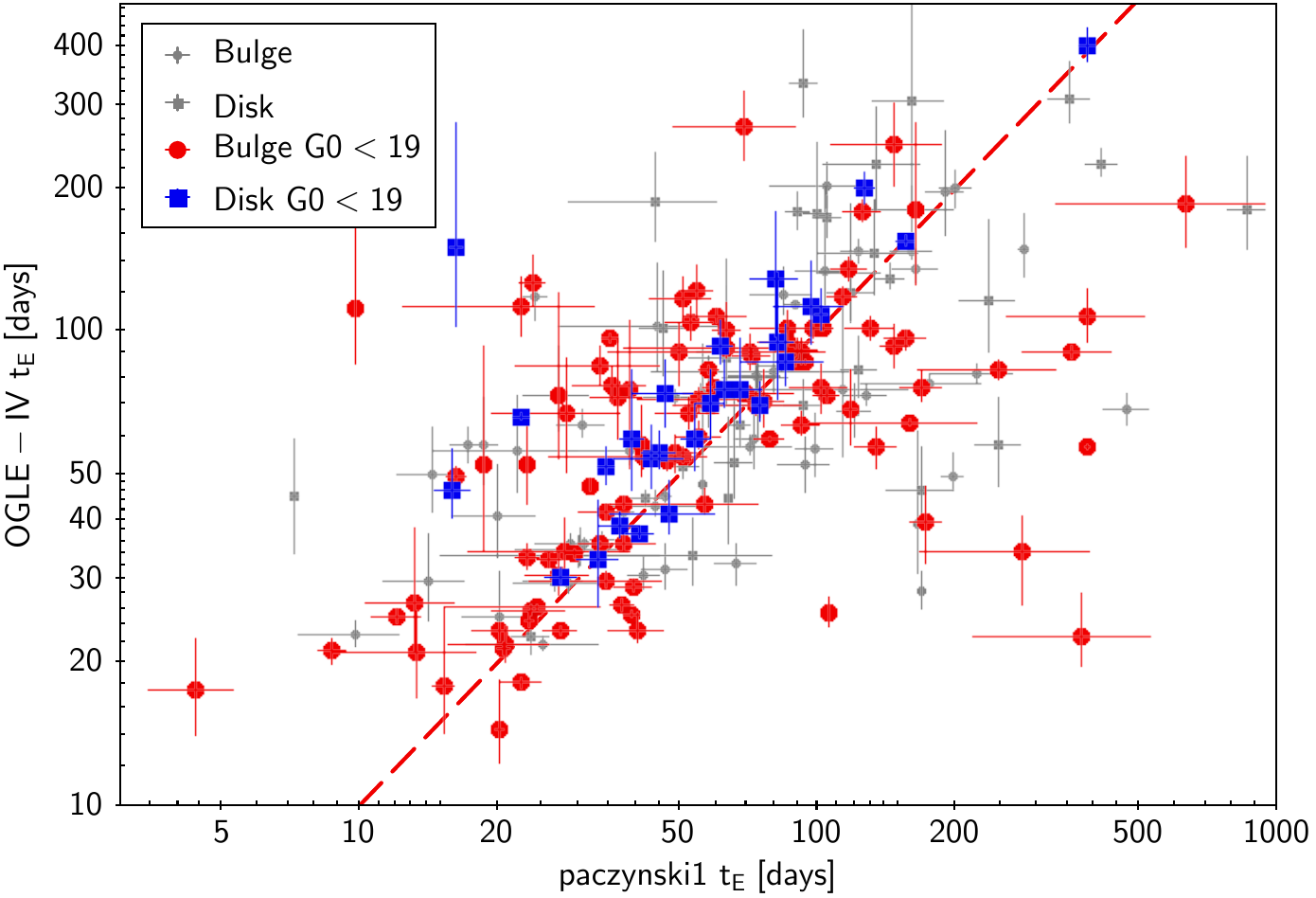}
    \caption{Comparison of timescales derived for \gdr{3} microlensing events also found by the OGLE-IV for bulge and disc. Parameter \texttt{paczynski1\_te} was used for \gaia events and $t_\mathrm{E(med)}$ for OGLE-IV events and their corresponding error bars.}
    \label{fig:te-ogle-gaia}
\end{figure}

The most important parameter of a standard, non-parallax single-point-source, single-point-lens microlensing event is its timescale, $t_{\rm E}$, as this contains information about the physical properties of the lens. 
Figure \ref{fig:te-ogle-gaia} compares the values of the Einstein timescale as derived by \gaia and OGLE for the same events found by both surveys. For \gaia, we used the \texttt{paczynski1\_te} parameter derived in the model with blending, as the OGLE timescales were also derived using a blended model. We note that the amount of blending seen in a ground-based survey should typically be greater than in a space-based survey such as \gaia; however, we assume that the microlensing model with blending yields the actual event timescale, which does not depend on blending. On the other hand, sparse \gaia light curves might not be the best suited for
obtaining robust measurements of the amount of blending. Indeed, most of the measurements with $t_E>100$~days that are outlying from the one-to-one relation to OGLE timescales in Figure \ref{fig:te-ogle-gaia} are fairly short-lasting events, but their blending, as derived from sparse \gaia data, was estimated to be high ($f_s\ll1$). Highly blended solutions should always be treated with caution and we recommend a more detailed study of these events and their light curves.

In general, the comparison of timescales reveals good agreement with the measurements of OGLE, especially for the brighter sample with $G_0<19$~mag (primarily Sample~A). The events from the disc show better correspondence than events from the bulge, primarily thanks to better cadence in \gaia observations of the disc fields.  
In a couple of cases, it can be seen that the error bars on the timescales of \gaia events are clearly underestimated. Typically, the timescale errors are about 7~days, but they span to 70~days in the case of Level~0, and up to 300 for Level~1, owing to the fact that the more complex blended model is less constrained by the sparse light curves of \gaia. 

\subsection{Contamination}
\label{sec:contam}

Microlensing events were identified in the \gaia DR3 data solely based on the shape of their light curves and the colours of the source star. In future \gaia data releases, when astrometric time-series become available, it will also be possible to recognise microlensing events as an astrometric deviation due to lensing \citep{2000DominikSahu, 2002Belokurov, 2018Rybicki}.
Until then, we have to rely on photometry only. 
The contamination with non-microlensing events, which can mimic the shape of the microlensing light curve, is particularly possible in the case of sparsely sampled light curves such as those from \gaia. 
The contamination with non-microlensing variables is mostly going to affect our Sample~A, as this latter was obtained in an automated fashion. Sample~B is composed of previously known and published microlensing events, and therefore their reliability is greater. 
As the events reported here are concentrated very near the Galactic plane (to within $\pm$20\deg), the main types of contaminants will be of Galactic origin. 

Nevertheless, we identified one candidate, \#257,
3388787058343659648,
in common with a  list of candidate quasi-stellar objects (QSOs; or `quasars')  from \cite{2019Shu}, who used \gdr{2} and AllWISE data to identify potential QSOs without knowledge of their light curves. The photometric evolution of this event clearly shows a slow rise with the amplitude of about 0.8 mag, but the event is not covered until the brightness returns to quiescence. Additional observations of this source from CRTS \citep{Drake2009}, covering about 8 years before \gaia started, show a flat light curve. Similarly, ZTF data from 3 years after \gaia also show a flat light curve, while for a QSO we might have expected at least low-level long-term variability. Further studies and future \gaia data are necessary to distinguish between the microlensing and QSO nature of this source. 

In a subsequent step, we cross-matched our candidate events with the catalogue of variables from OGLE, in particular the catalogue of nearly 250 000 long-period variables (LPVs), which show slowly varying large-amplitude periodic or semi-periodic variations in their light curves \citep{2013Soszynski}.
There was one object in common with our catalogue, namely event \#004 (4043236380953360512), which matched to OGLE-BLG-LPV-102568, and was identified from the OGLE light curve as an (OGLE small-amplitude red giant(OSARG; \citealt{2004Wray})-type variable. 
OSARGs are known for their small amplitudes, while here the event shows a very significant change in brightness of about 1 mag. We suggest that \#004 is a genuine microlensing event, in which the source happened to be a variable star of the OSARG type. 
Such events are fairly rare, but most often are just not being detected due to the prior variability in their baselines (but see \cite{2006Wyrzykowski}). This event was also not detected by any survey  other than \gaia, most likely because of the prior baseline variability, despite its high brightness and high amplitude. A very similar case is presented below in Section  \ref{sec:individual}, where a baseline variation of small amplitude is superimposed with the microlensing event (event \#007). 

The cross-match of microlensing events with the \gdr{2} catalogue of 550 000 variable star candidates \citep{2018Mowlavi} yielded another match, \#066 (4048971261782646528), classified in \gdr{2} as a potential LPV. The \gdr{3} light curve indeed shows a red variable object, but with an amplitude of nearly 2~mag and a non-repeating shape of the outburst; we suggest this again is a genuine microlensing event. The classification in \gdr{2} relied on a much shorter light curve and therefore an event without its full baseline could have fallen into the LPV selection. We note that this source is no longer considered as an LPV candidate in the newest DR3 LPV catalogue \citep{DR3-DPACP-171}.

Microlensing events can also be mimicked by other types of variable stars, with the main ones being 
Be-type outbursts \citep{2003Mennickent, 2005SabogalOGLEBe, 2002Keller, 2013Rivinius}, 
cataclysmic variables (CVs; \cite{1997DownesATLASCV}), 
and YSOs \citep{1994Andre, 2019Marton}. 
Automated ways to distinguish microlensing events from those common contaminants were recently discussed by \cite{2019Godines} and \cite{2022Gezer}.

We used Simbad\footnote{\url{http://simbad.u-strasbg.fr/simbad/}} in order to identify any potential overlaps between additional classes. 
We found two candidates, 
\#002, 4135677580466393216 and 
\#051, 443788525336160512, marked as potential LPVs in the catalogue of variable stars from the ATLAS survey \citep{2018Heinze}.
However, as the ATLAS data used for the classification spanned only two years and microlensing events were not considered in their search, we suspect these two events were incorrectly identified as LPVs and are actually microlensing events given the shape of their light curves and good matches to the microlensing models in \gdr{3} data.

Another event, 
\#137 (3012224323198211968), was classified as a potential YSO by \cite{2019Grossschedl} based on its VISTA/WISE archival observations. The \gaia light curve shows a double-peaked event, which can also be explained by a strong microlensing parallax. Such photometric behaviour is not uncommon in YSOs, but there is also a chance of microlensing occurring on a young and dusty star. Moreover, both CRTS and ZTF observations prior to and following the event show no further brightening episodes, which would be expected if the source were a YSO variable star. Further observations and a longer \gaia light curve are necessary to verify the nature of this event.

One of our events, \#118, 3355845866073702144, which was also discovered in real time as a transient by the ASAS-SN survey \citep{ASASSN} and dubbed ASASSN16-li, was marked as a candidate cataclysmic variable on the ASAS-SN web page\footnote{\url{https://www.astronomy.ohio-state.edu/asassn/transients.html}}. This transient was also detected by the OGLE disk survey \citep{Mroz2} as a potential (but weak) microlensing candidate. Both ASAS-SN and OGLE light curves cover the transient very sparsely, in particular, the pre-peak part of the event is not covered at all, hence the suspicion of a CV. Cataclysmic variables typically show a fast-rise-exponential-decline (FRED)-type light curve. \gaia data contain only one data point prior to the peak of the event, and so cannot be used to rule out the FRED shape. However, the long-term ASAS-SN and OGLE data do not show any other outbursts, as are typically seen in most CVs. We suspect that this is a genuine microlensing event  that was mistaken for a CV because of the sparse sampling, because to our knowledge no spectroscopic classification was performed. 

Relying on photometry only, we can use the known fact that microlensing events do not repeat. In the history of microlensing, so far the only few cases of repeating microlensing events were found when the lens was a very wide binary system or a planetary system and the lensing episodes were separated by 1-2 years \citep{2009SkowronWyrzyk, 2014Poleski}. Therefore, for our sample we can assume they should not show any further brightening episodes after the event. Repeating Galactic transients that could mimic microlensing are primarily CVs, YSOs, and Be-type outbursts.  
Here, we searched for outbursts reported some time after the end of \gdr{3} (from mid-2018) in surveys monitoring the sky where \gaia microlensing events are located. We searched among alerts from \gaia itself, the \gaia Science Alerts \citep{Hodgkin2021}, and found zero matches. The search among OGLE microlensing candidates from bulge and disc \citep{Mroz1, Mroz2} also yielded no matches. We also checked the microlensing catalogue from ZTF, which started operating after 2018 \citep{2021RodriguezZTF, 2022MedfordZTF}, and again found no matches to our events. We also extracted photometry from ZTF~DR9\footnote{\url{https://www.ztf.caltech.edu/ztf-public-releases.html}} \citep{2019Masci} and visually inspected all light curves with ZTF data covering the years 2018-2022. 
No variability or additional outbursts were found, lowering the chance that there are non-microlensing events within our sample.

In summary, we cross-matched \gaia microlensing events with known types of variables and potential repeaters and found six matches in total  with objects also classified as LPVs, QSOs, YSOs, and a CV. We are confident that four events matching the LPVs and a CV are actually genuine microlensing events, while the remaining two matches still have to be confirmed. With 363 events in total in our catalogue, the contamination is therefore estimated to  0.6\% (for two contaminants) and 1.7\% (for six contaminants). We note the contamination rate can be higher as we only explored a limited range of possible types of contaminants.

\subsection{Individual cases}
\label{sec:individual}




As listed in Table \ref{tab:xm}, 273 of 363 events in our catalogue have been independently identified by other large-scale surveys. Additionally, many of the sources with events identified during \gdr{3} and found only by \gaia were located in the fields monitored by various large-sky surveys after the event. As discussed above, publicly available ASAS-SN and ZTF time-series were used  in particular to check for any potential secondary outbursts and to rule out non-microlensing variable stars. Below, we discuss selected cases of events from our catalogue and show the microlensing model fit to the \gaia data combined with publicly available data from other surveys.

\begin{figure}
    \centering
    \includegraphics[width=\hsize]{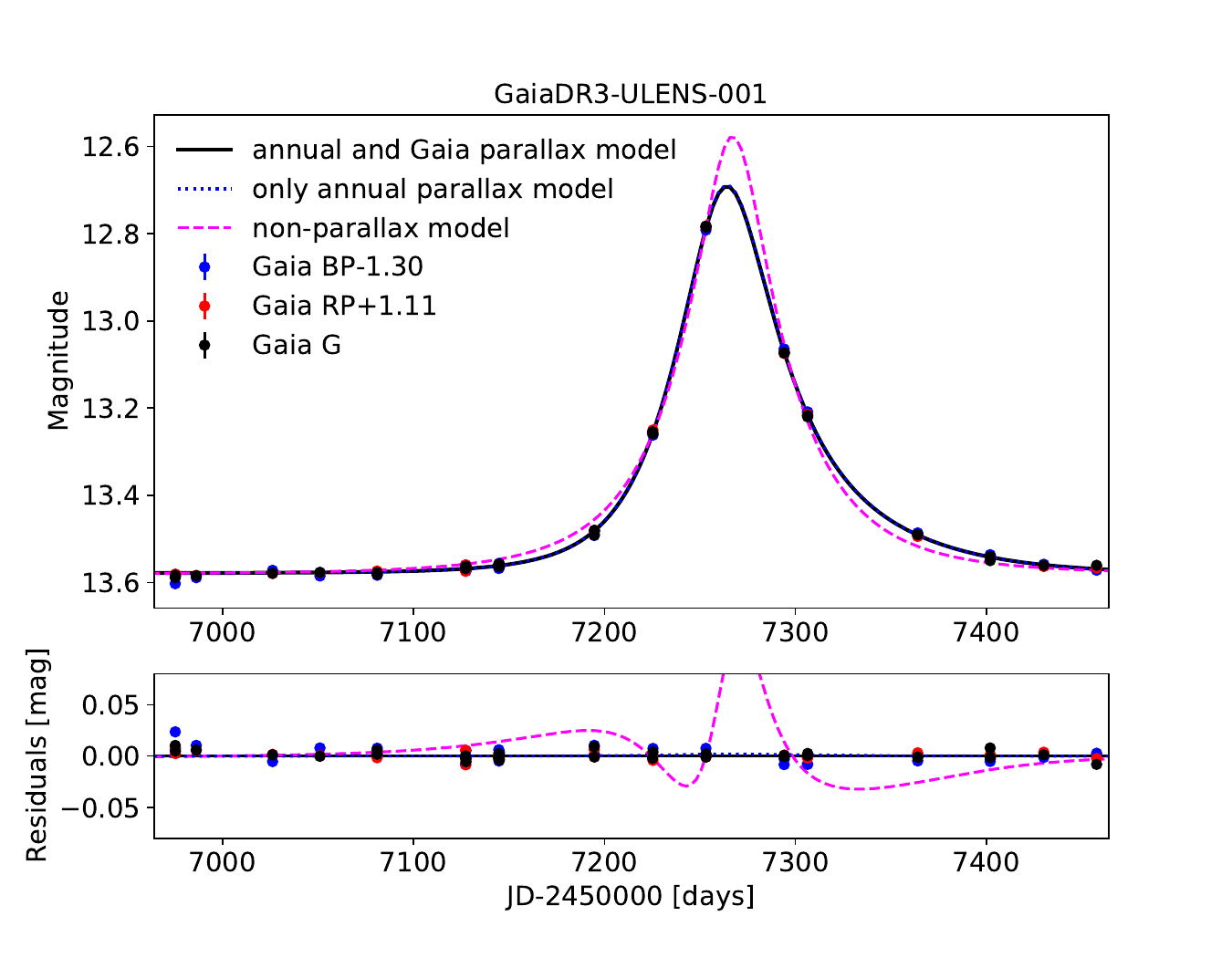}
    \caption{Data, models, and their residuals for GaiaDR3-ULENS-001 event, Gaia sourceid=6059400613544951552. Data error bars are smaller than the size of the marker. A standard microlensing model with blending is shown with the magenta dashed line, a model with annual parallax without space parallax is shown with a blue dotted line, while the black line shows the annual parallax model with \gaia space parallax effect. Models were obtained using the MulensModel code \citep{MulensPoleski}.}
    \label{fig:fitGaia1}
\end{figure}

The first example and the first entry in our catalogue, GaiaDR3-ULENS-001 (sourceid=6059400613544951552), has only \gaia data during the event.
Figure \ref{fig:fitGaia1} shows \gaia \gmag, \gbp, and \grp data for event \#001  which had a baseline magnitude of \gmag=13.58~mag, and was located in the Galactic disc at the Galactic longitude of 298\deg.
This part of the sky has also been monitored by OGLE and ASAS-SN, but neither of these surveys reported any transient at that location when it occurred in mid-2015.
Later  publicly available ASAS-SN data  show no variation in brightness. 
The light curve in all \gaia bands was modelled using the MulensModel open source code \citep{MulensPoleski} and three different models were fit. The standard model with blending is clearly not reproducing many of the data points, while the model with annual microlensing parallax fits the data much better. The figure also shows a comparison of the parallax model with and without the inclusion of the space parallax due to the Earth--\gaia distance. The space parallax is negligible for event \#001. 

Event \#001 is an example of a microlensing event found in \gaia DR3 data but previously missed by other surveys looking for microlensing events in the southern hemisphere. Such a bright event would have benefited from more detailed light curve coverage as well as spectroscopy in order to derive its microlensing parameters. In the case of such a bright event, the \gaia astrometric measurements are expected to have accuracy of better than 1 mas \citep{2018Rybicki}. This event will therefore be a perfect case study for future astrometric microlensing signal searches using \gaia's superb astrometric time series. 

\begin{figure}
    \centering
    \includegraphics[width=\hsize]{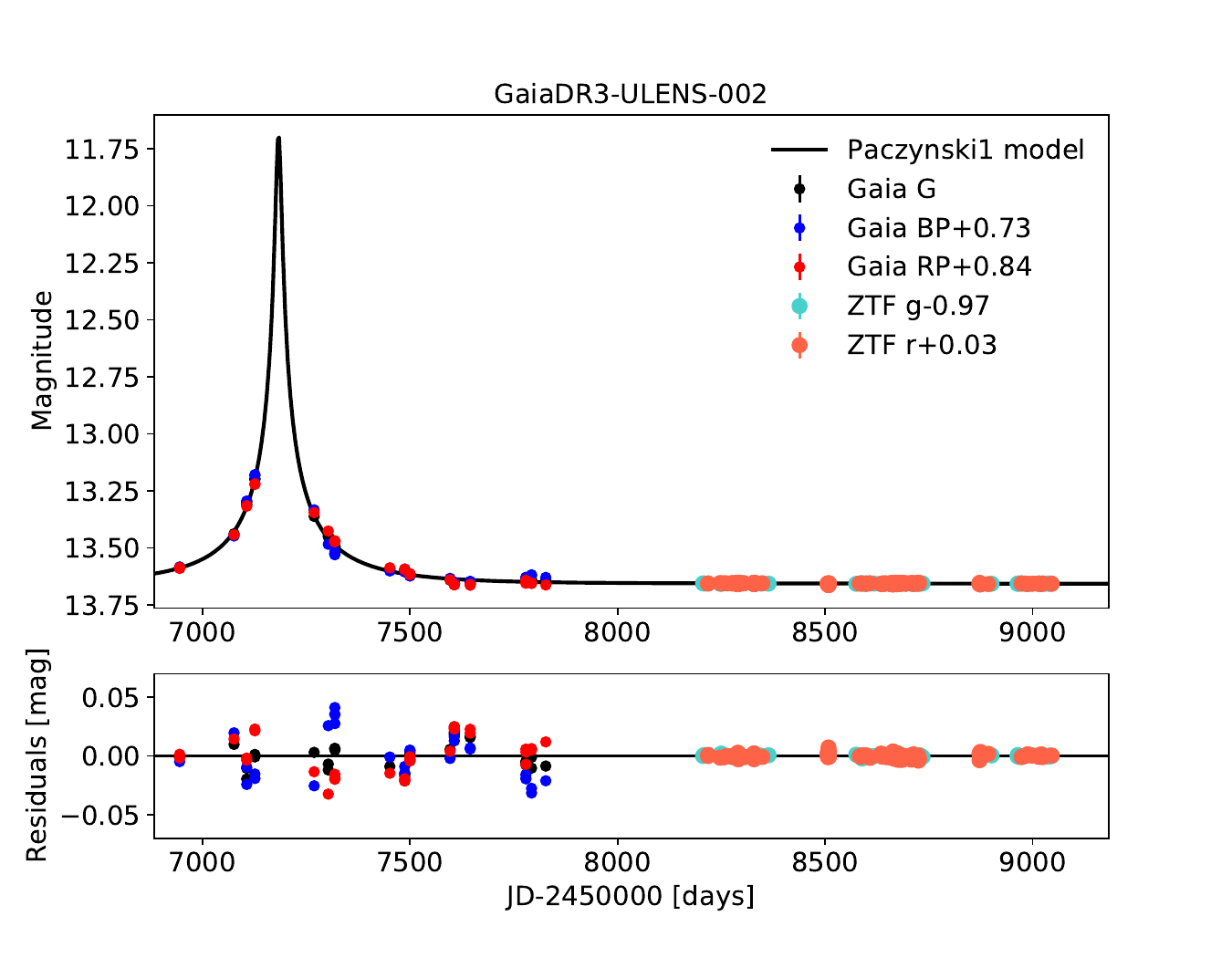}
    \caption{Data, Level~1 model, and its residuals for event GaiaDR3-ULENS-002/4135677580466393216, previously classified as a candidate LPV by the ATLAS survey. Data from the ZTF survey collected well after the event do not show any additional variability, which would usually be expected for a LPV star.}
    \label{fig:GaiaLPV}
\end{figure}

Figure \ref{fig:GaiaLPV} shows event GaiaDR3-ULENS-002  with \gaia sourceid=4135677580466393216. This star was previously classified as a LPV by the ATLAS survey \citep{2018Heinze}. Additional observations of the ZTF survey collected in the years 2018-2022, and therefore well after the increase in brightness, show no further large-scale variability, as would usually be expected for LPVs. This confirms that event \#002 is most likely a genuine microlensing event. 

\begin{figure}
    \centering
    \includegraphics[width=\hsize]{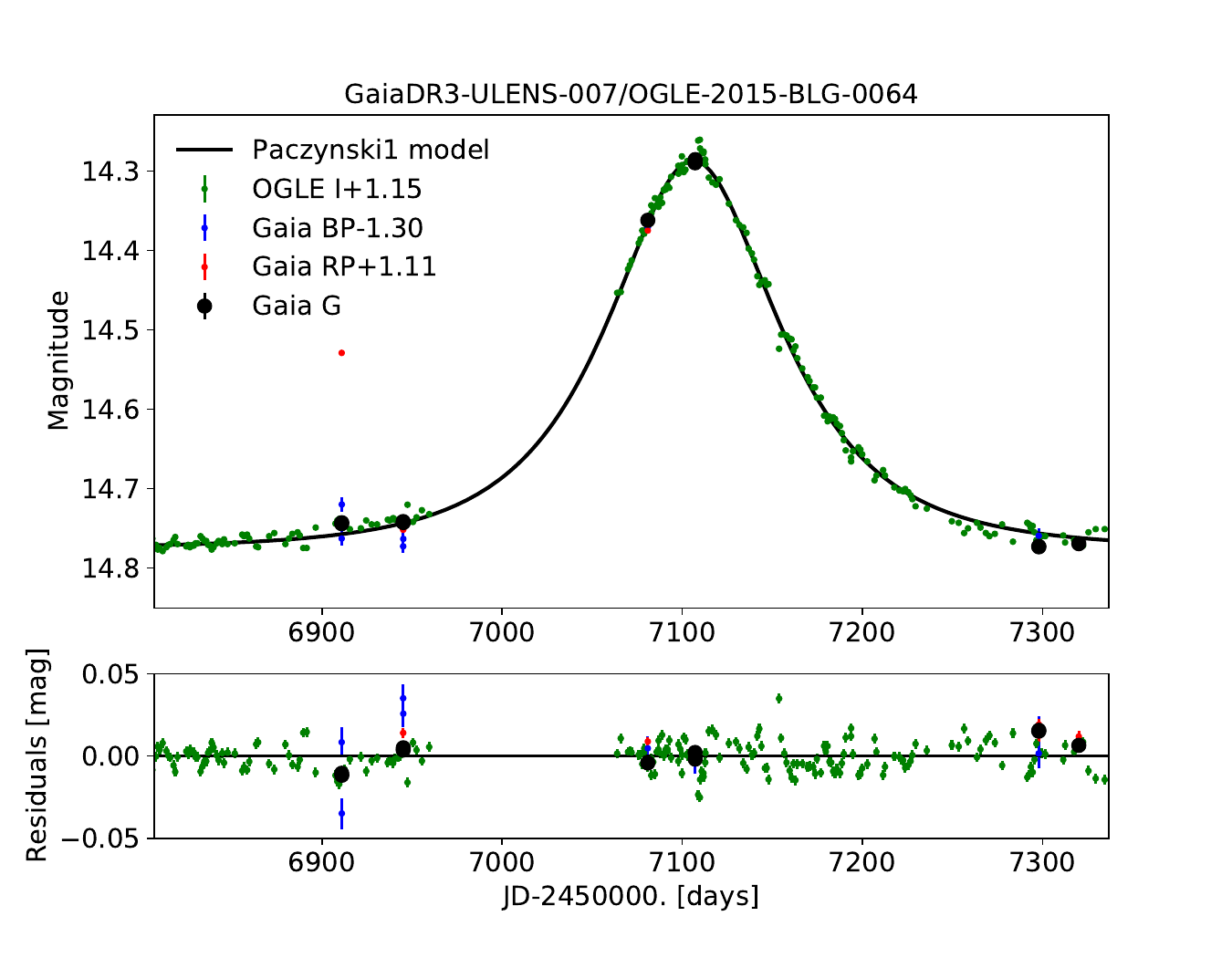}
    \caption{Data, microlensing model, and its residuals for event GaiaDR3-ULENS-007, Gaia sourceid=4042926314286962944. This event was also discovered in real-time by the OGLE-IV survey as OGLE-2015-BLG-0064. The OGLE data clearly show low-level irregular variability due to either source or lens variability.}
    \label{fig:fitGaia007}
\end{figure}

The next example we wish to present is GaiaDR3-ULENS-007 (\gaia sourceid=4042926314286962944). 
Figure \ref{fig:fitGaia007} shows \gaia \gmag, \gbp, and \grp data points together with the OGLE-IV data from their Early Warning System \citep{2015Udalski}, where its designation is OGLE-2015-BLG-0064. 
The standard microlensing model fit to both \gaia and OGLE data is also shown with its residuals at the bottom.  
We note that there is low-level variability present in the OGLE I-band data that is visible before and during the event. 
According to the study of variable baseline microlensing events by \cite{2006Wyrzykowski}, the change in the amplitude of the variation is related to the blending parameter and could indicate whether the variability comes from the source or the blend (including the lens). In case of a non-blended varying source, the variability amplitude should not change while amplified, but in the case of a strongly blended varying source, the observed amplitude should increase. In an opposite case of a variable blend (or lens), the variation amplitude should decrease with increasing magnification of  the source. In the case of GaiaDR3-ULENS-007/OGLE-2015-BLG-0064, the baseline amplitude is very small, of about 0.02 mag (peak-to-peak). 
The blending parameter obtained in the Level~1 fit for \gaia data only was about 0.12 in \gmag. For the \gaia-OGLE combined data set, we again used MulensModel \citep{MulensPoleski} and obtained a blending parameter of about 0.3 for \gmag and the OGLE I-band. If the source is variable and is contributing as little as 12\%-30\% of the light at the baseline, its variability amplitude should increase to about 0.03 mag if we take the microlensing amplification of about two in this case.
The observed amplitude during the event, seen more clearly in the residuals, is indeed somewhat larger, suggesting the source in this event is variable and its fully de-blended variation amplitude should be about 0.07 mag, as typically seen in OSARGs  (\citealt{2004Wray, 2009Soczynski, 2013Soszynski}). The red colour and bright magnitude of \#007, even after de-blending, is also consistent with the red giant star in the bulge, a location where many of the microlensing sources can be found.
In the case of a variable blend, the variability amplitude in this event (assuming A$\sim$2 obtained from the standard microlensing model) would decrease to about 0.01 mag. In fact, the irregularity of the variations makes it hard to measure the amplitude precisely, but we suspect the varying source scenario is more probable here. 

\begin{figure}
    \centering
    \includegraphics[width=\hsize]{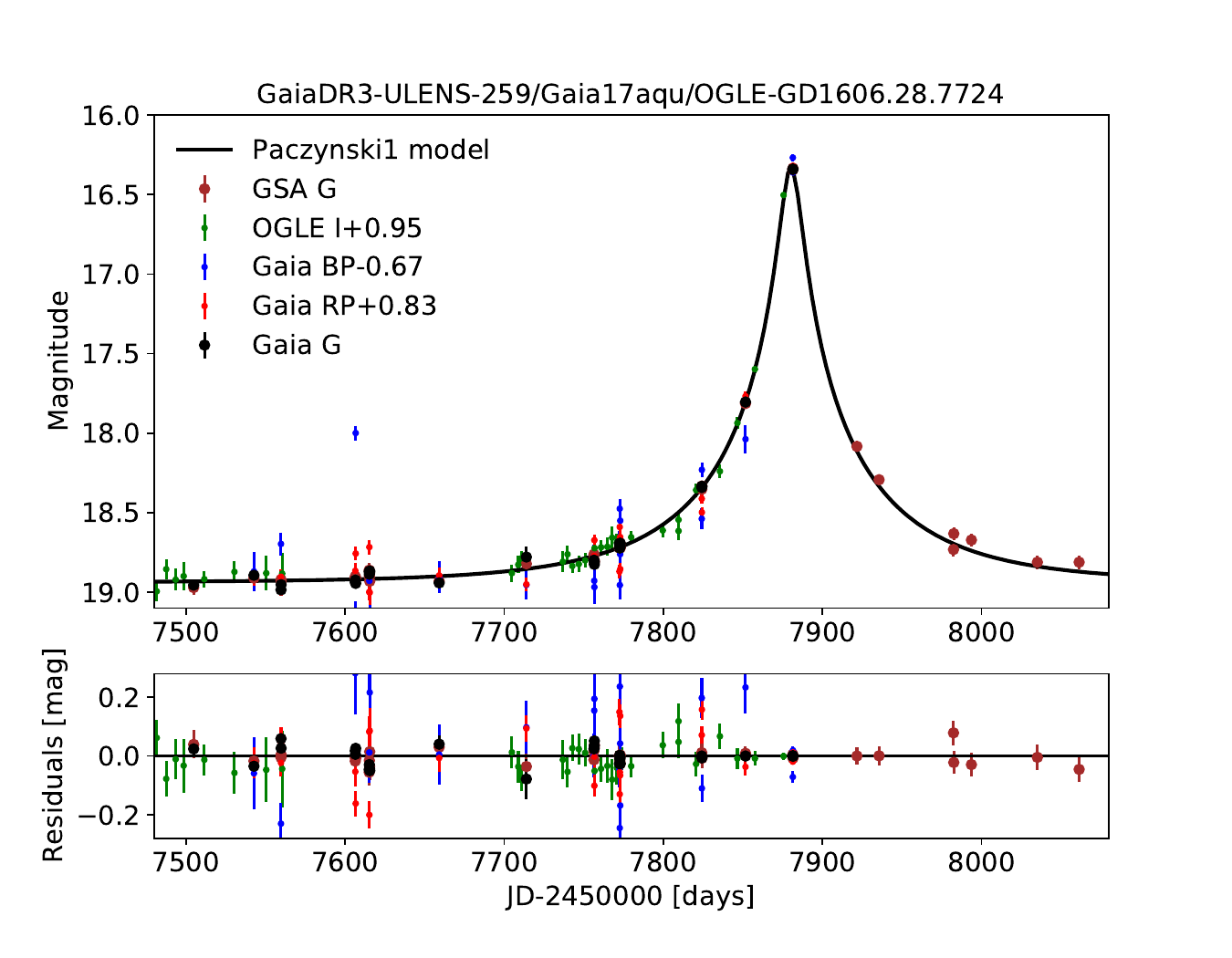}
    \caption{Data, the Level~1 microlensing model, and its residuals for event GaiaDR3-ULENS-259, \gaia sourceid=5599913394301352832. This event was discovered while ongoing by the GSA system as Gaia17aqu, and was also found by the OGLE survey.}
    \label{fig:fitGaia17aqu}
\end{figure}

There are seven microlensing events in our catalogue, which were alerted in near-real-time by the GSA system \citep{Hodgkin2021}. 
Figure \ref{fig:fitGaia17aqu} shows one of them, Gaia17aqu (or simply `Aqua'). 
The event was also discovered in a dedicated search for microlensing events in the OGLE-IV Galactic disc fields by \cite{Mroz2} and the OGLE I-band data are also shown in Figure \ref{fig:fitGaia17aqu}. 
\gaia DR3 data as well as those of OGLE cover only the rising part of the event, while GSA data\footnote{\url{https://gsaweb.ast.cam.ac.uk/alerts/alert/Gaia17aqu}} cover also the declining part of the light curve. The Level~1 Paczynski model (with blending) is shown in the figure and it reproduces all data sets well with the timescale of about 100 days.

\begin{figure}
    \centering
    \includegraphics[width=\hsize]{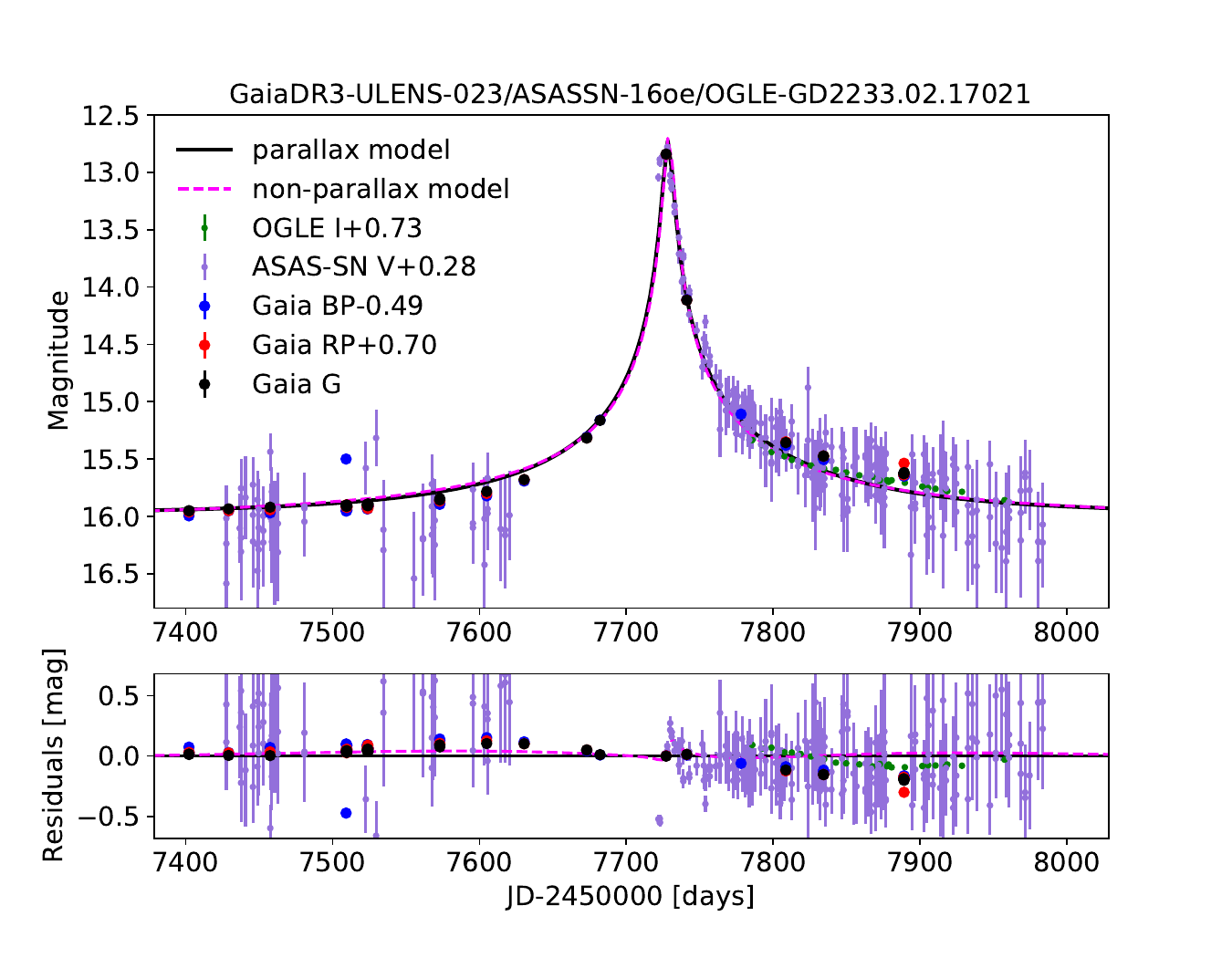}
    \caption{Data, microlensing models, and their residuals for event GaiaDR3-ULENS-023, \gaia sourceid=6059816980526317568. This event was discovered while ongoing by the ASAS-SN survey as ASASSN-16oe, and was also found by the OGLE survey. The microlensing models were obtained with the MulensModel code.}
    \label{fig:fitASASSN16oe}
\end{figure}

A long-lasting microlensing event, called ASASSN-16oe, was discovered in real-time by the ASAS-SN survey \citep{ASASSN} in 2016, which is within the time-span of \gaia DR3 data collection. The source observed by \gaia (sourceid=6059816980526317568) contained the microlensing event and was also detected by the OGLE survey \citep{Mroz2}. This source was added to our Sample~B and the event was dubbed GaiaDR3-ULENS-023.
Figure \ref{fig:fitASASSN16oe} shows the data from all surveys together with the parallax and standard models and their residuals obtained using MulensModel \citep{MulensPoleski}.
A simple parallax model fit to all the available data (\gaia, ASAS-SN, OGLE) yielded a good fit with the heliocentric timescale of about 177 days, blending in \gmag of about 0.65, and a length of the microlensing parallax vector of $\pi_{\rm E} = 0.18$. Those values, combined in Equation \ref{eq:mass} with $\thetaE=\tE / \murel$ and a typical value of relative source--lens proper motion of $\murel \sim3$ mas/yr, yield a mass for the lens of about 1~\Msun located at 2.5~\kpc, if the source distance is assumed to be 8~\kpc. Because of a high level of blending in this event, the lens can be explained with a regular main sequence star yielding all the blended light at that distance. A more detailed study is necessary in order to establish the most probable mass and distance of the lens \citep[e.g.][]{2016Wyrzykowski, 2020WyrzykowskiBH, 2021Mroz}. However, in the case of this bright and long-lasting event,   the actual mass and distance of the lens measurement should be possible with the use of the forthcoming \gaia astrometric time series \citep{2018Rybicki}.




\section{Summary}

We present the first catalogue of \gaia microlensing events from all over the sky. We find 363 unique events, of which 90 have not previously been reported.  \gaia DR3 contains complete light curves in \gmag, \gbp, and \grp \gaia bands as well as the parameters of microlensing models from Level~0 and Level~1 fits.
The catalogue is far from complete, as the relatively short time-span of the data (2014-2017) and a very large starting number of sources to investigate ($\sim$2~billion) required a very conservative approach in the search for potential microlensing events. The completeness, derived by comparison to the OGLE-IV results, is at a level of 30\% for the bulge region and up to 80\% in the disc regions for the brighter end of the sample. Our catalogue is also fairly complete for events with timescales longer than about 60 days, in particular in the disc regions. On the other hand, contamination of the catalogue is estimated to a very low level of below 1\%. 

The catalogue can be used in further statistical studies of the distribution of microlensing events in the Galaxy as well as for individual studies searching for black hole or planetary lenses. Some of the events at the time of publication are already 8~years old and can be targeted with high-resolution observations in order to see the source and lens separated, provided their relative proper motion is high \citep[e.g.][]{2007Kozlowski, 2021Fatima}. The remaining events will also be suitable for such studies within the next 5-10 years. 

Many of the microlensing events found in \gaia were also found by other surveys. The potential further work includes more detailed modelling of the \gaia and other data combined, with the inclusion of the annual and space microlensing parallax. Such studies could lead to the identification of potential dark lenses, including black holes and neutron stars \citep[e.g.][]{2002Mao, 2002Bennett, 2016Wyrzykowski, 2021Mroz}.

Future \gaia data releases will provide more photometric observations of the events published here, as well as new events from a much longer time baseline. Moreover, \gaia astrometric time-series available for photometrically identified microlensing events will yield complete solutions for lens masses and distances, and discoveries of dozens of black holes and neutron stars in the Milky Way. When combined with other \gaia data products, such as colour light curves, low-resolution spectroscopic time-series, and radial velocity measurements, \gaia will revolutionise studies of the populations of lenses in the Galaxy. 


\begin{acknowledgements}
This work has made use of data from the European Space Agency (ESA) mission \gaia (\url{https://www.cosmos.esa.int/gaia}), processed by the \gaia Data Processing and Analysis Consortium (DPAC, \url{https://www.cosmos.esa.int/web/gaia/dpac/consortium}). 
Funding for the DPAC has been provided by national institutions, some of which participate in the \gaia Multilateral Agreement, 
which include, 
for Switzerland, the Swiss State Secretariat for Education, Research and Innovation through the ESA Prodex program, the `Mesures d'accompagnement', the `Activit\'{e}s Nationales Compl\'{e}mentaires', the Swiss National Science Foundation, and the Early Postdoc.Mobility fellowship;
for Belgium, the BELgian federal Science Policy Office (BELSPO) through PRODEX grants;
for Italy, Istituto Nazionale di Astrofisica (INAF) and the Agenzia Spaziale Italiana (ASI) through grants I/037/08/0,  I/058/10/0,  2014-025-R.0, and 2014-025-R.1.2015 to INAF (PI M.G. Lattanzi);
for Hungary, the Lend\"ulet grants LP2014-17 and LP2018-7 from the Hungarian Academy of Sciences, and the NKFIH grants K-115709, PD-116175, PD-121203 from the Hungarian National Research, Development, and Innovation Office. 

L.W., K.K. and K.A.R. have been supported by the Polish National Science Centre (NCN) grants
Harmonia No. 2018/30/M/ST9/00311 and Daina No. 2017/27/L/ST9/03221 as well as 
the European Union's Horizon 2020 research and innovation programme under grant agreement No 101004719 (OPTICON-RadioNet Pilot, ORP) and
MNiSW grant DIR/WK/2018/12.

L.M.\ and E.P.\ have been supported by the J\'anos Bolyai Research Scholarship of the Hungarian Academy of Sciences.

We would like to thank the OGLE team for making their data and statistics available to the public.
We also thank numerous researchers and students involved in Gaia microlensing studies, whose comments and discussions helped in this work, in particular to Andrzej Udalski, Przemek Mr{\'o}z, Radek Poleski, Jan Skowron, Pawe{\l} Zieli{\'n}ski, Mariusz Gromadzki, Maja Jab{\l}o{\'n}ska, Piotr Trzcionkowski.

This work made use of software from Postgres-XL (\url{https://www.postgres-xl.org}), Java (\url{https://www.oracle.com/java/}), R~\citep{R-citation}, and TOPCAT/STILTS \citep{2020arXiv201210560T,2019ASPC..523...43T,2017arXiv170702160T,2005ASPC..347...29T}. 

This research has made use of the SIMBAD database,
operated at CDS, Strasbourg, France \citep{2000WegnerSIMBAD}.

\end{acknowledgements}

\bibliographystyle{aa}
\raggedbottom
\bibliography{GDR3_microlensing}

%
%


\clearpage
\onecolumn

\begin{appendix}

\section{Data fields in \gdr{3} for candidate microlensing events.}
\label{app:fields}

Table \ref{tab:parameters} describes all data fields available in table $ vari\_microlensing$ of \gdr{3} for microlensing event candidates.

\begin{table*}
\caption{Data fields available in the $vari\_microlensing$ table of \gdr{3}\label{tab:parameters}} 
\centering                  
\begin{tabular}{rl}
\hline\hline                 
{\verb source_id } & unique source identifier of the microlensing event candidate  \\
{\verb paczynski0_bp0 } & Level~0 \gbp baseline magnitude [mag] \\
{\verb paczynski0_bp0_error } &  Level~0 \gbp baseline magnitude error [mag] \\
{\verb paczynski0_chi2 } & Level~0 goodness of fit ($\chi^2$)  \\
{\verb paczynski0_chi2_dof } & Level~0 reduced goodness of fit ($\chi^2/dof$)  \\
{\verb paczynski0_g0 } & Level~0 \gmag baseline magnitude [mag] \\
{\verb paczynski0_g0_error } &  Level~0 \gmag baseline magnitude error [mag] \\
{\verb paczynski0_rp0 } &  Level~0 \grp baseline magnitude [mag]  \\
{\verb paczynski0_rp0_error } &  Level~0 \grp baseline magnitude error [mag]  \\
{\verb paczynski0_te } &  Level~0 timescale ($t_E$) [days]  \\
{\verb paczynski0_te_error } & Level~0 timescale ($t_E$) error [days] \\
{\verb paczynski0_tmax } &  Level~0 time of maximum ($t_{max}$) [Baricentric Julian Days-2455197.5] \\
{\verb paczynski0_tmax_error } & Level~0 time of maximum ($t_{max}$) [days]  \\
{\verb paczynski0_u0 } &  Level~0 impact parameter ($u_{0}$)  \\
{\verb paczynski0_u0_error } & Level~0 impact parameter ($u_{0}$) error  \\
{\verb paczynski1_bp0 } & Level~1 \gbp baseline magnitude [mag] \\
{\verb paczynski1_bp0_error } &  Level~1 \gbp baseline magnitude error [mag] \\
{\verb paczynski1_chi2 } & Level~1 goodness of fit ($\chi^2$)  \\
{\verb paczynski1_chi2dof } & Level~1 reduced goodness of fit ($\chi^2/dof$)  \\
{\verb paczynski1_fs_bp } &  Level~1 \gbp blending parameter  \\
{\verb paczynski1_fs_bp_error } &  Level~1 \gbp blending parameter error \\
{\verb paczynski1_fs_g } &  Level~1 \gmag blending parameter \\
{\verb paczynski1_fs_g_error } & Level~1 \gmag blending parameter error  \\
{\verb paczynski1_fs_rp } & Level~1 \grp blending parameter  \\
{\verb paczynski1_fs_rp_error } & Level~1 \grp blending parameter error  \\
{\verb paczynski1_g0 } & Level~1 \gmag baseline magnitude [mag]   \\
{\verb paczynski1_g0_error } &  Level~1 \gmag baseline magnitude error [mag]  \\
{\verb paczynski1_rp0 } & Level~1 \grp baseline magnitude [mag]  \\
{\verb paczynski1_rp0_error } & Level~1 \grp baseline magnitude error [mag]  \\
{\verb paczynski1_te } & Level~1 timescale ($t_E$) [days]  \\
{\verb paczynski1_te_error } & Level~1 timescale ($t_E$) error [days]  \\
{\verb paczynski1_tmax } &  Level~1 time of maximum ($t_{max}$) [Baricentric Julian Days-2455197.5] \\
{\verb paczynski1_tmax_error } & Level~1 time of maximum ($t_{max}$) error [days]  \\
{\verb paczynski1_u0 } &  Level~1 impact parameter ($u_{0}$) \\
{\verb paczynski1_u0_error } &  Level~1 impact parameter ($u_{0}$) error \\
\hline                       
\hline                       
\end{tabular}
\end{table*}

\section{Example of \gaia Archive queries} \label{app:queries}

This section describes queries in ADQL that can be used to obtain different data sets using the \texttt{vari\_microlensing} catalogue. More examples of queries can be found here: \url{https://www.cosmos.esa.int/web/gaia-users/archive/writing-queries#ADQLQueryExamples}.

\subsubsection*{Example 1, the entire catalogue}

The following query will download all of the 363 sources with their equatorial and Galactic coordinates, proper motions, and \texttt{ruwe} along with parameters of the best microlensing model fits for Level~0 and Level~1.

\begin{verbatim}
SELECT
s.source_id, s.ra, s.dec,
s.l, s.b, 
s.pmra, s.pmra_error, 
s.pmdec, s.pmdec_error, 
s.ruwe,
m.*
FROM gaiaedr3.gaia_source s
INNER JOIN gaiadr3.vari_microlensing m
   ON s.source_id = m.source_id
\end{verbatim}

\subsubsection*{Example 2, sources located in the Galactic bulge}

The following query will download all of the sources located only within 10\deg from the Galactic centre. It returns their equatorial and Galactic coordinates, proper motions, and \texttt{ruwe} along with parameters of the best microlensing model fits for Level~0 and Level~1.

\begin{verbatim}
SELECT
s.source_id, s.ra, s.dec,
s.l, s.b, 
s.pmra, s.pmra_error, 
s.pmdec, s.pmdec_error, 
s.ruwe,
m.*
FROM gaiaedr3.gaia_source s
INNER JOIN gaiadr3.vari_microlensing m
   ON s.source_id = m.source_id
WHERE s.l<10. OR s.l>350.
\end{verbatim}

\subsubsection*{Example 3, sources located outside of the Galactic bulge with Einstein timescales larger than 100~days}

The following query will download all of the sources located outside of the 10\deg from the Galactic centre with Einstein timescales of Level~0 larger than 100~days. It returns their equatorial and Galactic coordinates, proper motions, and \texttt{ruwe} along with parameters of the best microlensing model fits for Level~0 and Level~1.

\begin{verbatim}
SELECT
s.source_id, s.ra, s.dec,
s.l, s.b, 
s.pmra, s.pmra_error, 
s.pmdec, s.pmdec_error, 
s.ruwe,
m.*
FROM gaiaedr3.gaia_source s
INNER JOIN gaiadr3.vari_microlensing m
   ON s.source_id = m.source_id
WHERE s.l BETWEEN 10. AND 350.
AND m.paczynski0_te > 100.
\end{verbatim}

\subsubsection*{Accessing light curves} \label{app:lightcurves}

A light curve in VOTable format for a single source can be downloaded using a DataLink service. Instructions on how to do this can be found here: \url{https://www.cosmos.esa.int/web/gaia-users/archive/ancillary-data}. The output table has the following columns that are of interest: \texttt{source\_id}, \texttt{band}, \texttt{time}, \texttt{mag}, \texttt{flux\_over\_error}. In order to transform \texttt{flux\_over\_error} into uncertainty for a given band, multiply \texttt{flux\_over\_error} by $\frac{2.5}{ln(10)}$ and then
use formulae 9 and 10 from this paper (Section \ref{sec:data}).

\section{Selection of candidate microlensing events \label{app:selection}}
The selection of candidate microlensing at the Microlensing SOS was done by computing membership score and by applying a set of cuts on the available parameters. 

\subsubsection*{Membership score}
The initial score is set to 1 and then a series of tests are performed and penalties applied accordingly:

\begin{itemize}
    \item if the amplitude in \gmag-band is smaller than 0.25 mag, the score is multiplied by 0.6,
    \item if the amplitude in \gmag-band is between 0.25 and 0.5 mag, the score is multiplied by 0.9,
    \item if $\chi^2/dof$ of all fitting Levels is larger than 10, the score is multiplied by 0.5,
    \item if the lowest $\chi^2$ of all fit Levels is between 3 and 10, the score is multiplied by 0.9,
    \item if the time of maximum $t_{max}$ including the error bar of the model with lowest $\chi^2$ is not between the beginning and the end of the light curve in \gmag, the score is multiplied by 0.9,
    \item if the absolute value of the impact parameter $u_0$ including the error bar of the model with the lowest $\chi^2$ is not between 0 and 1.5, the score is multiplied by 0.7,
    \item if the absolute value of the impact parameter $u_0$ including the
error bar of the model with lowest $\chi^2$ is between 0 and 0.001, the score is multiplied by 0.9,
    \item if the model with the lowest $\chi^2$ has baseline magnitude $G_0$ below 20 mag
    , the score is multiplied by 0.8,
    \item if the Einstein time $t_E$ including the
error bar is not within 2 and 500~days, the score is multiplied by 0.2,
    \item if one of the parallax components is larger than 1, the score is multiplied by 0.95,
    \item  if both of the parallax components are larger than 1, the score is multiplied by 0.9.
\end{itemize}

\subsubsection*{Cuts and selection}

Table \ref{tab:cuts} presents the cuts used for selecting candidate microlensing events defining the {\bf Sample~A}. The cuts were identified in an iterative way, involving visual inspection of the candidates and preserving the largest number of known microlensing events. A more detailed description of the cuts and the parameters is available in the \gdr{3} Documentation.

\begin{table*}
\caption{Microlensing event selection cuts for {\bf Sample~A} \label{tab:cuts}} 
\centering                  
\begin{tabular}{p{0.5\linewidth} | p{0.5\linewidth}}     
\hline\hline                 
cut & description  \\  
\hline                       
\hline                       
$membership\ score>=0.85$ & membership score cut\\
\hline                       
$abs(\mathrm{\texttt{paczynski0\_u0}})<1.$ & only high amplification events\\
\hline                       
$20.<\mathrm{\texttt{paczynski0\_te}}<300.$ & timescale should be realistic and reasonable for DR3 duration\\
\hline                       
$\mathrm{\texttt{paczynski0\_g0}}<19.$ & magnitude cut \\
\hline                       
$skewness<0.$ & only events getting brighter over the baseline\\
\hline                       
$1.<(median_{BP}-median_{RP})<4.$ & colour cut to remove blue outburst and very red variables\\
\hline                       
$(median<17$ and $(max-min)>0.25)$ \\or \\$(median>17$ and $(max-min)>0.5)$ & amplitude cut as a function of the magnitude\\
\hline                       
$stddev_{G}<stddev_{RP}+0.04$ or $number\ points_{RP}<2$ & RP light curve robustness\\
\hline                       
$\mathrm{\texttt{paczynski0\_tmax}}-1*\mathrm{\texttt{paczynski0\_te}}>t\_min$ & at least 1 timescale within the data\\
\hline                       
$extractor\_number\_points\_in\_event>3$ & more than 3 outlying points identified by the Extractor\\
\hline                       
$duration > 135$ & distance in days between outlying points from Extractor\\
\hline                       
$max\_sigma\_in\_event>50$ & ``strength'' of the event from Extractor\\
\hline                       
$abs(\mathrm{\texttt{paczynski0\_u0}}/\mathrm{\texttt{paczynski0\_u0\_error}})<200$ & only fits with robust \texttt{paczynski0\_u0} determination selected\\
\hline                       
$\mathrm{\texttt{paczynski0\_u0\_error}}!=0$ and $\mathrm{\texttt{paczynski0\_te\_error}}>0$ & only fits with robust \texttt{paczynski0\_u0} and \texttt{paczynski0\_te} selected\\
\hline                       
$\mathrm{\texttt{paczynski2\_chi2\_dof}}<1.9$ & goodness of the parallax fit\\
\hline                       
$abs(\mathrm{\texttt{paczynski2\_parallax\_east}})<3.5$ \\and $abs(\mathrm{\texttt{paczynski2\_parallax\_north}})<3.5$ & only reasonable parallax vector selected\\
\hline                       
$(\mathrm{\texttt{paczynski0\_tmax}}<6,824.5$ or $\mathrm{\texttt{paczynski0\_tmax}}>7,189.5)$ \\ and $log(abbe)<0.8*log(-skewness)-0.6$  & \\
or & \\
$6,824.5<=\mathrm{\texttt{paczynski0\_tmax}}<=7,189.5$ \\ and $log(abbe)<1.2*log(-skewness)-0.84$  
 & skew-Abbe selection more strict for events in a period of times with more photometric outliers\\
\hline                       
\hline                       
\end{tabular}
\tablefoot{If no index is given to parameters, they refer to \gmag light curve.  \\}
\end{table*}

\section{List of events from \gaia DR3}
\label{app:table}

Table \ref{tab:all} contains a list of all 363 \gaia DR3 microlensing events, sorted by their baseline magnitude, with the brightest labelled with number 1. The table lists the \gaia source\_Id and the alias name, following the convention GaiaDR3-ULENS-NNN. The equatorial and galactic coordinates are given, the baseline magnitude in \gmag-band. The last column marks the method the event was found, with {\bf A} for {\bf Sample A}, {\bf B} for {\bf Sample B} and {\bf A+B} for events from both samples (see Section  \ref{sec:method}).
Data from the table are also available in the on-line version of the article. 

\onecolumn
\begin{footnotesize}
\begin{longtable}[]{|r|l|l|l|l|l|l|l|}
\caption[All events table]{\gaia DR3 microlensing events.} \label{tab:all} \\
\hline 
\multicolumn{1}{|c|}{\textbf{GaiaDR3-}}
& \multicolumn{1}{|c|}{\textbf{\gaia DR3}}
& \multicolumn{1}{c|}{\textbf{RA$_{J2000}$}}
& \multicolumn{1}{c|}{\textbf{Dec$_{J2000}$}} 
& \multicolumn{1}{c|}{\textbf{Gal.long.}}
& \multicolumn{1}{c|}{\textbf{Gal.lat.}} 
& \multicolumn{1}{c|}{\textbf{Baseline}} 
& \multicolumn{1}{c|}{\textbf{Method}} \\ 
\multicolumn{1}{|c|}{\textbf{ULENS-}}
& \multicolumn{1}{|c|}{\textbf{sourceid}}
& \multicolumn{1}{c|}{\textbf{[deg]}}
& \multicolumn{1}{c|}{\textbf{[deg]}} 
& \multicolumn{1}{c|}{\textbf{[deg]}}
& \multicolumn{1}{c|}{\textbf{[deg]}} 
& \multicolumn{1}{c|}{\textbf{\gmag [mag]}} 
& \multicolumn{1}{c|}{\textbf{}} \\ 
\hline
\endfirsthead
\multicolumn{8}{c}%
{{\bfseries \tablename\ \thetable{} -- continued from previous page}} \\
\hline 
\multicolumn{1}{|c|}{\textbf{GaiaDR3-}}
& \multicolumn{1}{|c|}{\textbf{\gaia DR3}}
& \multicolumn{1}{c|}{\textbf{RA$_{J2000}$}}
& \multicolumn{1}{c|}{\textbf{Dec$_{J2000}$}} 
& \multicolumn{1}{c|}{\textbf{Gal.long.}}
& \multicolumn{1}{c|}{\textbf{Gal.lat.}} 
& \multicolumn{1}{c|}{\textbf{Baseline}} 
& \multicolumn{1}{c|}{\textbf{Method}} \\ 
\multicolumn{1}{|c|}{\textbf{ULENS-}}
& \multicolumn{1}{|c|}{\textbf{sourceid}}
& \multicolumn{1}{c|}{\textbf{[deg]}}
& \multicolumn{1}{c|}{\textbf{[deg]}} 
& \multicolumn{1}{c|}{\textbf{[deg]}}
& \multicolumn{1}{c|}{\textbf{[deg]}} 
& \multicolumn{1}{c|}{\textbf{\gmag [mag]}} 
& \multicolumn{1}{c|}{\textbf{}} \\ 
\hline
\endhead
\hline \multicolumn{8}{|r|}{{Continued on next page}} \\ \hline
\endfoot
\hline \hline
\endlastfoot
\input{table-events-uniq-sort-mag-j2000}
\hline
\end{longtable}
\end{footnotesize}
\twocolumn

\section{Table with cross-match}
\label{app:xm}

Table \ref{tab:xm} lists \gaia DR3 events which have counterparts in other catalogues of microlensing events or real-time microlensing search programmes: OGLE-IV Bulge \citep{Mroz1}, OGLE-IV Disk \citep{Mroz2}, OGLE-IV's Early Warning System web-pages \citep{2015Udalski}, KMT-Net \citep{KMTNet}, MOA \citep{MOA}, ASAS-SN\citep{ASASSN} and \gaia Science Alerts \citep{Hodgkin2021}.
Data from the table are also available in the on-line version of the article.

\onecolumn
\begin{footnotesize}
\begin{longtable}[]{|r|l|}
\caption[Cross-match table]{\gdr{3} microlensing events cross-matched with other detections in OGLE, MOA, KMT-NET, ASAS-SN or Gaia Science Alerts surveys. There is 273 such events.} \label{tab:xm} \\
\hline 
\multicolumn{1}{|c|}{\textbf{GaiaDR3-}}
& \multicolumn{1}{c|}{\textbf{Cross-match}} \\ 
\multicolumn{1}{|c|}{\textbf{ULENS-}}
& \multicolumn{1}{c|}{\textbf{}} \\ 
\hline
\endfirsthead
\multicolumn{2}{c}%
{{\bfseries \tablename\ \thetable{} -- continued from previous page}} \\
\hline 
\multicolumn{1}{|c|}{\textbf{GaiaDR3-}}
& \multicolumn{1}{c|}{\textbf{Cross-match}} \\ 
\multicolumn{1}{|c|}{\textbf{ULENS-}}
& \multicolumn{1}{c|}{\textbf{}} \\ 
\hline
\endhead
\hline \multicolumn{2}{|r|}{{Continued on next page}} \\ \hline
\endfoot
\noalign{\vskip 2mm}
\multicolumn{2}{l}{\tablefoot{OGLE survey detections come either from the EWS (OGLE-year-field-nnnn) or catalogues published in \cite{Mroz1} and \cite{Mroz2} (BLGnnn.nn.nnnn or GDnnnn.nn.nnnnn). }} \\
\hline \hline
\endlastfoot
\input{table-sort-id-xm}
\hline
\end{longtable}
\end{footnotesize}
\twocolumn

\end{appendix}

\end{document}

%% file: table-events-uniq-sort-mag-j2000.tex
001 & 6059400613544951552 & 184.4363263 & -59.0294197 & 298.6005085 & 3.5586850 & 13.58 & A \\ 
002 & 4135677580466393216 & 258.0716963 & -16.8677137 & 6.0098740 & 12.9894820 & 13.64 & A \\ 
003 & 4084984459435880064 & 284.4366819 & -20.7757293 & 14.9165537 & -10.6404670 & 14.16 & A \\ 
004 & 4043236380953360512 & 268.4052228 & -33.1237924 & 357.2830138 & -3.6192437 & 14.38 & A \\ 
005 & 4046189531753775872 & 276.2347519 & -30.9712872 & 2.3615295 & -8.4309626 & 14.40 & A \\ 
006 & 5369575367669793536 & 175.9545194 & -51.7786254 & 292.4152033 & 9.7070281 & 14.74 & A \\ 
007 & 4042926314286962944 & 270.2539729 & -32.6895521 & 358.4346692 & -4.7498989 & 14.78 & A \\ 
008 & 6028496189222706432 & 255.7845506 & -31.2111200 & 352.9458509 & 6.3248291 & 15.00 & A \\ 
009 & 4103350800428736384 & 279.8111863 & -15.4369500 & 17.8014157 & -4.3444691 & 15.08 & A \\ 
010 & 4056251952898990848 & 269.5945985 & -29.9489827 & 0.5437424 & -2.9060128 & 15.16 & A \\ 
011 & 4093461453632265600 & 279.0502629 & -18.0503157 & 15.1331452 & -4.8801400 & 15.23 & A+B \\ 
012 & 5305778988625143296 & 143.5517537 & -57.8164443 & 279.0375096 & -4.4075008 & 15.46 & A \\ 
013 & 4052272339270449024 & 274.7611076 & -27.7293739 & 4.6769250 & -5.8077405 & 15.53 & A \\ 
014 & 4050176120351393792 & 270.6320911 & -30.1235644 & 0.8360192 & -3.7732156 & 15.61 & B \\ 
015 & 4053164279372999680 & 276.0098469 & -25.1043278 & 7.5357066 & -5.5861482 & 15.67 & A \\ 
016 & 5882824406244672256 & 234.7379625 & -56.0303213 & 324.8332656 & -0.4528306 & 15.96 & A \\ 
017 & 4041047008398705664 & 267.0329353 & -35.5803325 & 354.5860534 & -3.8936482 & 15.97 & A \\ 
018 & 4050208551598764032 & 271.1193843 & -29.8161981 & 1.3115071 & -3.9914357 & 16.00 & A+B \\ 
019 & 4159994512750671232 & 275.9289525 & -7.3020659 & 23.2666791 & 2.7836325 & 16.01 & A \\ 
020 & 4067907184112930944 & 265.7034019 & -25.5169917 & 2.5881853 & 2.3205060 & 16.03 & A \\ 
021 & 4041238259061973504 & 264.6744198 & -35.5066956 & 353.6436127 & -2.2195980 & 16.06 & A \\ 
022 & 5956630082001769984 & 269.5337814 & -42.5974274 & 349.4320429 & -9.0692191 & 16.08 & A \\ 
023 & 6059816980526317568 & 191.5840141 & -58.8941185 & 302.2714390 & 3.9710873 & 16.11 & B \\ 
024 & 5951752068649725312 & 262.4024369 & -46.2347526 & 343.6506395 & -6.5542396 & 16.13 & A \\ 
025 & 4108020361965432064 & 260.8780868 & -27.3787487 & 358.6732768 & 4.9403756 & 16.14 & A+B \\ 
026 & 5515709499618656768 & 127.4644053 & -47.1922367 & 264.6184588 & -4.7449336 & 16.17 & A \\ 
027 & 4039110875886635008 & 271.1335543 & -34.8929303 & 356.8557550 & -6.4587185 & 16.25 & A \\ 
028 & 4041760939405737344 & 266.8708058 & -34.1758834 & 355.7240287 & -3.0570121 & 16.29 & A \\ 
029 & 4063277591626468608 & 271.8813320 & -26.6791216 & 4.3856614 & -3.0549313 & 16.32 & B \\ 
030 & 4050909181279109888 & 273.1546856 & -27.8870166 & 3.8672279 & -4.6272229 & 16.33 & B \\ 
031 & 4062254461725875328 & 270.0523139 & -29.7174753 & 0.9422900 & -3.1359064 & 16.33 & B \\ 
032 & 3292310414861319936 & 71.4940480 & 8.2457040 & 189.7124147 & -23.1443290 & 16.33 & B \\ 
033 & 4104470897830990464 & 278.5822511 & -14.3406907 & 18.2360939 & -2.7875174 & 16.37 & A \\ 
034 & 5934751007588607616 & 244.2287289 & -51.5584142 & 331.9908631 & -0.6708695 & 16.39 & A \\ 
035 & 4040806872563108736 & 267.9705341 & -34.9748899 & 355.4993650 & -4.2436140 & 16.46 & A \\ 
036 & 1899263645288575232 & 333.5701007 & 33.1238321 & 88.7843066 & -19.1293885 & 16.47 & A \\ 
037 & 4043988927861984256 & 269.7853139 & -31.5969500 & 359.1920881 & -3.8662686 & 16.48 & B \\ 
038 & 4056260504334777856 & 269.6764775 & -29.8055925 & 0.7036549 & -2.8962750 & 16.51 & A \\ 
039 & 4043160166667499392 & 268.7876558 & -33.6170544 & 357.0164674 & -4.1432569 & 16.51 & A+B \\ 
040 & 6064734619321769856 & 207.0171663 & -55.3840033 & 310.9724261 & 6.6011920 & 16.52 & A \\ 
041 & 4039570020686594304 & 272.3027922 & -33.7608736 & 358.3209432 & -6.7678286 & 16.56 & A \\ 
042 & 4119903643141485824 & 266.0036138 & -19.4118124 & 7.9532570 & 5.2698599 & 16.57 & A \\ 
043 & 4041998223399082752 & 270.1778302 & -35.1540559 & 356.2447677 & -5.9003667 & 16.59 & A \\ 
044 & 5946723452194568704 & 264.7253653 & -50.3535242 & 340.9413001 & -10.0417012 & 16.60 & A \\ 
045 & 4045413207825559808 & 274.3978285 & -33.5364601 & 359.3393230 & -8.2074141 & 16.60 & A \\ 
046 & 4064632323134994816 & 272.8062077 & -26.1704776 & 5.2316184 & -3.5359947 & 16.61 & B \\ 
047 & 4055692507632670720 & 268.0851347 & -31.2949870 & 358.7251839 & -2.4599644 & 16.62 & A \\ 
048 & 4053054431284531968 & 274.4009350 & -25.8347569 & 6.2088590 & -4.6391281 & 16.65 & B \\ 
049 & 4049116805296535296 & 271.8379282 & -31.2615423 & 0.3424126 & -5.2319770 & 16.66 & B \\ 
050 & 4063140771207991168 & 271.0991224 & -27.3203639 & 3.4851742 & -2.7580522 & 16.67 & A \\ 
051 & 443788525336160512 & 57.0194916 & 52.7031377 & 147.9094846 & -1.3833894 & 16.68 & A \\ 
052 & 5960669348421847040 & 265.1301564 & -40.6199623 & 349.4853810 & -5.2296012 & 16.70 & A \\ 
053 & 4050164747272227584 & 270.5812067 & -30.3334375 & 0.6310845 & -3.8377637 & 16.70 & A+B \\ 
054 & 4077869107439976320 & 278.4208477 & -23.2662270 & 10.1877968 & -6.7097222 & 16.73 & A \\ 
055 & 4064540513876708096 & 273.1534553 & -26.7648570 & 4.8568517 & -4.0928488 & 16.75 & B \\ 
056 & 4065624361840345088 & 274.1892381 & -23.3639956 & 8.3030297 & -3.3059357 & 16.76 & B \\ 
057 & 4254983555662524288 & 283.5813926 & -4.6656679 & 29.1048953 & -2.7665401 & 16.79 & A \\ 
058 & 4105328207729395584 & 280.7047581 & -13.5224417 & 19.9060564 & -4.2487467 & 16.80 & A \\ 
059 & 4041267941519396352 & 264.8862278 & -35.1859206 & 354.0074049 & -2.1951870 & 16.85 & A \\ 
060 & 5962405713502000768 & 265.4393819 & -37.0974025 & 352.6188565 & -3.5844998 & 16.87 & A \\ 
061 & 4063286589604553600 & 271.6101851 & -26.6629815 & 4.2820978 & -2.8352249 & 16.89 & A \\ 
062 & 5933611054559826944 & 247.0982980 & -53.1454425 & 332.1007425 & -3.0324750 & 16.91 & A \\ 
063 & 5853767990462652032 & 213.3033358 & -63.5950879 & 311.8743962 & -2.1521389 & 16.94 & A \\ 
064 & 4043698789930892160 & 270.1587362 & -32.2749798 & 358.7574490 & -4.4763817 & 16.95 & B \\ 
065 & 5900412881430948608 & 226.8217997 & -51.5178025 & 323.3866456 & 5.8769005 & 16.98 & A \\ 
066 & 4048971261782646528 & 272.9183590 & -31.3252741 & 0.7277628 & -6.0742305 & 17.00 & A+B \\ 
067 & 4042928139682133120 & 270.2881994 & -32.5577055 & 358.5641559 & -4.7103806 & 17.02 & A \\ 
068 & 4053988127168351104 & 267.0739349 & -32.8775550 & 356.9254031 & -2.5331136 & 17.04 & B \\ 
069 & 4043375503532180608 & 269.0711792 & -32.4500107 & 358.1485840 & -3.7652406 & 17.08 & A \\ 
070 & 4091460990925014528 & 275.5313927 & -20.1550981 & 11.7261426 & -2.8974067 & 17.11 & B \\ 
071 & 4090523111881884416 & 273.9130130 & -22.3512268 & 9.0753924 & -2.6019945 & 17.12 & A \\ 
072 & 6057091226831797632 & 193.5817284 & -59.1361017 & 303.3032122 & 3.7335642 & 17.13 & A \\ 
073 & 4050046339277832832 & 272.4689862 & -30.0405762 & 1.6797474 & -5.1244076 & 17.14 & A \\ 
074 & 5958979330397499392 & 264.1271363 & -42.2729117 & 347.6769302 & -5.4647202 & 17.14 & A+B \\ 
075 & 5946702939434812800 & 265.5939850 & -49.2145748 & 342.2266449 & -9.9450824 & 17.16 & A \\ 
076 & 4041606110282426624 & 268.4350280 & -34.2213035 & 356.3451998 & -4.1934739 & 17.19 & B \\ 
077 & 4063274151469596928 & 271.7184311 & -26.7337580 & 4.2671704 & -2.9541322 & 17.19 & B \\ 
078 & 4042619211263272960 & 270.9847730 & -33.6993888 & 357.8479186 & -5.7757472 & 17.20 & B \\ 
079 & 4050127359632060672 & 271.8932501 & -29.7381780 & 1.7060521 & -4.5413865 & 17.23 & A \\ 
080 & 4043611005164862336 & 268.5469471 & -32.0885254 & 358.2389748 & -3.2004290 & 17.25 & A \\ 
081 & 4268189072623902976 & 288.5624318 & 2.9221359 & 38.1424521 & -3.7313608 & 17.27 & B \\ 
082 & 5989557607714349952 & 236.4613593 & -43.2927142 & 333.4771983 & 8.9988355 & 17.29 & A \\ 
083 & 4062324383827132928 & 269.6868712 & -29.4273326 & 1.0367233 & -2.7160153 & 17.30 & A \\ 
084 & 6028177846244738944 & 255.9928266 & -31.8704702 & 352.5246321 & 5.7855694 & 17.30 & A \\ 
085 & 4052203860288017024 & 273.6769567 & -27.9709295 & 4.0120583 & -5.0734779 & 17.31 & B \\ 
086 & 4118621539469673600 & 267.0133369 & -21.4765031 & 6.6695749 & 3.3937819 & 17.32 & B \\ 
087 & 4041534646190035456 & 267.4376260 & -34.6244616 & 355.5788166 & -3.6886289 & 17.37 & A+B \\ 
088 & 4043504794840743040 & 268.1614033 & -32.4895684 & 357.7274280 & -3.1218347 & 17.37 & A \\ 
089 & 4063987738711159680 & 270.4467027 & -26.5331095 & 3.8853809 & -1.8646211 & 17.38 & A \\ 
090 & 4050237070478829696 & 270.0699656 & -30.0015261 & 0.7025721 & -3.2896434 & 17.39 & B \\ 
091 & 4065785955602043264 & 272.0506217 & -24.9464191 & 5.9772280 & -2.3491989 & 17.39 & A \\ 
092 & 4050901140883623296 & 272.9654838 & -28.0545855 & 3.6396437 & -4.5598471 & 17.40 & B \\ 
093 & 5971477302907164032 & 253.7852905 & -36.7690218 & 347.5207022 & 4.2147304 & 17.40 & B \\ 
094 & 4053510694262196224 & 264.9520692 & -33.8109647 & 355.2016788 & -1.5105953 & 17.41 & B \\ 
095 & 6028136786357495936 & 256.2886509 & -31.8475311 & 352.6949212 & 5.5986369 & 17.42 & A+B \\ 
096 & 1829267601203551488 & 305.3910139 & 22.1809621 & 63.1316255 & -8.2590631 & 17.42 & A \\ 
097 & 4068509377297713408 & 267.5594056 & -23.6716987 & 5.0426374 & 1.8343494 & 17.42 & B \\ 
098 & 4053409539185281536 & 265.2059834 & -34.4377579 & 354.7814738 & -2.0211318 & 17.48 & B \\ 
099 & 5940987678007330816 & 250.5037981 & -48.4425257 & 337.0261825 & -1.4466471 & 17.49 & A \\ 
100 & 4064609130464186496 & 272.7319736 & -26.4511682 & 4.9527891 & -3.6119659 & 17.50 & A \\ 
101 & 4090456861958694784 & 274.5941539 & -22.5706540 & 9.1798581 & -3.2603976 & 17.55 & A+B \\ 
102 & 5989806337862480256 & 243.9410728 & -47.8559074 & 334.4268182 & 2.1240104 & 17.56 & B \\ 
103 & 5969060713789360000 & 249.8322853 & -40.8216865 & 342.4247810 & 3.9565907 & 17.56 & B \\ 
104 & 4052245775095993088 & 274.2059498 & -27.5994952 & 4.5620342 & -5.3122062 & 17.56 & B \\ 
105 & 4046296562307158784 & 274.2544336 & -31.3835109 & 1.2124290 & -7.1102658 & 17.56 & B \\ 
106 & 5857951284237224576 & 200.4698542 & -65.7444224 & 306.0551224 & -3.0574410 & 17.59 & A \\ 
107 & 6062116819613165952 & 200.4157781 & -58.4573281 & 306.8870844 & 4.1821170 & 17.59 & A \\ 
108 & 4040599133555482496 & 267.7088983 & -35.9483837 & 354.5493178 & -4.5531650 & 17.59 & B \\ 
109 & 4063547483074697600 & 269.3491914 & -27.3357397 & 2.7020446 & -1.4146956 & 17.59 & B \\ 
110 & 4059259353363569792 & 259.5895591 & -29.8185933 & 356.0126907 & 4.4914631 & 17.61 & A \\ 
111 & 4254940846504352640 & 283.2161231 & -5.0419195 & 28.6038404 & -2.6131505 & 17.63 & A+B \\ 
112 & 5958965925733552384 & 262.9667349 & -41.8291805 & 347.5873305 & -4.4996535 & 17.66 & A \\ 
113 & 4056324516377709184 & 269.0577079 & -29.6127381 & 0.6021889 & -2.3343562 & 17.67 & A \\ 
114 & 4061145771721121792 & 262.8249712 & -28.2275742 & 358.9201369 & 3.0363208 & 17.67 & B \\ 
115 & 4059548250037907200 & 261.7620984 & -29.0173858 & 357.7451697 & 3.3798476 & 17.71 & B \\ 
116 & 4052262473792354432 & 274.8619111 & -27.7150372 & 4.7313128 & -5.8801326 & 17.72 & A+B \\ 
117 & 1819571454860230400 & 298.4312248 & 15.3917600 & 53.8093713 & -6.2759832 & 17.73 & A \\ 
118 & 3355845866073702144 & 99.4166892 & 13.9551113 & 198.9501232 & 3.3306324 & 17.73 & A+B \\ 
119 & 4037380691305871616 & 270.5039509 & -36.5533798 & 355.1430158 & -6.8108670 & 17.73 & A \\ 
120 & 6030183600333074816 & 255.8594974 & -28.5190841 & 355.1526169 & 7.8921071 & 17.75 & A \\ 
121 & 4053893225594238720 & 265.3546739 & -33.5246112 & 355.6222160 & -1.6430896 & 17.75 & B \\ 
122 & 5254312773454650880 & 158.7228148 & -60.8680131 & 287.0649132 & -2.2744724 & 17.79 & A+B \\ 
123 & 5960494487427767040 & 261.7741444 & -39.0160049 & 349.4407582 & -2.1942944 & 17.82 & A \\ 
124 & 4068670073424544384 & 267.3609727 & -23.3558996 & 5.2205846 & 2.1524846 & 17.84 & B \\ 
125 & 4055674881038332928 & 268.2522962 & -31.4792946 & 358.6386626 & -2.6764402 & 17.85 & A \\ 
126 & 5864390372320262144 & 207.5090789 & -63.4576958 & 309.4227326 & -1.3268294 & 17.85 & B \\ 
127 & 4064792821720384640 & 272.3866551 & -26.2616693 & 4.9705014 & -3.2495582 & 17.86 & B \\ 
128 & 4043503351765043072 & 268.1198509 & -32.5218773 & 357.6817483 & -3.1080005 & 17.86 & B \\ 
129 & 4093948842138626688 & 273.9036943 & -20.9755081 & 10.2826516 & -1.9405510 & 17.87 & A \\ 
130 & 4065149544601352320 & 274.0410626 & -25.1339538 & 6.6757581 & -4.0227630 & 17.89 & A+B \\ 
131 & 4063574219313975296 & 269.4434323 & -27.1835496 & 2.8757834 & -1.4110711 & 17.92 & B \\ 
132 & 4041815777515891456 & 266.7597584 & -33.9288129 & 355.8883587 & -2.8506721 & 17.94 & A \\ 
133 & 4126633508902613504 & 252.6945648 & -21.7779794 & 358.8621809 & 14.2478890 & 17.95 & A \\ 
134 & 5962740239926319360 & 264.6305993 & -36.1399213 & 353.0883921 & -2.5270011 & 17.95 & B \\ 
135 & 5909419320449816832 & 267.8942932 & -63.7038382 & 329.6563000 & -17.8774726 & 17.97 & A \\ 
136 & 6661635838220317952 & 288.7314765 & -47.9782019 & 349.5217382 & -23.5592334 & 17.99 & A \\ 
137 & 3012224323198211968 & 85.6878701 & -9.3450754 & 213.6417900 & -19.4698544 & 18.00 & A \\ 
138 & 5836132614159541504 & 241.2056505 & -56.7787973 & 327.1387648 & -3.2897542 & 18.01 & B \\ 
139 & 4058941289651705600 & 259.8037421 & -30.9097048 & 355.2212356 & 3.7159046 & 18.01 & A \\ 
140 & 5982859447285974528 & 237.5862424 & -49.4631589 & 330.2113073 & 3.6935315 & 18.01 & B \\ 
141 & 5928438230256168832 & 249.6186558 & -57.0170251 & 330.2195255 & -6.7138346 & 18.01 & A \\ 
142 & 4110263189512077696 & 263.5394616 & -24.8825167 & 2.0827199 & 4.3158868 & 18.03 & A \\ 
143 & 6028476707251295488 & 255.1669862 & -30.9980304 & 352.7939735 & 6.8745986 & 18.03 & B \\ 
144 & 4116850741634492928 & 264.2705825 & -22.1296842 & 4.7761471 & 5.2232028 & 18.03 & B \\ 
145 & 4041835804931255808 & 266.3728811 & -33.7500823 & 355.8746238 & -2.4832268 & 18.04 & B \\ 
146 & 4041954689781627520 & 267.3630208 & -33.2712462 & 356.7117897 & -2.9431717 & 18.06 & A+B \\ 
147 & 4116613423221895552 & 264.0265519 & -23.4467857 & 3.5371400 & 4.7115858 & 18.06 & B \\ 
148 & 4058823431364942464 & 262.4470215 & -29.2527327 & 357.8798390 & 2.7511817 & 18.07 & B \\ 
149 & 4061084473952136832 & 265.4468053 & -26.3589413 & 1.7501594 & 2.0737903 & 18.07 & B \\ 
150 & 4146729184148567936 & 271.6896845 & -15.2721802 & 14.2701216 & 2.6485409 & 18.07 & A \\ 
151 & 4052759251109283584 & 275.4833650 & -26.6285173 & 5.9567735 & -5.8678536 & 18.10 & A+B \\ 
152 & 5851044534497774976 & 208.9291129 & -66.8117661 & 309.2106271 & -4.7268194 & 18.11 & A \\ 
153 & 2214532279378994816 & 357.6745127 & 69.9044944 & 117.7219614 & 7.6495224 & 18.11 & A \\ 
154 & 4043049837656829696 & 271.1497776 & -32.1075149 & 359.3147408 & -5.1268045 & 18.12 & B \\ 
155 & 4042922882640668544 & 270.3789088 & -32.6859422 & 358.4895173 & -4.8397995 & 18.13 & A+B \\ 
156 & 4043772908261758336 & 269.6591664 & -31.8952451 & 358.8790997 & -3.9208361 & 18.13 & B \\ 
157 & 4111395239809753856 & 262.1870809 & -23.5200710 & 2.5592456 & 6.0914192 & 18.14 & A \\ 
158 & 6733269879761239168 & 277.6758776 & -37.0713155 & 357.3095656 & -12.1642177 & 18.14 & B \\ 
159 & 4066359454004922368 & 273.4280538 & -23.5114232 & 7.8410908 & -2.7612510 & 18.14 & A+B \\ 
160 & 5980469521324165120 & 259.2930274 & -31.2017926 & 354.7299380 & 3.9067554 & 18.15 & A \\ 
161 & 6054706489138553728 & 188.7735491 & -62.0918408 & 301.0205901 & 0.7192392 & 18.16 & A \\ 
162 & 4065541653676268544 & 274.3487692 & -23.9847727 & 7.8237512 & -3.7276799 & 18.17 & B \\ 
163 & 4042333814993557888 & 268.9132839 & -34.5088596 & 356.2939688 & -4.6791999 & 18.17 & A \\ 
164 & 4178925766518902272 & 270.7344851 & -1.5572534 & 25.9361418 & 10.0482157 & 18.22 & A \\ 
165 & 4064067242840028928 & 269.8398302 & -26.1586438 & 3.9410134 & -1.2066784 & 18.24 & B \\ 
166 & 5972707141027232896 & 258.9272995 & -38.7666709 & 348.3848290 & -0.2303259 & 18.24 & A \\ 
167 & 4041152424074020352 & 265.3277727 & -35.9513341 & 353.5476212 & -2.9050249 & 18.24 & A+B \\ 
168 & 5933521890952818944 & 247.3385750 & -53.7472633 & 331.7628196 & -3.5509832 & 18.24 & B \\ 
169 & 4058485061049611264 & 264.1867294 & -30.1761843 & 357.9266938 & 0.9810429 & 18.25 & B \\ 
170 & 5977863129000915712 & 254.9885737 & -35.1240486 & 349.4207187 & 4.4703491 & 18.25 & A \\ 
171 & 4061956940776506752 & 265.0451994 & -25.5617241 & 2.2366989 & 2.8015561 & 18.26 & B \\ 
172 & 2046379503690071296 & 292.6789618 & 33.5108166 & 67.0297400 & 7.1918760 & 18.27 & A \\ 
173 & 4052251547519009152 & 274.0496310 & -27.4513041 & 4.6282303 & -5.1202905 & 18.27 & B \\ 
174 & 2073996517052513536 & 300.4303511 & 39.3734212 & 75.2493890 & 4.6788516 & 18.28 & A \\ 
175 & 4059051309459806208 & 261.1494453 & -30.1324210 & 356.5204941 & 3.2001105 & 18.29 & B \\ 
176 & 4040243987040723712 & 266.6236609 & -36.4556439 & 353.6629615 & -4.0602281 & 18.29 & B \\ 
177 & 5984758097700788992 & 241.2712500 & -47.6031209 & 333.2807177 & 3.5304936 & 18.30 & B \\ 
178 & 4090568050092602112 & 274.8290923 & -22.1125811 & 9.6871182 & -3.2366446 & 18.32 & B \\ 
179 & 5969213344108854912 & 251.0201117 & -39.8622421 & 343.7460714 & 3.9063263 & 18.34 & A+B \\ 
180 & 5870956376488659200 & 207.6610129 & -58.1113287 & 310.7058958 & 3.8639822 & 18.34 & A \\ 
181 & 4069391146859040640 & 269.3111481 & -23.1953382 & 6.2680048 & 0.6900547 & 18.37 & A \\ 
182 & 5962745015930138112 & 264.4097531 & -36.0880763 & 353.0369310 & -2.3484668 & 18.37 & B \\ 
183 & 4052042610131850624 & 277.3197534 & -26.3611762 & 6.9502111 & -7.2089360 & 18.37 & A \\ 
184 & 4096791702503683200 & 278.1534633 & -16.8257533 & 15.8346957 & -3.5602710 & 18.37 & A+B \\ 
185 & 4051359950839402368 & 274.4981001 & -28.6545384 & 3.7449944 & -6.0322559 & 18.39 & A \\ 
186 & 4053952118170529536 & 266.4911297 & -33.1603262 & 356.4302469 & -2.2610615 & 18.40 & B \\ 
187 & 4107328936697119232 & 259.2941796 & -29.3325566 & 356.2646229 & 4.9805318 & 18.41 & B \\ 
188 & 5864320239699873280 & 206.1492147 & -64.5226530 & 308.6110530 & -2.2386540 & 18.41 & A \\ 
189 & 4067943919094119936 & 265.9940280 & -25.0986050 & 3.0826089 & 2.3159390 & 18.42 & B \\ 
190 & 5962559885658176128 & 263.2992979 & -37.3678897 & 351.4779578 & -2.2890714 & 18.45 & A+B \\ 
191 & 4066190335405241216 & 272.9309166 & -24.4845947 & 6.7679335 & -2.8275151 & 18.45 & A+B \\ 
192 & 4053892503992268288 & 265.2309628 & -33.5649026 & 355.5334697 & -1.5768712 & 18.46 & B \\ 
193 & 4058603361633657088 & 262.4185974 & -30.3115913 & 356.9814792 & 2.1888248 & 18.46 & A+B \\ 
194 & 5256944518951207296 & 146.7764247 & -60.8177445 & 282.2670572 & -5.5730030 & 18.48 & A \\ 
195 & 4095830312689418368 & 271.9424380 & -18.0616657 & 11.9484674 & 1.0814379 & 18.48 & B \\ 
196 & 4058430321600567552 & 262.8396996 & -30.4251333 & 357.0862580 & 1.8228593 & 18.49 & B \\ 
197 & 4511290951726598656 & 281.4882183 & 16.9668436 & 47.5434923 & 8.8260696 & 18.49 & A \\ 
198 & 4060933948282842368 & 264.4758127 & -27.2098539 & 0.5680597 & 2.3568463 & 18.50 & B \\ 
199 & 4505592457837046144 & 280.2889447 & 12.9315238 & 43.3624144 & 8.1025887 & 18.50 & A \\ 
200 & 4042842648359086976 & 270.8783948 & -32.9919765 & 358.4265239 & -5.3553957 & 18.51 & A+B \\ 
201 & 5861169627804660096 & 186.6641376 & -65.2182060 & 300.3368129 & -2.4713988 & 18.52 & A+B \\ 
202 & 4077868965615513728 & 278.4587811 & -23.2755943 & 10.1951268 & -6.7450647 & 18.53 & A \\ 
203 & 4053910130560072704 & 265.7712325 & -33.3661346 & 355.9401137 & -1.8551335 & 18.53 & B \\ 
204 & 5877397384329104896 & 221.5741612 & -61.8966204 & 316.0204839 & -1.9793106 & 18.54 & A+B \\ 
205 & 5977665977162203136 & 255.4183781 & -35.2093177 & 349.5689039 & 4.1403689 & 18.55 & A \\ 
206 & 4053778184815517952 & 264.5381698 & -33.2013488 & 355.5335544 & -0.8936310 & 18.55 & B \\ 
207 & 5941348657100589824 & 248.7013705 & -47.0729817 & 337.2362231 & 0.3697771 & 18.56 & B \\ 
208 & 4063951867199966336 & 270.2247985 & -26.8641371 & 3.4994252 & -1.8559324 & 18.57 & A \\ 
209 & 4064181184057707136 & 270.2787371 & -25.7035632 & 4.5321502 & -1.3234119 & 18.57 & A \\ 
210 & 4058783097358064640 & 263.1718155 & -29.3457136 & 358.1483266 & 2.1713445 & 18.58 & B \\ 
211 & 5865028909305170176 & 202.9240268 & -63.6247581 & 307.3854205 & -1.1016564 & 18.58 & A \\ 
212 & 4053432770693996416 & 265.6344551 & -34.1037121 & 355.2523331 & -2.1458114 & 18.59 & B \\ 
213 & 4068873139592389376 & 267.4793217 & -23.0105532 & 5.5731132 & 2.2361845 & 18.61 & A \\ 
214 & 4063323530616960768 & 270.9486112 & -26.6014298 & 4.0468811 & -2.2890517 & 18.61 & B \\ 
215 & 4065065255833603968 & 274.2337533 & -25.7811786 & 6.1854205 & -4.4810869 & 18.63 & B \\ 
216 & 4116567308158603392 & 265.0202866 & -22.8762171 & 4.5087188 & 4.2403523 & 18.63 & B \\ 
217 & 4318866731734245504 & 295.5895723 & 16.3122565 & 53.2380609 & -3.4483504 & 18.63 & A \\ 
218 & 5870978096164685952 & 208.3506873 & -57.8367292 & 311.1265654 & 4.0457624 & 18.64 & A \\ 
219 & 4059802172898190976 & 260.3688199 & -28.7364750 & 357.2933063 & 4.5478290 & 18.65 & B \\ 
220 & 4096124500776996352 & 275.3708866 & -18.6687694 & 12.9693965 & -2.0671452 & 18.65 & A \\ 
221 & 4042284508764889600 & 269.0137070 & -34.6179441 & 356.2404416 & -4.8052667 & 18.66 & B \\ 
222 & 5884001227334516992 & 237.8343316 & -56.5594625 & 325.8686561 & -1.9225723 & 18.67 & A \\ 
223 & 5940488259212239488 & 248.8052843 & -49.5274975 & 335.4739259 & -1.3402478 & 18.67 & B \\ 
224 & 5965732938772861056 & 258.6423019 & -42.2448818 & 345.4308899 & -2.0805728 & 18.68 & A \\ 
225 & 4041307975063716608 & 265.7005799 & -35.1805268 & 354.3637083 & -2.7576021 & 18.68 & B \\ 
226 & 4061840461210116096 & 265.2607658 & -26.2210375 & 1.7792836 & 2.2882464 & 18.69 & B \\ 
227 & 4058004814930630912 & 262.3591087 & -30.9817501 & 356.3937766 & 1.8620678 & 18.70 & B \\ 
228 & 4100268594454909184 & 280.3173315 & -16.0416501 & 17.4812714 & -5.0524157 & 18.72 & A \\ 
229 & 5970981388806648960 & 251.5222616 & -38.5162735 & 345.0255266 & 4.4849943 & 18.72 & A+B \\ 
230 & 4050871488475400576 & 272.6729789 & -28.2520475 & 3.3422057 & -4.4272717 & 18.73 & B \\ 
231 & 6028410053760056576 & 255.3683853 & -31.3316812 & 352.6326028 & 6.5349870 & 18.73 & B \\ 
232 & 4096895979980830592 & 276.2916643 & -17.1857921 & 14.6910688 & -2.1478918 & 18.75 & B \\ 
233 & 4041902669088769280 & 267.4801630 & -33.4418828 & 356.6151461 & -3.1145675 & 18.76 & B \\ 
234 & 4067161612158838400 & 266.2195393 & -25.5506645 & 2.8033588 & 1.9060202 & 18.76 & A \\ 
235 & 5861647125082723200 & 193.4373693 & -65.2628395 & 303.1739466 & -2.3921771 & 18.77 & B \\ 
236 & 4068464640811687808 & 266.9737027 & -24.2389535 & 4.2804816 & 2.0017950 & 18.78 & B \\ 
237 & 5980759826773028608 & 258.7924340 & -30.3354372 & 355.1903571 & 4.7581095 & 18.78 & B \\ 
238 & 6029527672875998720 & 254.3123343 & -29.8863200 & 353.2277616 & 8.1370678 & 18.78 & A \\ 
239 & 4067933920317815808 & 265.8492725 & -25.3116082 & 2.8323558 & 2.3160490 & 18.79 & A+B \\ 
240 & 4060352787652332544 & 263.1659130 & -28.7975986 & 358.6055622 & 2.4742242 & 18.79 & A \\ 
241 & 4043141200181979392 & 268.5875724 & -33.5900499 & 356.9560801 & -3.9855480 & 18.80 & B \\ 
242 & 4049378115441315584 & 273.8580908 & -30.7063871 & 1.6577039 & -6.4951911 & 18.80 & B \\ 
243 & 4068349462749897088 & 266.2994753 & -23.9230249 & 4.2316761 & 2.6919627 & 18.81 & B \\ 
244 & 4058980940754769792 & 260.1012261 & -30.4782961 & 355.7225681 & 3.7517117 & 18.81 & A \\ 
245 & 6057162729441951360 & 193.4091801 & -58.7124212 & 303.2181457 & 4.1581056 & 18.82 & A \\ 
246 & 4320634776479693696 & 288.3573682 & 15.5360259 & 49.2335302 & 2.2912531 & 18.83 & A \\ 
247 & 5962474540419479040 & 264.8557139 & -36.7322801 & 352.6825964 & -2.9954043 & 18.83 & A \\ 
248 & 4043628764929570688 & 268.3669667 & -31.8911006 & 358.3324924 & -2.9690029 & 18.83 & B \\ 
249 & 4311179908525129088 & 282.6843949 & 9.1368055 & 40.9996588 & 4.3065008 & 18.85 & A \\ 
250 & 5982502415238669184 & 238.6085150 & -50.6298201 & 329.9810034 & 2.3747915 & 18.86 & A+B \\ 
251 & 5961927460317472512 & 262.1180142 & -39.2356138 & 349.4059653 & -2.5380687 & 18.87 & A \\ 
252 & 4055539469311609856 & 267.8331699 & -32.0415212 & 357.9726311 & -2.6551366 & 18.87 & B \\ 
253 & 4060412023879926144 & 263.7449633 & -28.1118721 & 359.4585551 & 2.4183364 & 18.88 & B \\ 
254 & 4062565073796700288 & 269.6173841 & -28.8609334 & 1.4982126 & -2.3814400 & 18.90 & A \\ 
255 & 4145925372472941568 & 271.9126788 & -15.7912663 & 13.9203002 & 2.2087043 & 18.91 & B \\ 
256 & 5931393098987729024 & 246.2183097 & -56.2729232 & 329.4902619 & -4.8372469 & 18.91 & A \\ 
257 & 3388787058343659648 & 78.1420400 & 13.4505753 & 188.9801077 & -14.8251729 & 18.92 & A \\ 
258 & 4050797615070566016 & 271.7191162 & -28.6009556 & 2.6311834 & -3.8593846 & 18.92 & B \\ 
259 & 5599913394301352832 & 116.0932117 & -28.4434223 & 243.9996825 & -2.1797752 & 18.94 & A+B \\ 
260 & 4252127887751777536 & 283.1794554 & -7.3671497 & 26.5116769 & -3.6349108 & 18.94 & A \\ 
261 & 4107297703717778688 & 258.8323075 & -29.6422735 & 355.7779904 & 5.1308321 & 18.95 & A+B \\ 
262 & 4053379891096922496 & 264.3814687 & -34.2998834 & 354.5357269 & -1.3716350 & 18.97 & B \\ 
263 & 4167774244713700864 & 262.4921958 & -8.1659576 & 15.9435442 & 13.9951464 & 18.97 & A \\ 
264 & 6033868957082637824 & 252.4820955 & -27.3937210 & 354.2065846 & 10.9407916 & 18.98 & A \\ 
265 & 4063248759576423936 & 271.5532851 & -26.8501954 & 4.0935600 & -2.8818337 & 18.99 & B \\ 
266 & 4100893013996898688 & 283.0916163 & -15.7178256 & 18.9702360 & -7.2960920 & 19.01 & B \\ 
267 & 5858050858769509888 & 201.1836388 & -65.1703815 & 306.4206645 & -2.5244209 & 19.06 & B \\ 
268 & 4060325815241951104 & 264.1173268 & -28.6325062 & 359.1962827 & 1.8618181 & 19.08 & B \\ 
269 & 4041281792784924928 & 265.6908267 & -35.5418934 & 354.0514810 & -2.9403699 & 19.08 & B \\ 
270 & 4060231733966232704 & 264.4790652 & -28.5588935 & 359.4288328 & 1.6331970 & 19.09 & B \\ 
271 & 5937557472277641088 & 251.7296056 & -50.2151296 & 336.2042344 & -3.2089269 & 19.09 & B \\ 
272 & 4058601093790902528 & 262.3608192 & -30.3921088 & 356.8867461 & 2.1860490 & 19.10 & B \\ 
273 & 4060009705621629696 & 264.3877009 & -29.6482630 & 358.4657249 & 1.1173850 & 19.10 & B \\ 
274 & 4041943836257278464 & 266.8802288 & -33.2807021 & 356.4958512 & -2.6018684 & 19.11 & B \\ 
275 & 5964077937551918720 & 254.8193304 & -45.0878700 & 341.4955974 & -1.5782302 & 19.11 & B \\ 
276 & 5874827413327237888 & 224.5879126 & -61.5869750 & 317.4355984 & -2.3421590 & 19.16 & B \\ 
277 & 5944591670951444224 & 247.5192868 & -42.9198069 & 339.7081212 & 3.8115889 & 19.16 & B \\ 
278 & 4040104834334974336 & 265.8576492 & -37.0968957 & 352.7943555 & -3.8684538 & 19.17 & B \\ 
279 & 5953319864813459200 & 260.0530609 & -43.9004978 & 344.6707385 & -3.8768375 & 19.20 & B \\ 
280 & 4041972281959897216 & 267.1848454 & -33.1022523 & 356.7804102 & -2.7284708 & 19.20 & B \\ 
281 & 4041682221271752576 & 267.7460167 & -33.9717257 & 356.2716334 & -3.5750520 & 19.21 & B \\ 
282 & 5970276223900786560 & 255.8049039 & -38.6010649 & 347.0664085 & 1.8263375 & 19.22 & B \\ 
283 & 4067117086266567680 & 266.1584137 & -25.9203596 & 2.4590897 & 1.7600529 & 19.22 & B \\ 
284 & 4061109904516035840 & 265.5054389 & -26.2074578 & 1.9067124 & 2.1089310 & 19.22 & B \\ 
285 & 5962697054058763776 & 263.9737431 & -36.6927566 & 352.3377466 & -2.3767062 & 19.22 & B \\ 
286 & 4061008783736781696 & 265.7825648 & -26.7868344 & 1.5441519 & 1.5932994 & 19.23 & B \\ 
287 & 4041546122348486912 & 266.9522307 & -34.5963319 & 355.3977891 & -3.3312211 & 19.25 & B \\ 
288 & 4061793388415544064 & 264.4527617 & -26.4140345 & 1.2305039 & 2.7993500 & 19.26 & B \\ 
289 & 5865115878111729408 & 200.7317435 & -63.7748718 & 306.4025100 & -1.1155151 & 19.27 & B \\ 
290 & 4064049685012894080 & 270.0333401 & -26.3194252 & 3.8877792 & -1.4372100 & 19.28 & B \\ 
291 & 5875465478011284608 & 229.1304472 & -61.0728283 & 319.5831859 & -2.9820985 & 19.29 & B \\ 
292 & 6029920267243932288 & 256.4340696 & -28.9850662 & 355.0792038 & 7.2097681 & 19.29 & B \\ 
293 & 4043812795636739200 & 270.7573594 & -31.9491642 & 359.2914624 & -4.7590857 & 19.31 & B \\ 
294 & 5861063421844974464 & 190.3513164 & -64.9059182 & 301.8677851 & -2.0549007 & 19.32 & B \\ 
295 & 4060376530237310080 & 263.4426601 & -28.4428715 & 359.0357556 & 2.4630811 & 19.33 & B \\ 
296 & 4254536642846675200 & 285.1751116 & -4.8791287 & 29.6355743 & -4.2789489 & 19.35 & B \\ 
297 & 5940833677625129472 & 249.4996873 & -49.3354686 & 335.9197382 & -1.5456783 & 19.37 & B \\ 
298 & 5967486075708158464 & 251.5272560 & -43.5605596 & 341.1828667 & 1.2109590 & 19.39 & B \\ 
299 & 5254164713048184192 & 161.6595814 & -60.7779843 & 288.2776961 & -1.5074606 & 19.41 & B \\ 
300 & 4060931336940814848 & 264.6497852 & -27.2646238 & 0.6043286 & 2.1968058 & 19.43 & B \\ 
301 & 5952994379322960384 & 258.6512536 & -45.0777922 & 343.1291090 & -3.7362420 & 19.46 & B \\ 
302 & 5932291262591322240 & 245.5826680 & -53.3436198 & 331.3248691 & -2.5213563 & 19.49 & B \\ 
303 & 4068560848199951360 & 267.4142708 & -23.5781787 & 5.0547817 & 1.9964602 & 19.51 & B \\ 
304 & 4041946924392819712 & 267.1120030 & -33.4274199 & 356.4698201 & -2.8433852 & 19.52 & B \\ 
305 & 4043602754521124224 & 268.2791404 & -32.2625389 & 357.9739153 & -3.0926412 & 19.54 & B \\ 
306 & 5875751076156980096 & 232.1402251 & -60.2554703 & 321.2669883 & -3.1024067 & 19.56 & B \\ 
307 & 4068552120771025152 & 267.2345688 & -23.7314823 & 4.8385648 & 2.0589500 & 19.56 & B \\ 
308 & 5932671414400913024 & 240.7127685 & -54.7195771 & 328.2975554 & -1.5643581 & 19.57 & B \\ 
309 & 4066693121448848384 & 272.7051515 & -22.6155035 & 8.3090903 & -1.7481540 & 19.57 & B \\ 
310 & 4067940826629400832 & 266.2870015 & -25.1348595 & 3.1902151 & 2.0707429 & 19.59 & B \\ 
311 & 4060071931141906688 & 263.9390906 & -29.4230961 & 358.4457957 & 1.5675481 & 19.60 & B \\ 
312 & 4053693114402109568 & 264.3047624 & -33.7152971 & 354.9951808 & -1.0045140 & 19.60 & B \\ 
313 & 4252208083428417024 & 283.4260287 & -7.1046541 & 26.8571522 & -3.7342959 & 19.61 & B \\ 
314 & 4108068087628927488 & 260.8122441 & -27.3216182 & 358.6878757 & 5.0208279 & 19.62 & B \\ 
315 & 5861145095000219904 & 189.9136213 & -64.4963691 & 301.6634513 & -1.6536459 & 19.64 & B \\ 
316 & 4055455219226699904 & 263.9980884 & -30.4539634 & 357.6048868 & 0.9687776 & 19.66 & B \\ 
317 & 4043093234064618240 & 271.3161066 & -31.6786351 & 359.7601605 & -5.0428975 & 19.66 & B \\ 
318 & 5338268560822425472 & 162.2437459 & -60.3139219 & 288.3209706 & -0.9634122 & 19.82 & B \\ 
319 & 4042335430079572480 & 269.1051390 & -34.4097249 & 356.4592976 & -4.7669692 & 19.83 & B \\ 
320 & 4203879874428941568 & 283.8803402 & -8.3215786 & 25.9720891 & -4.6853379 & 19.85 & B \\ 
321 & 4059443521610603136 & 261.8227707 & -29.6841245 & 357.2191397 & 2.9657481 & 19.85 & B \\ 
322 & 4037023521767456384 & 270.0894880 & -37.7201964 & 353.9532983 & -7.0863885 & 19.89 & B \\ 
323 & 4118603058326233856 & 267.3982118 & -21.2549663 & 7.0442273 & 3.1997564 & 19.91 & B \\ 
324 & 4058624355438550400 & 262.5552701 & -30.2487329 & 357.0990038 & 2.1248798 & 19.91 & B \\ 
325 & 5877273865374753152 & 219.2400503 & -62.8697034 & 314.6328332 & -2.4202255 & 19.93 & B \\ 
326 & 4041928271289275776 & 266.6457045 & -33.6365566 & 356.0896677 & -2.6183641 & 19.94 & B \\ 
327 & 4044021466548302208 & 268.9452222 & -31.7086935 & 358.7380941 & -3.3018391 & 19.94 & B \\ 
328 & 4043964322079279616 & 269.4002287 & -31.7925309 & 358.8589454 & -3.6789060 & 19.96 & B \\ 
329 & 4064003819077926016 & 270.2593267 & -26.3848730 & 3.9313389 & -1.6455239 & 19.97 & B \\ 
330 & 5334183595902105088 & 173.2937682 & -61.8896611 & 293.8530798 & -0.4052851 & 19.99 & B \\ 
331 & 4060858769126041344 & 264.9230327 & -27.6781376 & 0.3831575 & 1.7714696 & 20.00 & B \\ 
332 & 4060485828545912448 & 264.7081954 & -27.7289294 & 0.2388613 & 1.9054826 & 20.03 & B \\ 
333 & 5836614509548423680 & 239.6170816 & -55.1229348 & 327.5572054 & -1.4573336 & 20.04 & B \\ 
334 & 5861190587320895104 & 186.9168262 & -65.1816705 & 300.4389850 & -2.4250434 & 20.07 & B \\ 
335 & 4070219869407584768 & 268.1280807 & -22.6482848 & 6.1901924 & 1.9068253 & 20.08 & B \\ 
336 & 4092025491399809024 & 278.3416591 & -20.0957119 & 12.9983287 & -5.2116735 & 20.10 & B \\ 
337 & 5892416408213499776 & 216.6411098 & -57.1027661 & 315.6061919 & 3.4095751 & 20.10 & B \\ 
338 & 4042509534992606592 & 269.9602053 & -33.4604733 & 357.6393918 & -4.9140219 & 20.11 & B \\ 
339 & 5867019781264970496 & 213.1949603 & -59.5612100 & 313.0766512 & 1.6988310 & 20.14 & B \\ 
340 & 5932658533818012672 & 241.3674847 & -54.8080765 & 328.5215242 & -1.8812174 & 20.15 & B \\ 
341 & 4037003833628097920 & 269.4667028 & -37.9831402 & 353.4806264 & -6.7850622 & 20.19 & B \\ 
342 & 4041980395080048384 & 267.1631349 & -32.8566449 & 356.9819395 & -2.5865977 & 20.22 & B \\ 
343 & 4043408626233506944 & 267.8664022 & -33.1351658 & 357.0442820 & -3.2359067 & 20.23 & B \\ 
344 & 4068804248278621568 & 266.4129243 & -23.2555527 & 4.8567259 & 2.9500097 & 20.25 & B \\ 
345 & 4107306813251999360 & 258.4762371 & -29.6098498 & 355.6244981 & 5.4019286 & 20.26 & B \\ 
346 & 4070231689172730752 & 268.2801710 & -22.4708390 & 6.4144999 & 1.8759243 & 20.29 & B \\ 
347 & 5974617649856579456 & 263.0133009 & -36.6151121 & 351.9848738 & -1.6872397 & 20.33 & B \\ 
348 & 4060408862783337344 & 263.6466765 & -28.2434761 & 359.3008625 & 2.4201192 & 20.36 & B \\ 
349 & 5978422166252280192 & 255.8815186 & -33.5983108 & 351.0848963 & 4.8164961 & 20.37 & B \\ 
350 & 5977975730251951872 & 256.0464056 & -35.4189590 & 349.7144075 & 3.6060507 & 20.40 & B \\ 
351 & 4069377716512265728 & 268.6305482 & -23.2845096 & 5.8761169 & 1.1855468 & 20.43 & B \\ 
352 & 4064071061114240128 & 270.0497131 & -26.1174772 & 4.0704295 & -1.3497038 & 20.47 & B \\ 
353 & 4041642707503630208 & 267.6883747 & -34.2447902 & 356.0117680 & -3.6729389 & 20.49 & B \\ 
354 & 5836498614158612992 & 239.8515774 & -55.6005476 & 327.3480434 & -1.9069641 & 20.53 & B \\ 
355 & 5877288124678597760 & 219.3881478 & -62.7202007 & 314.7544098 & -2.3098646 & 20.59 & B \\ 
356 & 4059256780768511616 & 259.5160288 & -29.9342367 & 355.8809700 & 4.4775261 & 20.60 & B \\ 
357 & 4054189784460154624 & 265.0965831 & -32.7661462 & 356.1514651 & -1.0582661 & 20.68 & B \\ 
358 & 5962744333088467968 & 264.4586010 & -36.0892730 & 353.0570410 & -2.3824658 & 20.72 & B \\ 
359 & 4041164935329120128 & 265.1614358 & -35.9310040 & 353.4937286 & -2.7799171 & 20.72 & B \\ 
360 & 4054019738115896064 & 266.0015655 & -32.9409610 & 356.4035158 & -1.7965279 & 20.76 & B \\ 
361 & 5932749758903448448 & 240.2786925 & -54.1545852 & 328.4790714 & -0.9727373 & 20.76 & B \\ 
362 & 5938866067211518592 & 254.7067828 & -48.4276958 & 338.8201608 & -3.5826396 & 20.95 & B \\ 
363 & 4067300223660786304 & 266.9193778 & -25.0204118 & 3.5854498 & 1.6404974 & 20.97 & B \\ 

%% file: table-sort-id-xm.tex
007 & OGLE-2015-BLG-0064\\
008 & BLG896.14.123\\
010 & OGLE-2016-BLG-0133 KMT-2016-BLG-0485\\
011 & BLG586.01.37855\\
013 & OGLE-2015-BLG-1755\\
014 & BLG513.13.37217 OGLE-2015-BLG-0056\\
018 & BLG513.19.77649 OGLE-2016-BLG-1396 KMT-2016-BLG-0738\\
021 & OGLE-2016-BLG-0357\\
023 & GD2233.02.17021 ASASSN-16oe\\
025 & BLG921.22.94\\
026 & GD2110.04.10077\\
029 & BLG518.22.135316 OGLE-2015-BLG-1578 MOA-2015-BLG-338\\
030 & BLG523.06.68137 OGLE-2014-BLG-1948 MOA-2014-BLG-594\\
031 & BLG513.25.80350 OGLE-2016-BLG-1654 MOA-2016-BLG-457 KMT-2016-BLG-0698\\
032 & ASASSN-V J044558.57+081444.6\\
035 & OGLE-2015-BLG-0128 MOA-2015-BLG-004\\
037 & BLG507.19.46958 OGLE-2017-BLG-0033\\
039 & BLG509.32.88384 OGLE-2016-BLG-1790 KMT-2016-BLG-1370\\
041 & OGLE-2016-BLG-1571\\
043 & OGLE-2015-BLG-0145\\
046 & BLG522.16.33098 OGLE-2016-BLG-0385 MOA-2016-BLG-090 KMT-2016-BLG-1427\\
047 & KMT-2015-BLG-0157\\
048 & BLG529.14.92394 OGLE-2015-BLG-2098 MOA-2015-BLG-526\\
049 & BLG573.31.86116 OGLE-2016-BLG-0322 MOA-2016-BLG-028 KMT-2016-BLG-2093\\
050 & OGLE-2014-BLG-1991\\
053 & BLG513.05.95050 OGLE-2017-BLG-0175 MOA-2017-BLG-071 KMT-2017-BLG-0595\\
054 & MOA-2015-BLG-033\\
055 & BLG523.31.89572 OGLE-2016-BLG-0085 KMT-2016-BLG-1422\\
056 & BLG527.15.55557 OGLE-2015-BLG-0473 MOA-2015-BLG-082\\
059 & OGLE-2015-BLG-1466\\
061 & OGLE-2016-BLG-1864 KMT-2016-BLG-1559\\
062 & GD1127.32.33414\\
064 & BLG515.25.78 OGLE-2016-BLG-1866 MOA-2016-BLG-586 KMT-2016-BLG-1536\\
066 & BLG526.16.27934 OGLE-2015-BLG-0862\\
067 & OGLE-2015-BLG-0149 KMT-2016-BLG-1533\\
068 & BLG660.18.34262 OGLE-2017-BLG-0140 KMT-2017-BLG-0872\\
069 & MOA-2015-BLG-158 KMT-2016-BLG-1346\\
070 & BLG572.15.1672\\
071 & MOA-2017-BLG-051\\
073 & OGLE-2015-BLG-1579 MOA-2015-BLG-398\\
074 & BLG935.14.35174\\
076 & BLG503.26.102570 OGLE-2017-BLG-0144 MOA-2017-BLG-001\\
077 & BLG518.23.90006 OGLE-2016-BLG-0079 MOA-2016-BLG-023 KMT-2016-BLG-0260\\
078 & BLG516.11.82635 OGLE-2014-BLG-1945\\
079 & OGLE-2015-BLG-0729 MOA-2015-BLG-167 KMT-2015-BLG-0615\\
080 & BLG535.18.21961 OGLE-2015-BLG-0255 MOA-2015-BLG-051 KMT-2015-BLG-0138\\
081 & DG1068.02.19241\\
083 & OGLE-2016-BLG-0125\\
085 & BLG523.03.48248 OGLE-2015-BLG-0877 MOA-2015-BLG-216\\
086 & BLG637.25.39934 OGLE-2015-BLG-0456\\
087 & BLG503.23.135537 OGLE-2017-BLG-0143 MOA-2017-BLG-032 KMT-2017-BLG-1090\\
088 & OGLE-2016-BLG-0293 MOA-2016-BLG-147 KMT-2016-BLG-1296\\
089 & OGLE-2016-BLG-0985 KMT-2016-BLG-1569\\
090 & BLG513.16.48941 OGLE-2015-BLG-2131\\
091 & OGLE-2015-BLG-2112 MOA-2016-BLG-031 KMT-2016-BLG-0259\\
092 & BLG524.32.133709 OGLE-2015-BLG-1479\\
093 & BLG889.16.283\\
094 & BLG661.12.38593 OGLE-2016-BLG-1299 KMT-2016-BLG-1101\\
095 & BLG897.28.315\\
097 & BLG639.30.78829 OGLE-2017-BLG-0260\\
098 & BLG609.28.76 OGLE-2017-BLG-0207 KMT-2017-BLG-1397\\
099 & GD1108.21.19118\\
101 & BLG527.30.75497 OGLE-2014-BLG-1587 MOA-2014-BLG-528\\
102 & GD1137.01.17177\\
103 & GD1102.30.23357\\
104 & BLG523.08.47678 OGLE-2016-BLG-1762\\
105 & Gaia17bej\\
108 & BLG599.22.1393 OGLE-2015-BLG-0233\\
109 & BLG645.04.59638 OGLE-2016-BLG-0025\\
110 & OGLE-2016-BLG-1820\\
111 & DG1029.23.41205\\
113 & OGLE-2016-BLG-1260 MOA-2016-BLG-433 KMT-2016-BLG-0526\\
114 & BLG612.10.58071 OGLE-2015-BLG-1656\\
115 & BLG613.25.83084 OGLE-2017-BLG-0081 KMT-2017-BLG-2162\\
116 & BLG531.21.154747 OGLE-2016-BLG-1366\\
118 & GD1793.08.3677 ASASSN-16li\\
121 & BLG661.19.35266 OGLE-2015-BLG-0224\\
122 & GD1382.18.3097\\
124 & BLG638.06.43147 OGLE-2015-BLG-0183\\
125 & OGLE-2017-BLG-0065 MOA-2017-BLG-019 KMT-2017-BLG-1553\\
126 & GD1262.09.28593\\
127 & BLG518.27.142401 OGLE-2017-BLG-0060 KMT-2017-BLG-1228\\
128 & BLG535.11.49818 OGLE-2017-BLG-0503 MOA-2017-BLG-174 KMT-2017-BLG-1116\\
130 & BLG545.09.7367 OGLE-2016-BLG-0336 MOA-2016-BLG-092 KMT-2016-BLG-1406\\
131 & BLG645.04.20686 MOA-2017-BLG-004\\
132 & OGLE-2016-BLG-1641 MOA-2016-BLG-534 KMT-2016-BLG-1070\\
134 & BLG610.13.87104 OGLE-2016-BLG-0464\\
138 & GD1159.28.34506\\
139 & BLG499.05.1914\\
140 & GD1160.13.1131\\
142 & OGLE-2015-BLG-0642\\
143 & BLG885.27.55275\\
144 & BLG621.18.15694 OGLE-2016-BLG-1883\\
145 & BLG603.29.135144 OGLE-2016-BLG-1861\\
146 & BLG502.24.72636 OGLE-2016-BLG-0407\\
147 & BLG622.19.48941 OGLE-2016-BLG-0361\\
148 & BLG613.12.125877 OGLE-2014-BLG-1521\\
149 & BLG652.11.51046 OGLE-2016-BLG-0313 KMT-2016-BLG-0248\\
151 & BLG530.17.93206 OGLE-2015-BLG-0347\\
154 & BLG515.19.84730 OGLE-2016-BLG-0828\\
155 & BLG515.06.136119 OGLE-2016-BLG-0266 MOA-2016-BLG-003 KMT-2016-BLG-1530\\
156 & BLG507.10.125254 MOA-2014-BLG-526\\
158 & Gaia16bnn\\
159 & BLG544.21.82617 OGLE-2015-BLG-0427\\
162 & BLG528.32.1070\\
165 & BLG645.26.75287 OGLE-2014-BLG-1059\\
167 & BLG610.19.30823 OGLE-2016-BLG-0356 KMT-2016-BLG-1629\\
168 & GD1127.14.39807\\
169 & BLG654.02.28137 OGLE-2017-BLG-0100\\
171 & BLG652.30.94823 KMT-2017-BLG-0851\\
173 & BLG523.09.72801 OGLE-2015-BLG-0607\\
175 & BLG615.12.122 OGLE-2016-BLG-0689\\
176 & BLG605.20.37253 OGLE-2016-BLG-0311\\
177 & GD1144.30.787\\
178 & BLG656.12.1967 MOA-2015-BLG-052\\
179 & GD1093.07.26917\\
182 & BLG610.15.14750 OGLE-2014-BLG-1608\\
184 & BLG585.07.7110\\
185 & OGLE-2016-BLG-0399\\
186 & BLG660.12.67091 OGLE-2016-BLG-1873 KMT-2016-BLG-1051\\
187 & BLG498.16.2520 OGLE-2017-BLG-0323\\
189 & BLG633.07.23251 OGLE-2017-BLG-0516\\
190 & BLG673.28.27627\\
191 & BLG545.32.45815 OGLE-2015-BLG-0684\\
192 & BLG661.11.40559 OGLE-2017-BLG-0242 KMT-2017-BLG-0929\\
193 & BLG662.21.79580 OGLE-2015-BLG-1841\\
195 & BLG769.23.732\\
196 & BLG662.19.60408 OGLE-2015-BLG-1949\\
198 & BLG675.25.90179 OGLE-2014-BLG-0993\\
200 & BLG516.29.58361 OGLE-2016-BLG-1484 MOA-2016-BLG-583 KMT-2016-BLG-1524\\
201 & GD1304.24.3051\\
202 & MOA-2017-BLG-002\\
203 & BLG661.17.20507 OGLE-2015-BLG-0670 KMT-2016-BLG-1088\\
204 & GD1219.15.2476\\
206 & BLG661.32.332 OGLE-2016-BLG-0370\\
207 & GD1114.11.1805\\
208 & OGLE-2015-BLG-1419 MOA-2015-BLG-425\\
209 & OGLE-2016-BLG-1133\\
210 & BLG613.08.71969 OGLE-2017-BLG-0079 KMT-2017-BLG-0766\\
211 & GD1268.32.3277\\
212 & BLG603.25.85063 OGLE-2017-BLG-0157 KMT-2017-BLG-0924 Gaia17bdk\\
214 & BLG646.03.17834 OGLE-2016-BLG-0027 KMT-2016-BLG-0261\\
215 & BLG529.15.95592 OGLE-2015-BLG-1079\\
216 & BLG625.12.100189 OGLE-2016-BLG-0365\\
219 & BLG614.16.62771 OGLE-2016-BLG-0688\\
220 & BLG571.25.2358\\
221 & BLG509.06.60012\\
223 & GD1116.12.4538\\
225 & BLG604.25.52475 OGLE-2016-BLG-1382 KMT-2016-BLG-1599\\
226 & BLG652.12.71950 OGLE-2016-BLG-0871 KMT-2016-BLG-0943\\
227 & BLG662.05.6243 OGLE-2015-BLG-0211\\
229 & GD2150.13.1140\\
230 & BLG519.09.44226\\
231 & BLG885.18.55350\\
232 & BLG570.28.2627\\
233 & BLG502.23.135466 OGLE-2016-BLG-0150 MOA-2016-BLG-019 KMT-2016-BLG-2501\\
235 & GD1292.21.4478\\
236 & BLG633.26.16850 OGLE-2016-BLG-0091 KMT-2016-BLG-0250\\
237 & BLG617.05.57777 OGLE-2017-BLG-0165\\
239 & BLG714.08.34194\\
240 & KMT-2016-BLG-0868\\
241 & BLG502.08.145185 OGLE-2017-BLG-0015 MOA-2017-BLG-070\\
242 & BLG526.28.2567 OGLE-2016-BLG-0495\\
243 & BLG632.05.19651 OGLE-2017-BLG-0355 KMT-2017-BLG-2002\\
244 & BLG499.12.6674\\
248 & BLG535.27.9299 OGLE-2017-BLG-0132 KMT-2017-BLG-1550\\
250 & GD1161.09.16949\\
252 & BLG535.22.22503 OGLE-2017-BLG-0543 KMT-2017-BLG-0140\\
253 & BLG653.30.42787 OGLE-2015-BLG-0190\\
255 & BLG767.14.27556\\
258 & BLG519.06.147922 OGLE-2016-BLG-0081 KMT-2016-BLG-1479\\
259 & GD1606.28.7724 Gaia17aqu\\
261 & BLG617.22.38934 OGLE-2015-BLG-0455\\
262 & BLG609.32.43624 OGLE-2016-BLG-0534\\
265 & BLG518.15.121322 OGLE-2015-BLG-0254 MOA-2015-BLG-171\\
266 & Gaia17bbi\\
267 & GD1274.05.9905\\
268 & BLG653.11.27375 KMT-2016-BLG-0852\\
269 & BLG604.16.38216 OGLE-2015-BLG-0158\\
270 & BLG653.18.3068 OGLE-2016-BLG-0190 KMT-2016-BLG-0892\\
271 & GD1109.08.2085\\
272 & BLG662.22.1326 OGLE-2016-BLG-1811 KMT-2016-BLG-0813\\
273 & BLG654.18.70902 OGLE-2017-BLG-0085 KMT-2017-BLG-0736\\
274 & BLG660.02.75339 OGLE-2015-BLG-0194\\
275 & GD1082.31.26433\\
276 & GD1213.04.5682\\
277 & GD1111.25.22363\\
278 & BLG605.07.15081 OGLE-2015-BLG-0667\\
279 & Gaia17ahl\\
280 & BLG660.08.86517 OGLE-2015-BLG-0808\\
281 & BLG502.05.67258 OGLE-2016-BLG-1509 KMT-2016-BLG-1386\\
282 & BLG998.08.18494\\
283 & BLG683.07.41367\\
284 & BLG652.11.33676 OGLE-2017-BLG-0286 KMT-2017-BLG-0956\\
285 & BLG672.08.61265\\
286 & BLG675.26.43531 OGLE-2016-BLG-0041\\
287 & BLG603.01.136791 OGLE-2014-BLG-1803 MOA-2014-BLG-565\\
288 & BLG667.01.7909 OGLE-2016-BLG-0122 KMT-2016-BLG-0961\\
289 & GD1273.06.52572\\
290 & BLG646.16.38801 OGLE-2016-BLG-0059 KMT-2016-BLG-1575\\
291 & GD1202.12.28001\\
292 & BLG894.03.9531\\
293 & BLG515.29.111770 OGLE-2014-BLG-1664\\
294 & GD1297.13.4851\\
295 & BLG653.24.13071\\
296 & DG1041.09.1613\\
297 & GD1116.09.42264\\
298 & GD1096.05.24355\\
299 & GD1369.23.25570\\
300 & BLG675.24.114167 OGLE-2016-BLG-0428 KMT-2016-BLG-0978\\
301 & GD1909.29.25476\\
302 & GD1134.05.55989\\
303 & BLG639.31.35102 OGLE-2017-BLG-0261 KMT-2017-BLG-2017\\
304 & BLG502.25.124534 OGLE-2017-BLG-0264 KMT-2017-BLG-0879\\
305 & BLG535.19.77280 OGLE-2015-BLG-0284 MOA-2015-BLG-104\\
306 & GD1188.07.5777\\
307 & BLG639.32.1758 OGLE-2014-BLG-1058\\
308 & GD1157.19.6827\\
309 & BLG543.16.58089 OGLE-2016-BLG-1754 Gaia17bbs\\
310 & BLG633.05.33870 OGLE-2016-BLG-1855 KMT-2016-BLG-1215\\
311 & BLG654.29.32704 OGLE-2017-BLG-0087 KMT-2017-BLG-0747\\
312 & BLG661.16.9749 OGLE-2017-BLG-0213 KMT-2017-BLG-0678\\
313 & DG1030.04.3259\\
314 & BLG921.23.4257 KMT-2017-BLG-0653\\
315 & GD1297.23.38557\\
316 & BLG654.03.51419 OGLE-2015-BLG-1600\\
317 & BLG514.01.66114 OGLE-2015-BLG-0252\\
318 & GD1368.05.3561\\
319 & BLG509.13.122856 OGLE-2015-BLG-0141\\
320 & DG1031.02.9498\\
321 & BLG615.17.48284 OGLE-2017-BLG-0113 KMT-2017-BLG-1942\\
322 & BLG537.17.11402\\
323 & BLG637.31.2736\\
324 & BLG662.21.34275\\
325 & GD1232.09.44403\\
326 & BLG603.28.35150 KMT-2016-BLG-2061\\
327 & BLG507.14.185211 OGLE-2017-BLG-0032 MOA-2017-BLG-021 KMT-2017-BLG-0104\\
328 & BLG507.12.63267 OGLE-2017-BLG-0031 KMT-2017-BLG-0098\\
329 & BLG646.15.2528 OGLE-2015-BLG-0925\\
330 & GD1338.14.2320\\
331 & BLG675.14.31799 OGLE-2016-BLG-1475 KMT-2016-BLG-0970\\
332 & BLG675.07.100337 OGLE-2016-BLG-1862 KMT-2016-BLG-2291\\
333 & GD1157.14.5178\\
334 & GD1304.23.13626\\
335 & BLG638.18.94340 OGLE-2016-BLG-0053\\
336 & BLG563.14.46178\\
337 & GD1227.05.19355\\
338 & BLG508.01.34340 OGLE-2017-BLG-0539\\
339 & GD1242.02.14463 Gaia16aua\\
340 & GD1149.07.11095\\
341 & BLG537.12.70927\\
342 & BLG660.17.94309 OGLE-2017-BLG-0817 KMT-2017-BLG-0871\\
343 & BLG502.29.100629 OGLE-2017-BLG-0095 MOA-2017-BLG-160 KMT-2017-BLG-1123\\
344 & BLG632.21.71948 OGLE-2017-BLG-0563 KMT-2017-BLG-1073\\
345 & BLG617.24.41328 OGLE-2016-BLG-0231\\
346 & BLG642.16.64346\\
347 & BLG672.22.32355 OGLE-2015-BLG-1558\\
348 & BLG653.30.92569 OGLE-2015-BLG-0191\\
349 & BLG898.13.20247\\
350 & BLG900.29.16170\\
351 & BLG643.23.58194 OGLE-2015-BLG-1555\\
352 & BLG646.25.18413 OGLE-2017-BLG-0441 KMT-2017-BLG-1371\\
353 & BLG503.30.103160 OGLE-2017-BLG-0145 MOA-2017-BLG-095 KMT-2017-BLG-2618\\
354 & GD1158.31.47588\\
355 & GD1232.17.23871\\
356 & BLG617.09.66418\\
357 & BLG679.04.38335 KMT-2017-BLG-0867\\
358 & BLG610.14.97071 OGLE-2016-BLG-1727\\
359 & BLG610.20.3895 OGLE-2016-BLG-1408 KMT-2016-BLG-1630\\
360 & BLG660.15.22580 OGLE-2017-BLG-0307 KMT-2017-BLG-0856\\
361 & GD1156.03.7788\\
362 & GD1092.18.44649\\
363 & BLG633.01.52040 OGLE-2017-BLG-0116 KMT-2017-BLG-1029\\